\documentclass[a4paper,11pt]{article}
\pdfoutput=1 

\usepackage{jheppub1} 

\usepackage[T1]{fontenc} 

\usepackage{amsmath,amssymb,amsfonts,graphicx,subfigure,slashed,dsfont,braket}

\def\bea{\begin{eqnarray}}
\def\eea{\end{eqnarray}}
\def\nn{\nonumber}
\def\ba{\begin{array}}
\def\ea{\end{array}}
\def\nn{\nonumber}
\def\Tr{\text{Tr}}

\def\sgn{\text{sgn}}

\def\Pf{\text{Pf}}

\newlength{\fighskip} \fighskip=2pt
\newlength{\figvskip} \figvskip=3pt

\newcommand*{\figbox}[2]{{
  \def\figscale{#1}
  \def\arraystretch{0.8}
  \arraycolsep=0pt
  \begin{array}{c}
    \vbox{\vskip\figscale\figvskip
      \hbox{\hskip\figscale\fighskip
        \includegraphics[scale=\figscale]{#2}}}
  \end{array}}}

\title{\boldmath Holographic measurement and quantum teleportation in the SYK thermofield double}

\author[1]{Stefano Antonini }
\author[2]{, Brianna Grado-White }
\author[2]{, Shao-Kai Jian }
\author[2]{, Brian Swingle }

\affiliation[1]{University of Maryland, College Park, MD, 20742, USA}

\affiliation[2]{Department of Physics, Brandeis University, Waltham, Massachusetts 02453, USA}

\emailAdd{santonin@umd.edu}
\emailAdd{bgradowhite@brandeis.edu}
\emailAdd{skjian@brandeis.edu}
\emailAdd{bswingle@brandeis.edu}

\abstract{According to holography, entanglement is the building block of spacetime; therefore, drastic changes of entanglement will lead to interesting transitions in the dual spacetime.
In this paper, we study the effect of projective measurements on the Sachdev-Ye-Kitaev (SYK) model's thermofield double state, dual to an eternal black hole in Jackiw-Teitelboim (JT) gravity.
We calculate the (Renyi-2) mutual information between the two copies of the SYK model upon projective measurement of a subset of fermions in one copy.
We propose a dual JT gravity model that can account for the change of entanglement due to measurement, and observe an entanglement wedge phase transition in the von Neumann entropy.
The entanglement wedge for the unmeasured side changes from the region outside the horizon to include the entire time reversal invariant slice of the two-sided geometry as the number of measured Majorana fermions increases.
Therefore, after the transition, the bulk information stored in the measured subsystem is not entirely lost upon projection in one copy of the SYK model, but rather teleported to the other copy.
We further propose a decoding protocol to elucidate the teleportation interpretation, and connect our analysis to the physics of traversable wormholes. }

\begin{document} 
\maketitle
\flushbottom

\newpage

\section{Introduction} 

The Anti de-Sitter/Conformal Field Theory (AdS/CFT) correspondence \cite{Maldacena:1997re,Witten:1998qj,Gubser:1998bc,Aharony:1999ti} is one of the most promising frameworks for describing quantum gravity. Within AdS/CFT, and more generally holography, quantum entanglement is deeply interconnected with spacetime geometry, with the structure of boundary entanglement dictating the structure of spacetime~\cite{Ryu2006a,Ryu2006b,Hubeny:2007xt,Swingle:2009bg,VanRaamsdonk:2010pw,Maldacena:2013xja,Engelhardt:2014gca,Dong:2016eik,Harlow:2016vwg}. It is therefore reasonable to expect that radical changes in the entanglement structure of the boundary theory, such as those arising from a projective measurement performed on a subsystem of the boundary, lead to radical modifications of both the dual geometry and the way bulk information is encoded in the boundary theory. A comprehensive description of this phenomenon in AdS$_3$/CFT$_2$ and related tensor network models was recently given in \cite{Antonini:2022sfm} (see also \cite{numasawa2016epr} for earlier work). There, it was found that performing a projective measurement on a boundary subregion $A$ teleports the bulk information contained in $A$ into the complementary unmeasured region $A^c$. In particular, while the projective measurement destroys the entanglement structure of $A$, much of the bulk information encoded in $A$ (i.e. much of the entanglement wedge of $A$) is not erased. Instead, if the mesaurement outcome is known, this information becomes accessible from $A^c$ after the measurement, thanks to the pre-measurement entanglement resource between the two.\footnote{For this reason, we should think of the projective measurement studied in \cite{Antonini:2022sfm} as postselection. We will see in Section \ref{sec:teleportation} that knowledge of the measurement outcome is necessary to implement the teleportation protocol in our SYK setup.}

In this paper, we study a related question in lower-dimensional holography. We consider two copies (the ``right'' and ``left'' sides) of the SYK model \cite{Sachdev:1992fk,kitaev, maldacena2016remarks, maldacena2016conformal} in a thermofield double (TFD) state \cite{maldacena2003eternal,Maldacena:2018lmt,Su:2020zgc}, perform a projective measurement on a subset of $M<N$ Majorana fermions $\psi$ in the right side (where $N$ is the total number of Majorana fermions in each copy), and ask how such a measurement affects the dual two-sided black hole in JT gravity \cite{Teitelboim:1983ux,Jackiw:1984je}. In particular, in analogy with the analysis of Kourkoulou and Maldacena \cite{kourkoulou2017pure} \footnote{See also \cite{Antonini:2021xar} for a generalization to the complex SYK model.}, we employ the fermion parity $M_{k}=-i2\psi_{2k-1}\psi_{2k}$ as our measurement operator. Our results reduce to the ones obtained in \cite{kourkoulou2017pure} in the special case $M=N$, i.e. when all the Majorana fermions are measured in the left side~\footnote{We also study two-sided measurement. See Appendix~\ref{append:syk_two-side}.}.

To better understand how the bulk changes upon boundary measurement, we first study the entanglement structure of the post-measurement state by computing the Renyi-2 mutual information between the left side and the unmeasured fermions in the right side. We find that it is always non-vanishing unless all the fermions in the right side are measured. In other words, the left and right sides always remain entangled with each other, independently of how many fermions are measured. This can be understood as a consequence of the all-to-all interaction of the SYK model (see Section \ref{sec:SYKMI}). 

We then study the holographic dual theory, JT gravity coupled to bulk matter. The bulk matter is described by $N$ copies of a CFT, which should be understood as the bulk dual of the $N$ Majorana fermions in the SYK model. The effect of a measurement performed on a subset of $M$ fermions in the left boundary~\footnote{Note that in our JT gravity analysis we assume the measurement occurs in the left side unlike in the SYK analysis (where it is performed in the right side). This is useful to make a direct comparison with the analysis of Kourkoulou and Maldacena \cite{kourkoulou2017pure}. Clearly, given the symmetry of the TFD state, the two cases are equivalent.
} is to create an end-of-the-world (ETW) brane in the left asymptotic region. In particular, such a brane is visible only to the bulk matter dual to the $M$ measured Majorana fermions, which are now described by a boundary conformal field theory (BCFT)~\cite{cardy1989boundary}. On the other hand, the brane is transparent to the remaining bulk matter, dual to the $N-M$ unmeasured fermions. In this model, we reproduce the results obtained in the SYK model by computing the post-measurement mutual information between the two sides via a prescription for the quantum extremal surface formula proposed in \cite{nezami2021quantum} (and later discussed in \cite{chandrasekaran2022quantum}). This prescription states that the entanglement wedge of a small number of Majorana fermions $K\ll N$ is given by a near-boundary region with spatial size of order $\epsilon$, where $\epsilon$ is the UV cutoff of the bulk theory. A complete analysis of this new prescription will be explored in future work \cite{newqes}. 

Our JT gravity analysis shows that by increasing the number $M$ of measured Majorana fermions in the left boundary, the entanglement wedge of the right boundary can grow to include not just the right asymptotic region, but extend beyond the black hole horizon to include part of the left asymptotic region. Specifically, when the majority of the left Majorana fermions are measured, i.e. $M/N = m >m^*$, a phase transition occurs such that the entanglement wedge of the right system jumps to contain the entire time reversal symmetric slice up to an ETW brane sitting near the left boundary.
The transition can be interpreted as a quantum teleportation process \cite{Antonini:2022sfm}: a projective measurement of most of the left system teleports the bulk information it contains into the right system. Intuitively, this is possible because in the TFD state, a large subset $M \sim N$ of the left fermions is mostly entangled with the right system, and very weakly entangled with the remaining $N-M$ unmeasured fermions on the left. This large entanglement resource between the two sides allows the teleportation to take place.
On the other hand, if only a small subset $m<m^*$ of the left Majorana fermions are measured, the bulk information contained in the $M$ measured fermions is teleported into the unmeasured $N-M$ fermions in the same side. Therefore, the quantum extremal surface is still located at the bifurcation point. 
See Fig.~\ref{fig:summary} for an illustration.
We also construct a simple toy model (see Appendix~\ref{append:random_tfd}) using Haar random unitaries to show that the critical value of $M$ is determined by the pre-measurement entanglement between the two sides. 
Specifically, if the entanglement between the two sides is large, information can be teleported more easily and the transition occurs at a smaller value of $m^*$.

We then proceed to make this bulk teleportation interpretation more precise. 
First, we build a decoding operator able to reconstruct from the right system information that, before the measurement, was contained in the left system. Crucially, the decoding operator depends on the measurement outcome, as expected. 
This measurement-decoding protocol can be understood as a quantum channel acting on the
pre-measurement state. 
Second, we study correlation functions between operators inserted in the two sides in this setup. This quantity is closely related to teleportation fidelity.
In the large-$q$ limit (where $q$ is the number of directly interacting fermions, see Section \ref{sec:pathint}), we find that at low-temperature the left-right correlation function is enhanced to nearly its maximum value by this quantum channel.
We also relate our results to traversable wormhole setups, where the measurement-decoding channel effectively plays the role of the two-sided deformation that makes the wormhole traversable~\cite{gao2017traversable,maldacena2017diving,brown2019quantum,nezami2021quantum,gao2021traversable}. 
Finally, we provide an encoding method that reaches nearly perfect fidelity in this teleportation protocol.
By constructing such an explicit decoding protocol, we put the teleportation and entanglement wedge transition induced by projective measurements on a firmer ground.

\begin{figure}
    \centering
    \subfigure[$m<m_*$]{
    \includegraphics[width=0.45\textwidth]{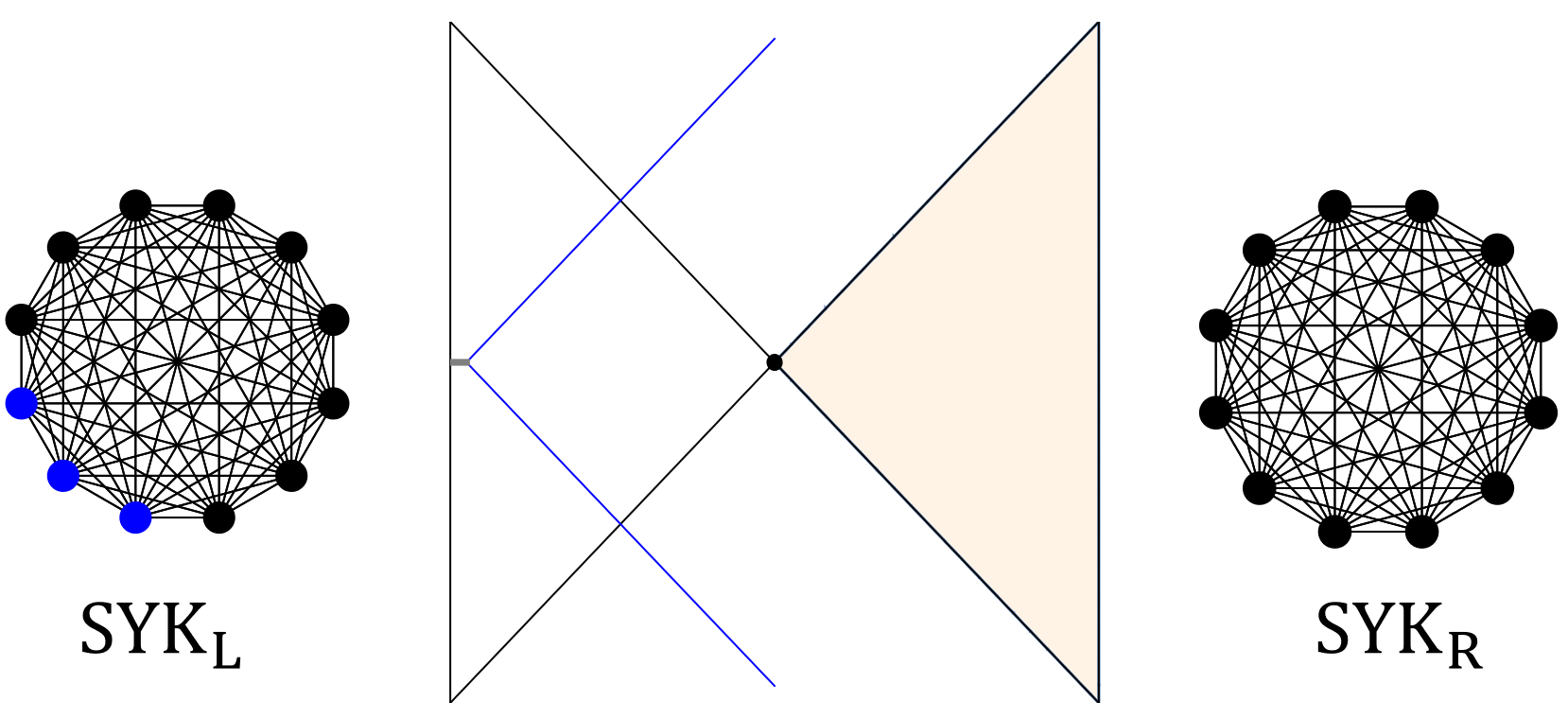}} \qquad
    \subfigure[$m>m_*$]{
    \includegraphics[width=0.46\textwidth]{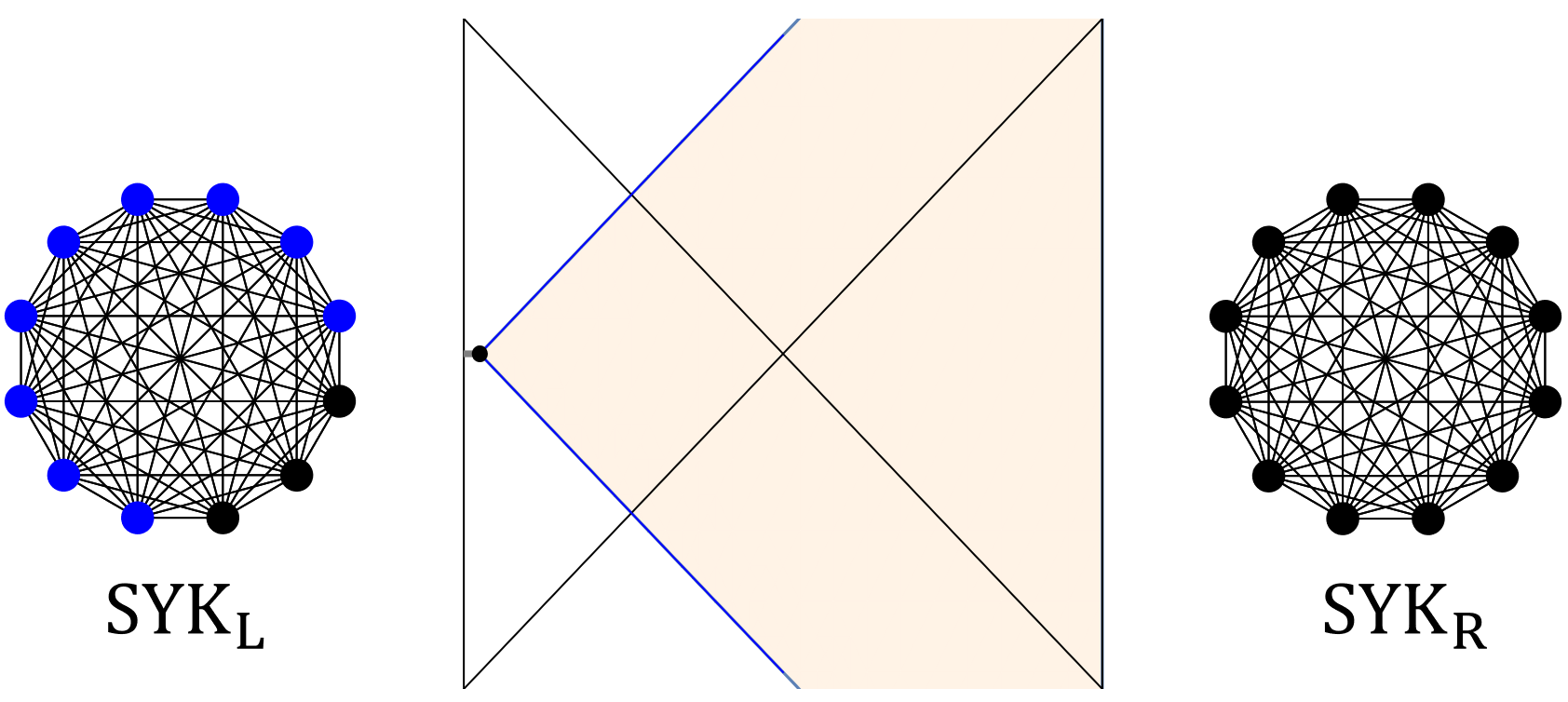}}
    \caption{Measurement-induced entanglement wedge transition.
    Two copies of the SYK model, denoted by left and right sides, are in a thermofield double state, which is dual to a two-sided eternal $AdS_2$ black hole. 
    A subset of the left Majorana fermions are measured, as indicated by blue dots. 
    An end-of-the-world brane is created in the left asymptotic region by the measurement.
    The entanglement wedge of all of the right Majorana fermions is indicated by the tan region.
    A small gray segment on the left side indicates the cutoff region (see discussion in Section~\ref{sec:jt_meas}).
    (a) When the number of measured Majorana fermions is less than a critical value, $M/N = m < m_*$, the quantum extremal surface is located at the bifurcation surface.
    (b) When the number of measured Majorana fermions exceeds a critical value, $M/N = m > m_*$, the entanglement wedge includes nearly the entire time reversal invariant slice. Note that the transition from (a) to (b) is a sharp phase transition in the bulk von Neumann entropy calculation, while it is a crossover in the SYK Renyi-2 entropy calculation; these are compared in Figure~\ref{fig:syk_partial-measure}. }
    \label{fig:summary}
\end{figure}

The main advantage the lower-dimensional toy-models studied here is the large amount of control we have on both sides of the duality. For instance, we are able to explicitly compute quantities in a boundary theory (the SYK model) which can be regarded as a candidate UV completion of the dual bulk theory. We can then compare the results with those obtained in the dual gravitational system in the appropriate limits. On the other hand, thanks to the simplicity of (1+1)-dimensional JT gravity, it is possible to analytically compute the bulk generalized entropy in the presence of bulk matter. Both these features would be remarkably cumbersome to obtain in higher-dimensional theories. Thus, the present analysis allows us to explicitly realize and further extend the constructions described in \cite{Antonini:2022sfm}, gaining much additional insight into the physical processes taking place in both the boundary and bulk theory when projective measurements are performed. 
Moreover, the generalized entropy prescription proposed here is a first step towards a more general framework for entanglement in setups that include non-unitary operations such as measurements.

The rest of the paper is organized as follows. In Section \ref{sec:measurement} we define the SYK model and its TFD state and describe the measurement procedure. We then compute the correlation functions in the post-measurement state using Euclidean path integral techniques in the saddle point approximation (analytically in the large-$q$ limit and numerically for any value of $q$). We then proceed to compute the post-measurement Renyi-2 mutual information between the two sides using a similar path integral procedure. In Section \ref{sec:JT} we define the bulk dual theory using JT gravity coupled to matter and study the effects of the projective measurement on the geometry. We compute the post-measurement mutual information between the two sides using the QES prescription and describe the entanglement wedge phase transition and its interpretation in terms of bulk teleportation. In Section \ref{sec:teleportation} we study the teleportation protocol and its fidelity in detail, and make the connection between our results and traversable wormholes. In section \ref{sec:discussion} we draw our conclusions and discuss future directions. Technical details of our analysis are contained in Appendices \ref{append:renyi_SYK}, \ref{append:JT}, \ref{append:fidelity}, and \ref{append:correlation_function}. In Appendix \ref{append:random_tfd} we give a simple toy model of the bulk teleportation using Haar random unitaries. In Appendix \ref{app:KM} we compare our results to the analysis of \cite{kourkoulou2017pure,zhang2020entanglement}. In Appendix \ref{append:syk_two-side} we report SYK model results for a measurement performed on subsets of fermions in both sides of the TFD state.

\section{Partial measurement of a SYK thermofield double}
\label{sec:measurement}

In this section, we investigate the entanglement structure of a thermofield double state of two copies of the SYK model (the ``left'' and the ``right'') after a measurement is performed on a subset of fermions in one of the two copies (which we will conventionally identify as the ``right''). A similar analysis for a measurement performed on a subset of fermions in both sides is reported in Appendix \ref{append:syk_two-side}. We start by defining the measurement operator and describing a Euclidean path integral representation of the measurement procedure. We then introduce the customary bilocal field formulation of the SYK model \cite{kitaev,Sachdev:2015efa,maldacena2016remarks} and study the saddle point solution for the post-measurement system. Finally, we compute the post-measurement Renyi-2 mutual information between the left and the remaining (unmeasured) fermions on the right. We find that the mutual information decreases as we increase the fraction of measured fermions, but is always non-vanishing unless all the right fermions are measured. The analysis of the present section serves also as a benchmark for the results obtained in the holographic dual description of Section \ref{sec:JT}, where we find an analogous behavior for the mutual information in JT gravity coupled to $N$ copies of a CFT.

\subsection{Partial measurement and path integral}
\label{sec:pathint}

The SYK model is defined by the all-to-all interacting Hamiltonian
\bea
    H[\psi] = i^{q/2} \sum_{i_1<...<i_q} J_{i_1,...,i_q} \psi_{i_1} ... \psi_{i_q},
\eea
where $\psi_i, i=1,...,N$ denotes a Majorana fermion (satisfying $\{\psi_i, \psi_j\} = \delta_{ij}$ and $\psi_i^\dag = \psi_i$), and $J_{i_1,...,i_q}$ gives the coupling between the Majorana fermions $i_1,...,i_q$.  $J_{i_1,...,i_q}$  is a random Gaussian variable with average and variance given by
\bea \label{eq:gaussian}
    \overline{J_{i_1,...,i_q}} = 0, \quad \overline{J_{i_1,..,i_q} J_{i_1',...,i_q'}} = \prod_{k=1}^q \delta_{i_k,i_k'} \frac{J^2}{q N^{q-1}},
\eea
where $J$ sets the overall interaction strength. 

We are interested in studying the post-measurement properties of the TFD state of two copies of the SYK model. The unnormalized TFD state, which is defined in a doubled Hilbert space, is given by 
\bea \label{eq:tfd}
    |TFD\rangle = e^{- \frac\beta4 (H_L + H_R)} |\infty\rangle
\eea
where $\beta$ is a parameter representing the inverse temperature of the thermal system obtained by tracing out one of the two sides, $H_L = H[\psi_L]$, $H_R = H[i \psi_R]$ (with $\psi_L$ and $\psi_R$ Majorana fermion operators on the left and right Hilbert space respectively), and $\ket{\infty}$ is the thermofield double state at infinite temperature defined by $ (\psi_{L,i} + i \psi_{R,i}) | \infty \rangle = 0$ \cite{Maldacena:2018lmt}.

Given the thermofield double state (\ref{eq:tfd}), we now want to perform a projective measurement on $M$ of the $N$ Majorana fermions in the right system. 
Inspired by the local projective measurement considered in $AdS_3/BCFT_2$~\cite{numasawa2016epr,Antonini:2022sfm} that retains bulk geometry, the measurement operator of interest here will be the fermion parity operator $M_{R,k}$ formed by the $2k-1$ and $2k$ right Majorana fermions in analogy with the analysis of Kourkoulou and Maldacena \cite{kourkoulou2017pure}:
\bea \label{eq:measurement_operator}
    M_{R,k} =- i 2 \psi_{R,2k-1} \psi_{R, 2k}
    \label{eq:parity}
\eea
where the factor of 2 accounts for the chosen normalization of the Majorana fermions and we assume for simplicity that $M$ is an even integer. Intuitively, if we divide Majorana fermions on the right side in pairs $(\psi_{R,2k-1},\psi_{R,2k})_{k=1,...,N/2}$ and identify each pair with a complex fermion $\tilde{\psi}_{R,k}$, the fermion parity (\ref{eq:parity}) corresponds to the Pauli-$Z$ operator $\sigma_{R,k}^z$ for the $k$-th complex fermion on the right side. The raising and lowering operators for the complex fermions are then
\begin{equation}
    \sigma_{R,k}^+=\sqrt{2}\left(\psi_{R,2k-1}+i\psi_{R,2k}\right), \quad \sigma_{R,k}^-=\sqrt{2}\left(\psi_{R,2k-1}-i\psi_{R,2k}\right).
    \label{eq:raiselower}
\end{equation}

To perform a projective measurement on the first $M$ fermions on the right side we must measure the fermi parity (\ref{eq:parity}) for $k=1,...,M/2$. The measurement operators for different $k$ commute. 
The eigenvalues of each operator are $r_k = \pm 1$~\footnote{Corresponding to a spin-up ($r_k=+1$) or spin-down ($r_k=-1$) state for the complex fermion.}. Therefore, the measurement record is given by a string of binary numbers $\textbf{r} = (r_1,...,r_{M/2})$. 
In our analysis below we will study how properties of the post-measurement state change when varying the number of measurement operators applied on the right side---given by $M/2$. More precisely, since we are interested in studying the SYK model in the large-$N$ limit (where it admits a semiclassical bulk dual spacetime), we define $m\equiv M/N$, take the large-$N$ limit keeping $m$ fixed, and then vary the value of $m$.

Suppose we measure the fermion parity (\ref{eq:parity}) for $k=1,...,M/2$ in the TFD state (\ref{eq:tfd}) and get a measurement outcome $\textbf{r}=(r_1,..., r_{M/2})$. 
The post-measurement state $|R_\textbf{r}(m) \rangle$ for the $M$ measured right Majorana fermions is an eigenstate of all the $M/2$ fermion parity operators~\footnote{We take the post-measurement state for the $M$ measured Majorana fermions to be normalized, i.e. $
    \langle R_\textbf{r}(m) | R_\textbf{r} (m) \rangle = 1$.}:
\bea
    M_{R,k} | R_\textbf{r}(m) \rangle = r_k |R_\textbf{r}(m)\rangle, \quad k=1,...,M/2.
    \label{eq:eigen}
\eea
Using the raising and lowering operators introduced in equation (\ref{eq:raiselower}),  equation (\ref{eq:eigen}) can be rewritten in the form
\bea
    \psi_{R,2k-1} | R_\textbf{r}(m) \rangle=-ir_k\psi_{R,2k} | R_\textbf{r}(m) \rangle, \quad k=1,...,M/2,
    \label{eq:raisebc}
\eea
which will be useful for identifying the correct boundary conditions in the path integral formalism.
Therefore, the full post-measurement (unnormalized) state of the two-sided system can be obtained by inserting the projection operator $| R_\textbf{r}(m) \rangle \langle R_\textbf{r}(m) |$: 
\bea \label{eq:state_one}
    |\Psi_\textbf{r}(m) \rangle = \left( \mathds 1^{(1-m)} \otimes | R_\textbf{r}(m) \rangle \langle R_\textbf{r}(m) | \right) |TFD \rangle
\eea
where we use $\mathds 1^{(1-m)}$ to denote the identity operator acting on the unmeasured Majorana fermions on the right side~\footnote{Here we are omitting the identity operator which acts on all the Majorana fermions of the left side.}. 
The Born probability of observing this state is $p_{\textbf{r}}(m) = \frac{ \langle \Psi_\textbf{r}(m) |\Psi_\textbf{r}(m) \rangle }{\langle TFD|TFD \rangle }$.
Note that the simultaneous eigenstates of all the fermion parity operators (\ref{eq:parity}) for $k=1,...,M/2$ form a complete basis of the measured subspace, i.e.
\bea
    \sum_{\textbf{r}} | R_\textbf{r}(m) \rangle \langle R_\textbf{r}(m)| = \mathds 1^{(m)},
\eea
which leads to $\sum_{\textbf{r}} p_\textbf{r} (m) = 1$, as expected.

\begin{figure}
    \centering
    \subfigure[]{
    \includegraphics[height=0.3\textwidth]{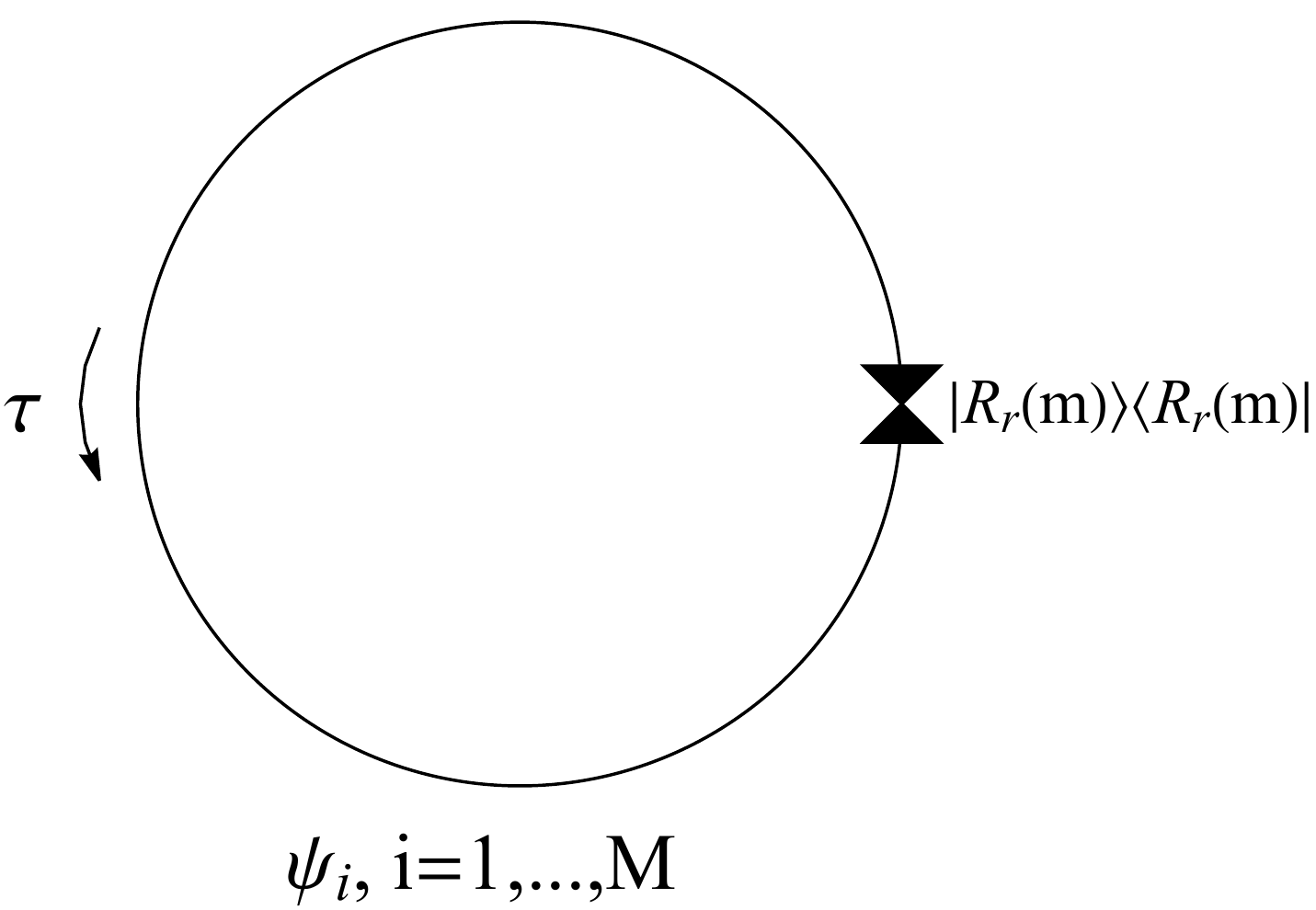}} \quad 
    \subfigure[]{
    \includegraphics[height=0.3\textwidth]{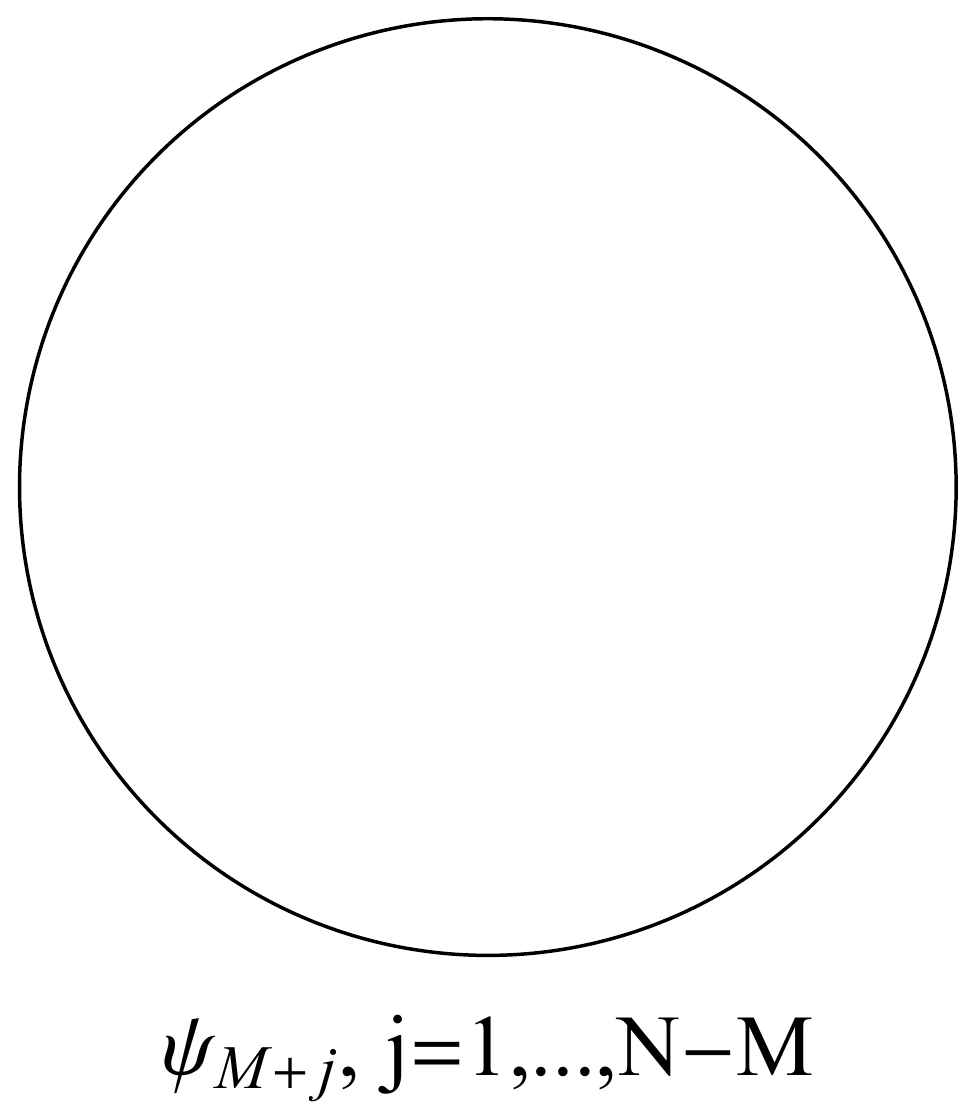}}
    \caption{Different boundary conditions for Majorana fields in the imaginary time contour. 
    (a) The measurement operator $|R_\textbf{r} (m) \rangle \langle R_\textbf{r}(m) |$ is inserted on the contour at $\tau = 0$, which is indicated by two small triangles, for $\psi_i, i=1,...,M$. 
    (b) The conventional contour for $\psi_{j+M}, j=1,...,N-M$.}
    \label{fig:syk_bc}
\end{figure}

Now that we have fully characterized the measurement operator and the post-measurement state, we can compute the amplitude $\overline{\langle \Psi_{\textbf{r}}(m)| \Psi_{\textbf{r}}(m) \rangle}$ using a Euclidean path integral representation. Here, as above, the overline indicates averaging over the Gaussian couplings~(\ref{eq:gaussian}). The amplitude can be written as
\bea
\begin{aligned}
     \overline{\langle \Psi_{\textbf{r}}(m)| \Psi_{\textbf{r}}(m) \rangle   } &= \overline{\langle TFD| \left( | R_\textbf{r}(m) \rangle \langle R_\textbf{r}(m) | \right) |TFD \rangle} \\
    &= \overline{ \Tr_R\left[e^{-\beta H_R} |R_\textbf{r} (m) \rangle \langle R_\textbf{r}(m) | \right] } =  \int D\psi e^{-I},
    \end{aligned}
\eea
where in the second equality we used the properties of the TFD state (\ref{eq:tfd}), $\Tr_R$ denotes the trace over the right system, and the action appearing in the Euclidean path integral is given by~\footnote{Note that this action is valid only in the large-$N$ limit, in which replica non-diagonal terms arising when averaging over the gaussian couplings are suppressed by $N^{1-q}$.}
\bea \label{eq:majorana_action}
    -I = -\int d\tau \sum_j \frac12 \psi_j(\tau) \partial_\tau \psi_j(\tau) +  \int d\tau_1 d\tau_2 \frac{J^2}{2q N^{q-1}} \left( \sum_j \psi_j(\tau_1) \psi_j(\tau_2) \right)^q.
\eea
The insertion of the projection operator $|R_\textbf{r} (m) \rangle \langle R_\textbf{r}(m)|$ in the trace corresponds to the following boundary conditions in the path integral:
\bea \label{eq:boundary-condition}
\begin{aligned}
    & \psi_{2k-1}(0) = -i r_k \psi_{2k}(0), \quad \psi_{2k-1}(\beta) = i r_k \psi_{2k}(\beta), \quad k=1,...,M/2, \\
    & \psi_{M+j}(0) = - \psi_{M+j}(\beta), \quad j=1,...,N-M
\end{aligned}
\eea
where we inserted the projection operator at Euclidean time $\tau = 0$ (corresponding to the right system). The boundary conditions for the measured Majorana fermions---i.e. $\psi_k$ with $ 1 \le k \le M$---reported in the first line of equation (\ref{eq:boundary-condition}) can be obtained from equation (\ref{eq:raisebc}) and its complex conjugate~\footnote{The boundary conditions can be easily derived by noticing that the Euclidean path integral prepares the ket state for the right system at $\tau=0$ and the bra state for the right system at $\tau=\beta$.}.
On the other hand, the unmeasured Majorana fermions $\psi_k$ with $ M +1 \le k \le N $ satisfy the conventional anti-periodic boundary conditions for fermionic fields, reported in the second line of equation (\ref{eq:boundary-condition}). 
The boundary conditions on the Euclidean time circle are illustrated in Fig.~\ref{fig:syk_bc}. 

If $m=0$---which means no fermions are measured--- the path integral clearly computes the usual thermal partition function with inverse temperature $\beta$.
On the other hand, if $m=1$---which means that the whole right system is measured---the path integral can be interpreted as preparing a Kourkoulou-Maldacena state~\cite{kourkoulou2017pure} at $\tau=\beta/2$ (see Appendix \ref{app:KM}).

It is now useful to introduce bilocal collective fields, in terms of which the equations of motion derived in the saddle point approximation of the Euclidean path integral are substantially simplified \cite{kitaev,Sachdev:2015efa,maldacena2016remarks}. The form of the boundary conditions (\ref{eq:boundary-condition}) suggests that we should introduce different types of bilocal fields. First, we introduce the ``diagonal'' bilocal fields
\bea \label{eq:bilocal}
\begin{aligned}
    & \tilde{G}_{11}(\tau_1, \tau_2) = \frac2M \sum_{k=1}^{M/2} \psi_{2k-1}(\tau_1) \psi_{2k-1}(\tau_2), \qquad
    \tilde{G}_{22}(\tau_1, \tau_2) = \frac2M \sum_{k=1}^{M/2} \psi_{2k}(\tau_1) \psi_{2k}(\tau_2),\\
    & \tilde{G}_{33}(\tau_1, \tau_2) = \frac{1}{N-M} \sum_{j=1}^{N-M} \psi_{M+j}(\tau_1) \psi_{M+j}(\tau_2).
    \end{aligned}
    \eea
For later convenience, we also introduce the following ``off-diagonal'' bilocal fields (similar to the ones introduced in \cite{kourkoulou2017pure}):
\begin{equation}
    \begin{aligned}
        \tilde{G}_{12}(\tau_1,\tau_2)=\frac2M \sum_{k=1}^{M/2}r_k \psi_{2k-1}(\tau_1) \psi_{2k}(\tau_2), \quad \tilde{G}_{21}(\tau_1,\tau_2)=\frac2M \sum_{k=1}^{M/2}r_k \psi_{2k}(\tau_1) \psi_{2k-1}(\tau_2).
    \end{aligned}
    \label{eq:offdiagonal}
\end{equation}
In terms of these bilocal fields (the arguments of which are in the range $\tau_{1,2}\in (0,\beta)$), the large-$N$ action becomes
\bea \label{eq:SYK-action}
    && -I = \int d\tau_1 d \tau_2 \Big[ -  \frac12 \Big( \sum_{k=1}^{M/2} \psi_{2k-1}(\tau_1) (\partial - \tilde{\Sigma}_{11}) \psi_{2k-1}(\tau_2) + \sum_{k=1}^{M/2} \psi_{2k}(\tau_1) (\partial - \tilde{\Sigma}_{22}) \psi_{2k}(\tau_2) \nn \\ 
    && + \sum_{j=M+1}^N \psi_{j}(\tau_1) (\partial - \tilde{\Sigma}_{33}) \psi_{j}(\tau_2) \Big)  
    -  \frac12 \left( \frac{M}2 \tilde{\Sigma}_{11} \tilde{G}_{11} + \frac{M}2 \tilde{\Sigma}_{22} \tilde{G}_{22} + (N-M) \tilde{\Sigma}_{33} \tilde{G}_{33} \right) \nn \\
    && + \frac{J^2}{2q N^{q-1}} \left( \frac{M}2 \tilde{G}_{11}(\tau_1, \tau_2) + \frac{M}2 \tilde{G}_{22}(\tau_1, \tau_2) + (N-M) \tilde{G}_{33}(\tau_1, \tau_2) \right)^q \Big],
\eea
where $\partial = \delta(\tau_1 - \tau_2) \partial_{\tau_2}$, and we have introduced the Lagrange multiplier fields $\tilde{\Sigma}_{11}(\tau_1,\tau_2)$, $\tilde{\Sigma}_{22}(\tau_1,\tau_2)$ and $\tilde{\Sigma}_{33}(\tau_1,\tau_2)$ (which play the role of the self-energy).  
It is easy to check that integrating out the Lagrange multiplier fields  enforces the definitions~(\ref{eq:bilocal}) of the bilocal fields.

Before deriving the equations of motion, it is useful to define a new Majorana field~\cite{zhang2020entanglement}
\bea \label{eq:chi}
    \chi_k(s) = 
    \begin{cases}
    \psi_{2k-1}(s), \quad & 0<s<\beta, \\
    i r_k \psi_{2k}(2\beta -s ), \quad & \beta< s < 2\beta. 
    \end{cases}
\eea
In the rest of the present subsection and in the next subsection we denote by $s$ coordinates in the range $(0,2\beta)$ and by $\tau$ coordinates in the range $(0,\beta)$.
The advantage of introducing this new field is that the boundary conditions (\ref{eq:boundary-condition}) reduce to the conventional anti-periodic boundary conditions for $\chi_k(s)$, i.e. $\chi_k(0)=-\chi_k(2\beta)$. 
In terms of the new field $\chi_k(s)$ the first line of equation~(\ref{eq:SYK-action}) becomes
\bea
    -   \frac12 \int_0^{2\beta} ds_1 \int_0^{2\beta} ds_2 \chi_{k}(s_1) \left( \delta(s_1-s_2) \partial_{s_2} -  \hat{\tilde{\Sigma}}(s_1,s_2)  \right) \chi_{k}(s_2),
\eea
where we used $r_k^2=1$ and introduced a compact notation for the self-energy $\hat {\tilde{\Sigma}}(s_1, s_2)$ (with $0<s_{1,2} < 2\beta$)
\bea \label{eq:compact_sigma}
    \hat {\tilde{\Sigma}}(s_1, s_2) = \left[ \ba{cccc} \tilde{\Sigma}_{11}(s_1,s_2) & 0 \\ 0 & -\tilde{\Sigma}_{22}(2\beta - s_1, 2\beta - s_2)\ea \right].
\eea
In this matrix notation, the range of the $s_i$ coordinates for the $\hat{\tilde{\Sigma}}_{a_1a_2}(s_1,s_2)$ element is $(a_i-1) \beta < s_i < a_i \beta$ with $a_i=1,2$, for $i=1,2$, respectively. More explicitly, $\hat{\tilde{\Sigma}}_{11}(s_1,s_2)=\tilde{\Sigma}_{11}(s_1,s_2)$ with $0<s_1,s_2<\beta$ and $\hat{\tilde{\Sigma}}_{22}(s_1,s_2)=-\tilde{\Sigma}_{22}(2\beta-s_1,2\beta-s_2)$ with $\beta<s_1,s_2<2\beta$. The ranges $0<s_1<\beta$, $\beta<s_2<2\beta$ and $\beta<s_1<2\beta$, $0<s_2<\beta$ correspond to the (vanishing) off-diagonal elements $\hat{\tilde{\Sigma}}_{12}(s_1,s_2)$ and $\hat{\tilde{\Sigma}}_{21}(s_1,s_2)$, respectively.

Now both $\psi_j, j=M+1,...,N$ and $\chi_{k}, k=1,...,M/2$ satisfy the conventional anti-periodic boundary conditions typical of fermionic fields. 
We can then integrate out the Majorana fermions to get
\bea
    - \frac{I}N &=&   \frac{m}2 \log \Pf \left( \partial -  \hat {\tilde{\Sigma}}  \right) + (1-m)\log \Pf (\partial - \tilde{\Sigma}_{33}) \\ 
    &&     -  \int d\tau_1 d\tau_2 \frac12 \left( \frac{m}2 \tilde{\Sigma}_{11} \tilde{G}_{11} + \frac{m}2 \tilde{\Sigma}_{22} \tilde{G}_{22} + (1-m) \tilde{\Sigma}_{33} \tilde{G}_{33} \right) \nn \\
    && + \int d\tau_1 d\tau_2 \frac{J^2}{2q} \left( \frac{m}2 \tilde{G}_{11} + \frac{m}2 \tilde{G}_{22} + (1-m) \tilde{G}_{33} \right)^q
\eea
where $\Pf$ indicates the Pfaffian.
Owing to the large $N$ structure of the action, we can derive the equations of motion in the saddle point approximation, which are given by the Schwinger-Dyson equations (here and in the following we drop the $\tilde{}$ to indicate on-shell solutions)
\bea \label{eq:eom}
    \hat G(s_1, s_2) &=&  \left( \partial -  \hat \Sigma  \right)^{-1} (s_1,s_2), \quad 
    G_{33}(\tau_1,\tau_2) = (\partial - \Sigma_{33})^{-1}(\tau_1,\tau_2) ,  \\
    \Sigma_{ii}(\tau_1,\tau_2) &=& J^2\left( \frac{m}2 G_{11}(\tau_1,\tau_2) + \frac{m}2 G_{22}(\tau_1,\tau_2) + (1-m) G_{33}(\tau_1,\tau_2) \right)^{q-1}, \quad i = 1,2,3. \nn  
\eea
where we have introduced a compact notation for $\hat{G}$ analogous to the one introduced in equation (\ref{eq:compact_sigma}) for the self-energy:\footnote{Note that, although $\hat{\Sigma}$ is diagonal, $(\partial-\hat{\Sigma})^{-1}$ is not, and therefore $\hat{G}$ contains off-diagonal components.}
\bea \label{eq:compact_G}
    \hat G(s_1, s_2 ) = \left[ \ba{cccc} G_{11}(s_1,s_2) & i G_{12} (s_1, 2\beta - s_2) \\ i G_{21} ( 2\beta - s_1,s_2) & -G_{22}(2\beta - s_1, 2\beta - s_2)\ea \right].
\eea
Note that for the off-diagonal components, $\hat{G}_{a_1a_2}(s_1,s_2)$ with $a_1\neq a_2$, the matrix notation introduced after equation (\ref{eq:compact_sigma}) implies that the ranges of the $s_1$ and $s_2$ coordinates are different (i.e. $0<s_1<\beta$, $\beta<s_2<2\beta$ for $\hat{G}_{12}$ and $\beta<s_1<2\beta$, $0<s_2<\beta$ for $\hat{G}_{21}$).
It is not hard to see that $\hat G(s_1, s_2) = \frac2{M} \sum_{k=1}^{M/2} \chi_k (s_1) \chi_k(s_2)$. 

$\hat{G}$ and $G_{33}$ are on-shell Green's functions. In the next subsection, we will compute $\hat{G}$ and $G_{33}$ by solving the Schwinger-Dyson equations (\ref{eq:eom}) analytically in the large-$q$ limit and numerically for any value of $q$.

\subsection{Saddle point solution}

The action (\ref{eq:majorana_action}) is $O(N)$-symmetric. In particular, it is invariant under the action of a subgroup of $O(N)$, the ``Flip group'' \cite{kourkoulou2017pure}, the elements of which flip the sign of any number of even-indexed Majorana fermions. In other words, the action of an element of the Flip group is given by $\psi_{2k-1} \rightarrow \psi_{2k-1}, \psi_{2k} \rightarrow - \psi_{2k}$ for one or more given values of $k$ with $1\leq k\leq N/2$. In terms of the complex fermions $\tilde{\psi}_k$ associated with pairs of Majorana fermions, an element of the Flip group flips the sign of one or more spins. Note that both the diagonal and the-off diagonal bilocal fields (\ref{eq:bilocal}) and (\ref{eq:offdiagonal}) are invariant under the Flip group as well as the on-shell Green's functions.

The invariance under the Flip group clearly implies that all our results are independent of the measurement outcome $\textbf{r}=(r_1,...,r_{M/2})$: the Flip group acting on any number of the first $M$ Majorana fermions can map a given measurement outcome to any other measurement outcome. One consequence of this invariance is that the on-shell solutions $\hat G(s_1, s_2)$ and $G_{33}(\tau_1,\tau_2)$ in our setup involving measurement are related to the thermal Green's function $G_{th}(\tau_1,\tau_2)$. In fact, the on-shell Green's function $G_{11}(\tau_1,\tau_2)$ can be written as
\bea
    G_{11}(\tau_1, \tau_2) &=& \frac2M \sum_k \frac1Z \Tr\left[ e^{-\beta H[\psi]} \mathcal{T} \left( \psi_{2k-1}(\tau_1)  \psi_{2k-1}(\tau_2) \right)  |R_\textbf{r}(m) \rangle \langle R_\textbf{r} (m) | \right],
\eea
where $ Z =  \Tr\left[ e^{-\beta H[\psi]} |R_\textbf{s}(m) \rangle \langle R_\textbf{s} (m) | \right]$ is the normalization factor, $\psi(\tau) = e^{H[\psi] \tau} \psi  e^{-H[\psi] \tau}$, and $\mathcal T$ denotes imaginary time ordering.
Using the invariance under measurement outcome, we can then write  
\bea
    G_{11}(\tau_1, \tau_2) &=& \frac2M \sum_k \frac1Z \left( \frac1{2^{M/2}} \sum_{\textbf{r}} \Tr\left[ e^{-\beta H[\psi]} \mathcal{T} \left( \psi_{2k-1}(\tau_1)  \psi_{2k-1}(\tau_2) \right) |R_\textbf{r}(m) \rangle \langle R_\textbf{r} (m) | \right] \right)  \nn\\ 
    &=&  \frac2M \sum_k \frac{  \Tr\left[ e^{-\beta H[\psi]} \mathcal{T} \left( \psi_{2k-1}(\tau_1)  \psi_{2k-1}(\tau_2) \right)  \right]}{\Tr\left[ e^{-\beta H[\psi]} \right]} \nn \\ &=& \frac1N \sum_i  \frac{  \Tr\left[ e^{-\beta H[\psi]} \mathcal{T} \left( \psi_{i}(\tau_1)  \psi_{i}(\tau_2) \right)  \right]}{\Tr\left[ e^{-\beta H[\psi]} \right]}=G_{th}(\tau_1,\tau_2)
    \label{eq:greenth}
\eea
where to go from the first to the second line we have used the completeness relation for the measured subspace $\sum_\textbf{r} |R_\textbf{r} (m) \rangle \langle R_\textbf{r}(m)| = \mathds 1^{(m)}$ and $ Z = \frac1{2^{M/2}} \sum_{\textbf{r}} \Tr\left[ e^{-\beta H[\psi]} |R_\textbf{r}(m) \rangle \langle R_\textbf{r} (m) | \right] = \frac1{2^{M/2}}  \Tr\left[ e^{-\beta H[\psi]} \right]
$.
The equality leading from the second to the third line is true in virtue of the $O(N)$ symmetry of the model.
An analogous relation can be obtained for $G_{22}(\tau_1, \tau_2)$ and $G_{33}(\tau_1,\tau_2)$. 
Therefore, $G_{ii}, i = 1,2,3$ are all given by the usual imaginary time correlation.

The Schwinger-Dyson equations~(\ref{eq:eom}) can then be rewritten in a simpler form~\footnote{We omit here and in the following analysis the equations for $G_{33}$ and $\Sigma_{33}$, which can be solved in complete analogy with the $11$ component.}
\bea 
  \hat G = (\partial - \hat \Sigma)^{-1}, \quad \hat \Sigma(s_1, s_2) = \frac{\mathcal J^2}{q} P(s_1, s_2) [2\hat G(s_1, s_2)]^{q-1}, \label{eq:sd}
  \eea
  where we defined $\mathcal{J}^2=qJ/2^{q-1}$ for later convenience and
  \bea
     P(s_1, s_2) = \theta(\beta - s_1) \theta(\beta - s_2) + \theta(s_1 -\beta) \theta(s_2 - \beta). \label{eq:P}
\eea
The boundary conditions now read
\bea 
    \hat G(s,s) = \frac12, \quad \hat G(s_1, s_2) = - \hat G(s_2, s_1), \quad \hat G(s_1 + 2\beta, s_2) = - \hat G(s_1, s_2).
\eea
The first and second equations are a consequence of the commutation relations and the definition of the time ordering operator, whereas the third condition implements the anti-periodic boundary conditions for the fermionic field $\chi(s)$.

\begin{figure}
    \centering
    \includegraphics[width=0.3\textwidth]{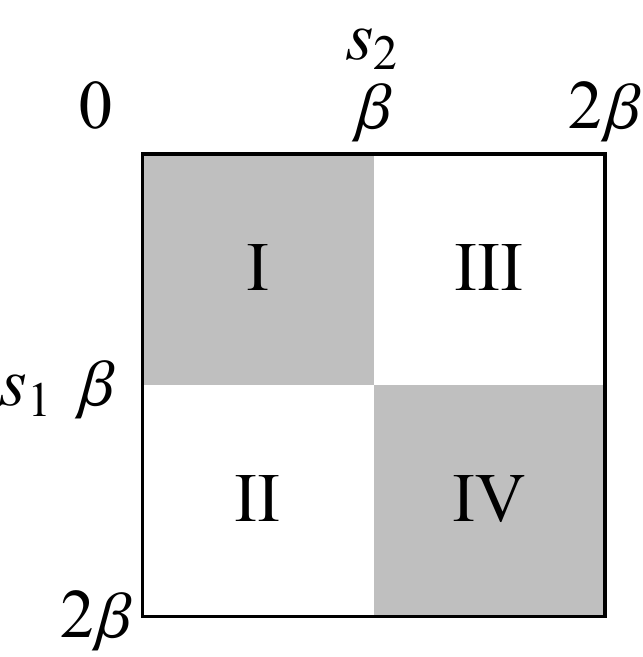}
    \caption{The domain of the bilocal Green's function $\hat{G}(s_1, s_2)$. $P(s_1, s_2)$ is nonzero in the shaded regions I and IV. According to our matrix notation, $\hat{G}_{11}(s_1,s_2)=G_{11}(s_1,s_2)$ denotes the Green's function in region I, $\hat{G}_{21}(s_1,s_2)=iG_{21}(2\beta-s_1,s_2)$ denotes the Green's function in region II, $\hat{G}_{12}(s_1,s_2)=iG_{12}(s_1,2\beta-s_2)$ denotes the Green's function in region III, and $\hat{G}_{22}(s_1,s_2)=-G_{22}(2\beta-s_1,2\beta-s_1)$ denotes the Green's function in region IV.}
    \label{fig:2D_measure}
\end{figure}

The Schwinger-Dyson equations can be solved analytically in the large-$q$ limit~\cite{maldacena2016remarks}.
Using the ansatz $\hat G(s_1, s_2) = \hat{G}_0(s_1, s_2) \left( 1 + \frac1q g(s_1, s_2) +...\right)$, where we suppressed terms of higher order in $q$ and introduced the bare propagator 
\bea
\hat{G}_0(s_1,s_2)=\frac{1}{2}\left[ \ba{cccc} \sgn(s_1-s_2) & -1 \\ 1 & \sgn(s_1-s_2)\ea \right],
\eea
the Schwinger-Dyson equations (\ref{eq:sd}) at leading order in $q$ give an equation for $g(s_1,s_2)$ \cite{maldacena2016remarks}:
\bea
    \partial_{s_1} \partial_{s_2} [\hat{G}_0(s_1, s_2) g(s_1, s_2)] = \mathcal J^2 P(s_1, s_2) [2 \hat{G}_0(s_1, s_2)]^{q-1} e^{g(s_1, s_2)}.
    \label{eq:liou}
\eea
Note that the measurement introduces a factor $P(s_1, s_2)$ on the right-hand side which is not present in the conventional Liouville equation \cite{maldacena2016remarks}. 
Given the expression~(\ref{eq:P}) for $P(s_1,s_2)$, we can divide the domain into four regions as shown in Fig.~\ref{fig:2D_measure}.
In region I and IV, the Liouville equation (\ref{eq:liou}) takes the form~\footnote{Here, we used the fact that $q$ is an even number in order for the Hamiltonian to be a bosonic operator.}
\bea
    \partial_{s_1} \partial_{s_2} g(s_1, s_2) = 2 \mathcal J^2 e^{g(s_1, s_2)}.
\eea
The relationship (\ref{eq:greenth}) between the diagonal Green's functions in our setup and the thermal Green's function implies that we can assume the diagonal elements of $\hat{G}(s_1,s_2)$ (i.e. the Green's function in regions I and IV) are time translation invariant. Therefore in region I and IV, $g(s_1,s_2)=g(s_{12})$ where $s_{12} \equiv s_1 - s_2$. The solution of the Liouville equation is then
\begin{equation}
    g_\text{I}(s_{12})=g_{\text{IV}}(s_{12})=2\log\left[\frac{\sin\gamma}{\sin(\alpha |s_{12}| + \gamma)}\right]
\end{equation}
where $\alpha=\mathcal{J}\sin\gamma$ and $\gamma$ is a constant to be determined by imposing the anti-periodic boundary condition. At leading order in $q$ we then obtain the following result for the diagonal elements of $\hat{G}(s_1,s_2)$:
\bea \label{eq:solution_I}
    \hat{G}_{11}(s_1, s_2) = \hat{G}_{22}(s_1, s_2)=\frac12 \sgn(s_{12}) \left( \frac{\sin \gamma}{\sin(\alpha |s_{12}| + \gamma)} \right)^{2/q}.
\eea
In regions II and III, the Liouville equation (\ref{eq:liou}) takes the form $\partial_{s_1} \partial_{s_2} g(s_1, s_2) = 0$, which is solved by the ansatz $g(s_1, s_2) = f_1(s_1) + f_2(s_2)$. The form of $f_1(s_1)$ and $f_2(s_2)$ can be obtained by matching the solutions for $\hat{G}(s_1,s_2)$ at the boundaries between regions I,IV and regions II,III. This yields the following off-diagonal elements of $\hat{G}(s_1,s_2)$:
\bea \label{eq:solution_II}
\begin{aligned}
    &\hat{G}_{21}(s_1, s_2) = \frac12 \left( \frac{\sin \gamma}{\sin (\alpha (s_1 - \beta) + \gamma)} \frac{\sin \gamma}{\sin (\alpha (\beta-s_2) + \gamma)} \right)^{2/q}, \\
    &\hat{G}_{12}(s_1, s_2) = - \frac12 \left( \frac{\sin \gamma}{\sin (\alpha (s_2 - \beta) + \gamma)} \frac{\sin \gamma}{\sin (\alpha (\beta-s_1) + \gamma)} \right)^{2/q}.
    \end{aligned}
    \label{eq:offfdiagresult}
\eea
Finally, imposing the anti-periodic boundary condition $\hat{G}(s_1+2\beta,s_2)=-\hat{G}(s_1,s_2)$, the constant $\gamma$ is implicitly determined by 
\bea \label{eq:constant}
   \gamma + \frac{\mathcal{J} \beta \sin\gamma}2 = \frac\pi2. 
\eea
Note that our results for the off-diagonal Green's function (\ref{eq:offfdiagresult}) are in agreement with the $m=1$ results of \cite{kourkoulou2017pure}.

\begin{figure}
    \centering
    \subfigure[$G_{33}$]{\includegraphics[width=0.4\textwidth]{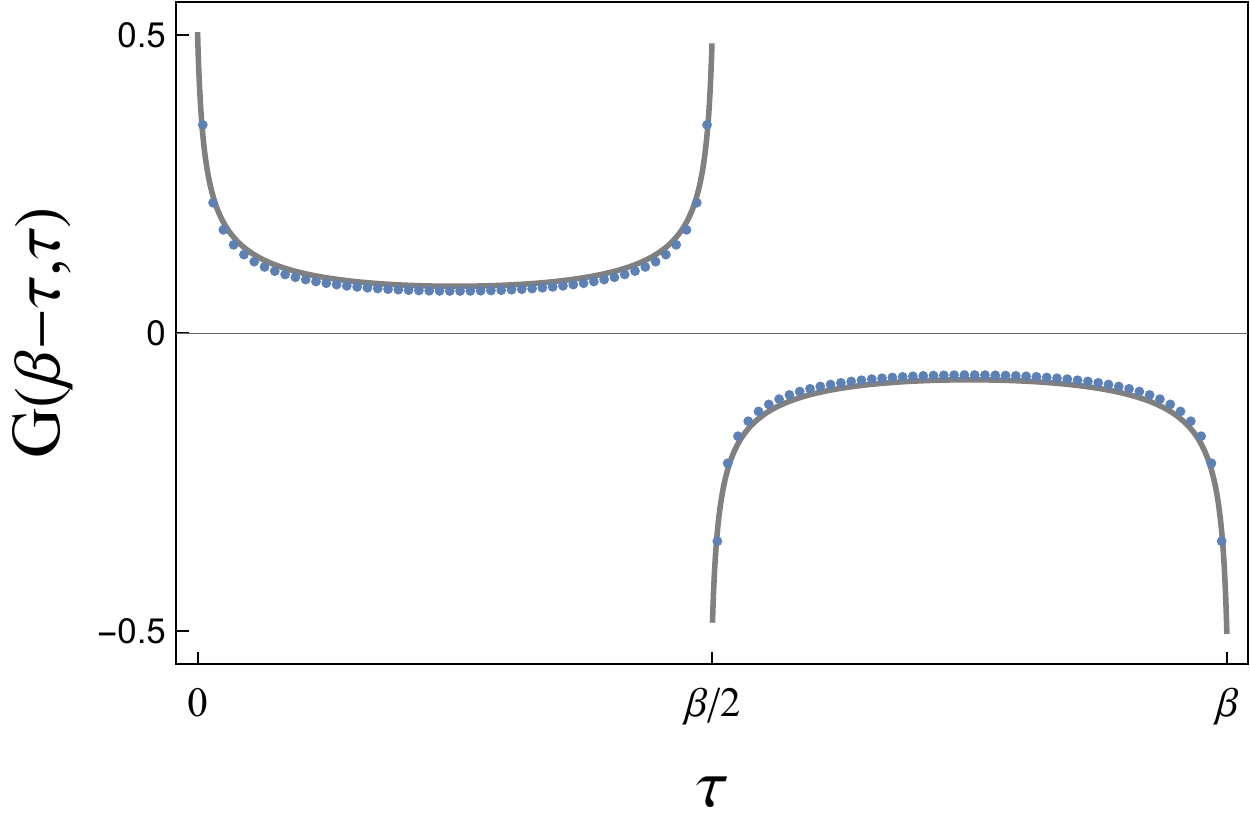}} \quad \quad
    \subfigure[$G_{11,22}$]{\includegraphics[width=0.4\textwidth]{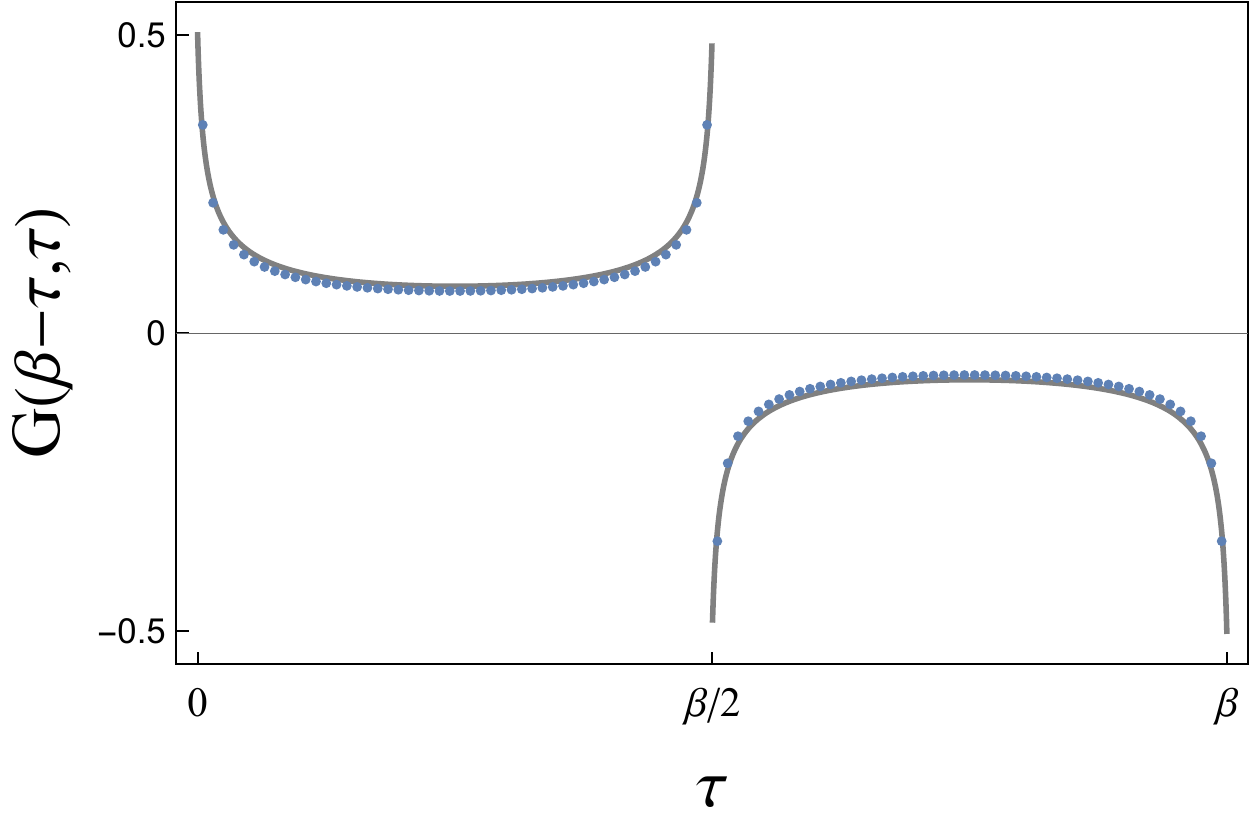}} \\
    \subfigure[$G_{21}$]{\includegraphics[width=0.4\textwidth]{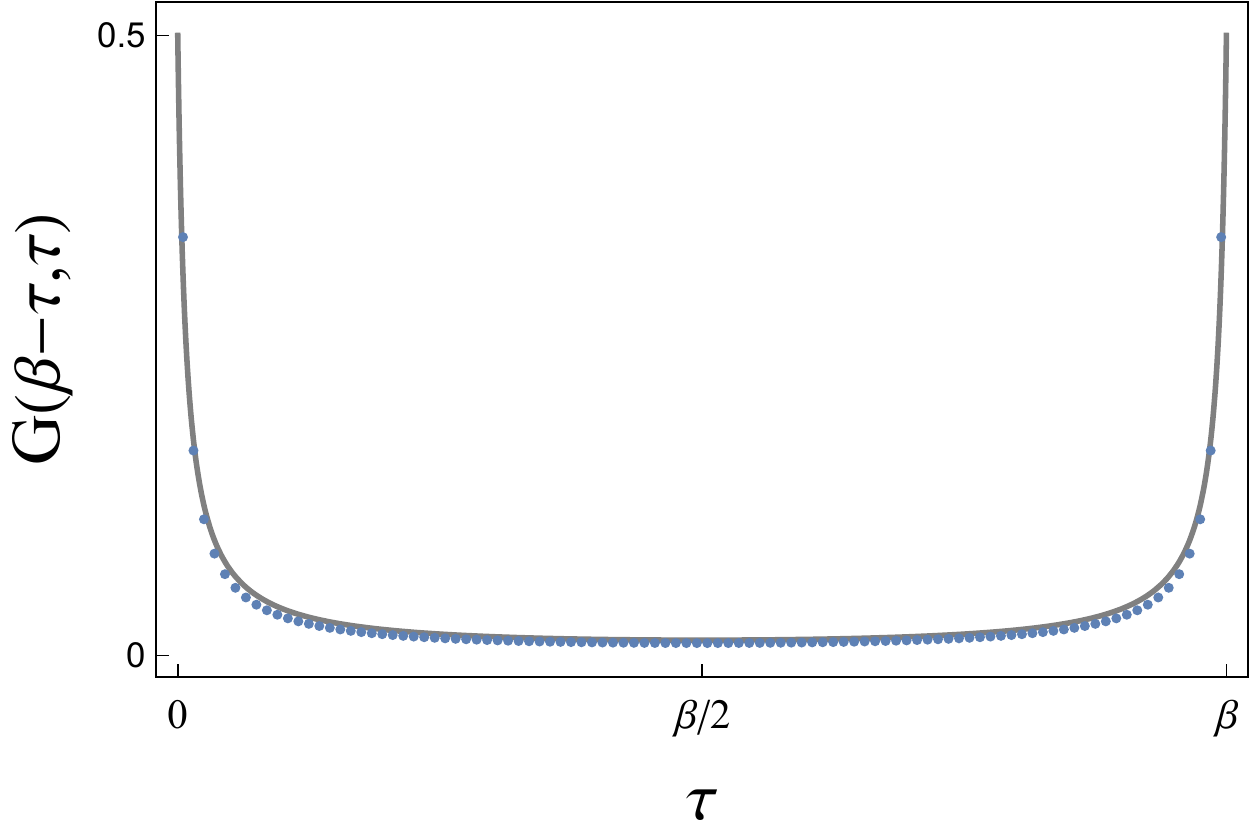}}
    \caption{A comparison between numerical (represented by blue dots) and large-$q$ analytical (represented by gray curves) solutions for the diagonal components of the Green's function.
    The parameters are $\beta \mathcal J = 20$, $q=4$. 
    The imaginary time contour with length $\beta$ is discretized into $100$ intervals. (a) The unmeasured Majorana fermions' Green's function $G_{33}(\tau_1,\tau_2)$. (b) The measured Majorana fermions' Green's functions $G_{11}(\tau_1,\tau_2)$, $G_{22}(\tau_1,\tau_2)$. Note that $G_{11}(\tau_1,\tau_2)=G_{22}(\tau_1,\tau_2)=G_{33}(\tau_1,\tau_2)$ as expected by the flip symmetry argument in equation (\ref{eq:greenth}). 
    (c) The measured Majorana fermions' Green's function $G_{21}(\tau_1, \tau_2)$. }
    \label{fig:green_funcion}
\end{figure}

Alternatively, the Schwinger-Dyson equation~(\ref{eq:sd}) can be solved numerically by iteration for any value of $q$.
Given that our results are independent of the measurement outcome, we take for simplicity $r_k = 1$ for all $k=1,...,M/2$.
A comparison of the numerical results with the large $q$ solution described above is shown in Fig.~\ref{fig:green_funcion}, where the dots are numerical results and the gray curves are the analytical large-$q$ solutions derived above.
Note that, as expected, the numerical solutions for the diagonal Green's functions of the measured Majorana fermions $G_{11}(\tau_1,\tau_2)$, $G_{22}(\tau_1,\tau_2)$ and the ummeasured Majorana fermions $G_{33}(\tau_1,\tau_2)$ are identical.

\subsection{Mutual information}
\label{sec:SYKMI}

In order to set the stage for the holographic calculation carried out in Section \ref{sec:JT}, we now compute the post-measurement Renyi-2 mutual information between the left side and the unmeasured Majorana fermions on the right side. The SYK mutual information computed in the present subsection will serve as a benchmark to test our results for the holographic mutual information, which will give us information about the effect of the measurement on the wormhole dual to the SYK thermofield double state. The choice of computing the Renyi-2 mutual information in the SYK setup rather than the Von Neumann mutual information is due to the fact that it can be easily computed in the saddle point approximation using a Euclidean path integral for two replicas of our system with appropriate boundary conditions. The Renyi-2 entropy is also known to provide a good qualitative approximation of the von Neumann entropy for the SYK model (see the discussion and comparison between Renyi-2 entropy and von Neumann entropy from exact diagonalization~\cite{liu2018quantum} in Ref.~\cite{zhang2020subsystem}). This is confirmed by our comparison between the SYK Renyi-2 mutual information and the holographic von Neumann entropy computed in the bulk dual setup in Section \ref{sec:JT}.

As we discussed in the previous subsection, the unnormalized post-measurement state of the two-sided system is $|\Psi_\textbf{r}(m) \rangle$ given by~(\ref{eq:state_one}), and the corresponding unnormalized density matrix is $\Psi_\textbf{r}(m) \equiv | \Psi_\textbf{r}(m) \rangle \langle \Psi_\textbf{r}(m)|$. 
Because $|\Psi_\textbf{r}(m) \rangle $ is a pure state, we have $S_L(\Psi_\textbf{r}(m) ) = S_R(\Psi_\textbf{r}(m) )$, and $S_{LR}(\Psi_\textbf{r}(m) ) = 0$. 
Here $S_L$ ($S_R$) denotes the entanglement entropy of the left (right) side, and $S_{LR}$ denotes the entanglement entropy of the full two-sided system.
Therefore, the mutual information is twice the entanglement entropy of the left system
\bea
    I_{LR}(\Psi_\textbf{r}(m) ) = S_L(\Psi_\textbf{r}(m) ) + S_R(\Psi_\textbf{r}(m)) - S_{LR}(\Psi_\textbf{r}(m)) = 2 S_L(\Psi_\textbf{r}(m)).
\eea

We focus our attention on the Renyi-2 entanglement entropy---denoted by $S^{(2)}(\Psi_\textbf{r}(m))$. This can be obtained via numerical evaluation of a path integral for two replicas of our system, with appropriate boundary conditions which we will derive below. 
The Renyi-2 entropy of the left system is defined by
\bea \label{eq:renyi_one}
    e^{-S_L^{(2)}(\Psi(m))} = \frac{ \overline{\Tr_L \left[( \Tr_R[\Psi(m)])^2 \right]}}{ \overline{ \left(\Tr[\Psi(m)] \right)^2}}, 
\eea
where we used the fact that disorder-averaging the ratio is equivalent to disorder-averaging the numerator and denominator separately up to $1/N$-suppressed corrections. We have also omitted the subscript $\textbf{r}$ indicating the measurement outcome due to the invariance under measurement outcome discussed above. In particular, in the following calculation we set $r_k = 1, k=1,...,M/2$ for simplicity. 

\begin{figure}
    \centering
    \subfigure[$i=1,...,M$]{\includegraphics[height=0.4\textwidth]{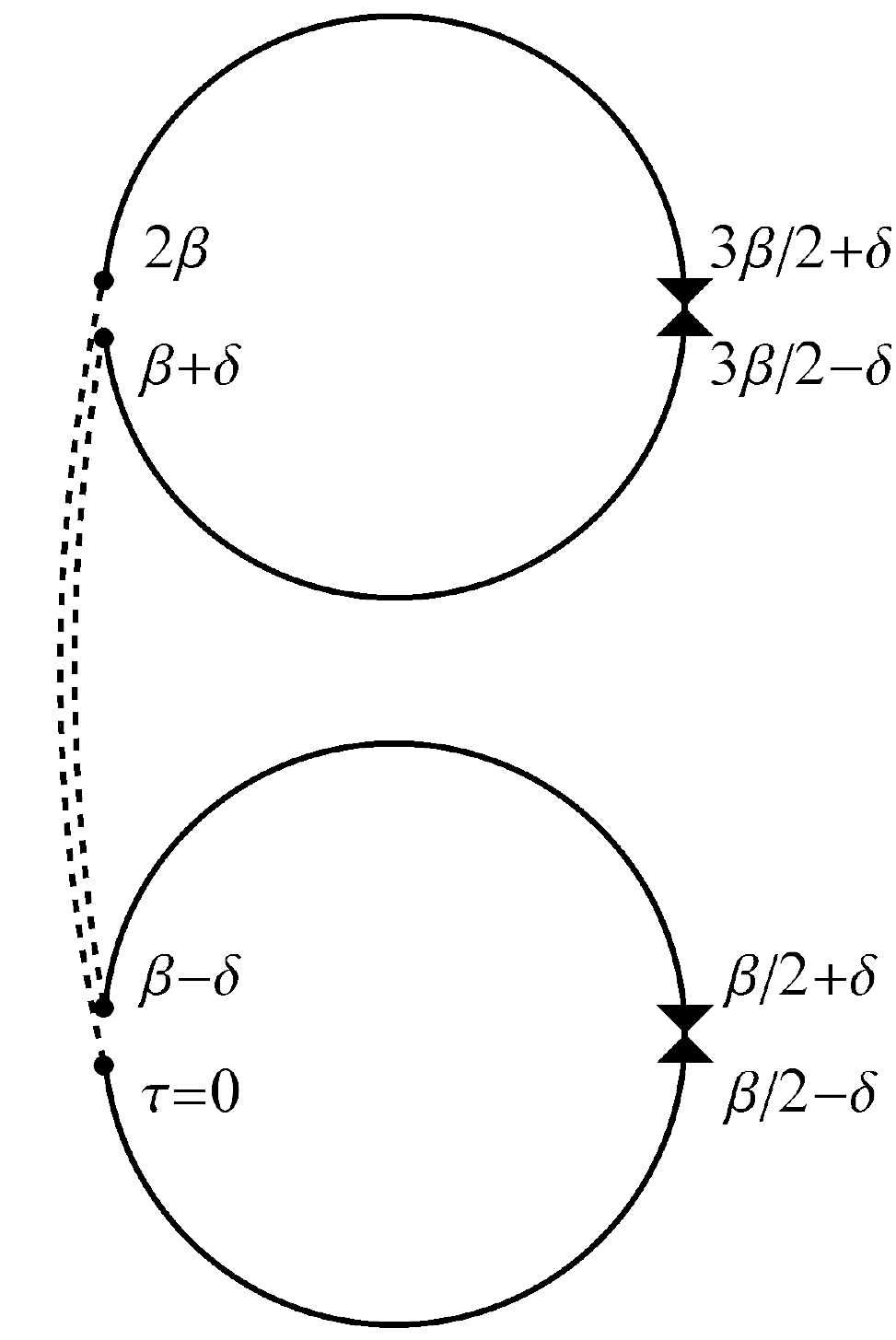}} \quad \quad \quad \quad 
    \subfigure[$i=M+1,...,N$]{\includegraphics[height=0.4\textwidth]{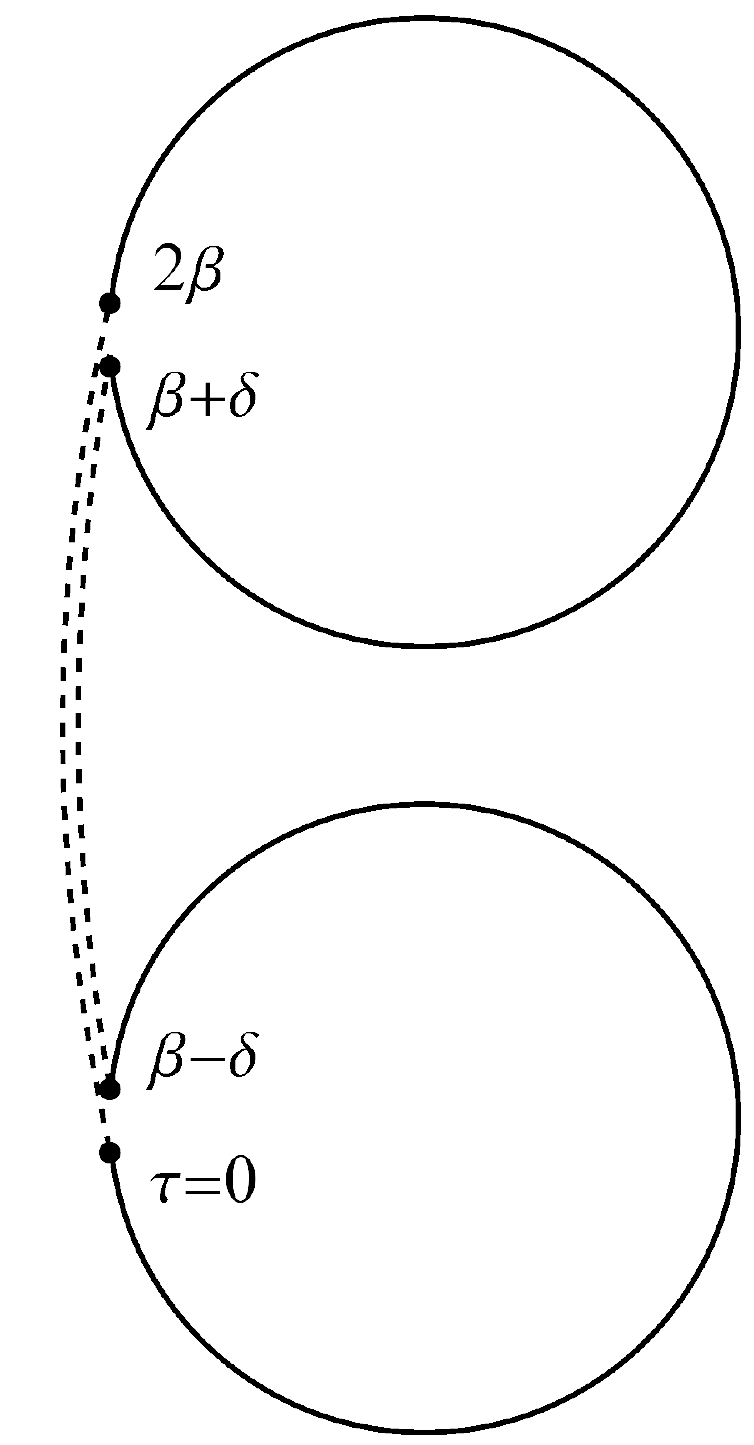}}
    \caption{Imaginary time contour $\tau \in(0,2\beta) $ for the Euclidean path integral computing the numerator of equation (\ref{eq:renyi_one}). The points
    $\tau =0$, $\beta \pm \delta$, $\beta/2 \pm \delta $, $3\beta/2 \pm \delta $ and $2\beta$ where boundary conditions must be imposed are labeled in the figure. 
    The twist operator inserted in the left system is indicated by the dashed line.
    (a) The effect of the measurement operator on $\psi_i, i=1,...,M$ is indicated by the triangles, corresponding to the second and third lines of the boundary conditions~(\ref{eq:renyi_one_bc}). At the insertion point of the twist operator, the boundary conditions are given by the first line of~(\ref{eq:renyi_one_bc}).
    (b) There is no measurement operator for $\psi_i, i=M+1,...,N$. 
    The boundary conditions are then simply set by the first line of ~(\ref{eq:renyi_one_bc}).}
    \label{fig:contour_one}
\end{figure}

Equation~(\ref{eq:renyi_one}) can be expressed in terms of Euclidean path integrals computing the numerator and denominator separately. Because the Renyi-2 entropy involves the square of the density matrix, we extend the imaginary time contour to $\tau \in (0,2\beta)$ with $\tau\in (0,\beta)$ corresponding to the first replica and $\tau\in (\beta,2\beta)$ corresponding to the second replica. With a change of notation from the previous subsections, we consider the left system to be at $\tau=0,\beta,2\beta$ (for the two replicas) and the right system where the measurement is performed to be at $\tau=\beta/2,3\beta/2$.
When computing the numerator of (\ref{eq:renyi_one}), a twist operator must be inserted in the left system, setting anti-periodic boundary conditions at $\tau=0,2\beta$ for the Majorana fermions with period $2\beta$; there is no twist operator for the denominator. Additionally, as in the previous subsections, the measurement operator inserted in the right side sets different boundary conditions for measured and unmeasured Majorana fermions. See Fig.~\ref{fig:contour_one} for a representation of the boundary conditions in the path integral computing the numerator of (\ref{eq:renyi_one}). 

The path integral for the numerator reads
\bea
    \overline{\Tr_L \left[( \Tr_R[\Psi(m)])^2 \right]} = \int D\psi  e^{-I},
\eea
where the action is given by
\bea \label{eq:syk_action_one}
 -I = -\int_0^{2\beta} d\tau \sum_j \frac12 \psi_j(\tau) \partial_\tau \psi_j(\tau) +  \int_0^{2\beta} d\tau_1 \int_0^{2\beta} d\tau_2 \frac{J^2}{2q N^{q-1}} \left( \sum_j \psi_j(\tau_1) \psi_j(\tau_2) \right)^q, \nn \\
\eea
and the boundary conditions are
\bea
\begin{aligned}
    &
    \psi_i(0) = -\psi_i(2\beta), \quad \psi_{i}(\beta_-) =  \psi_{i}(\beta_+), \quad 
    i=1,...,N, \\
    & \psi_{2k-1}\left(\frac{\beta_-}2\right) = i \psi_{2k}\left(\frac{\beta_-}2\right), \quad \psi_{2k-1}\left(\frac{\beta_+}2\right) = -i \psi_{2k}\left(\frac{\beta_+}2\right),  \quad k = 1,..., M/2, \\
    & \psi_{2k-1}\left(\frac{3\beta_-}2\right) = i \psi_{2k}\left(\frac{3\beta_-}2\right), \quad \psi_{2k-1}\left(\frac{3\beta_+}2\right) = -i \psi_{2k}\left(\frac{3\beta_+}2\right),  \quad k = 1,..., M/2.
    \end{aligned}
    \label{eq:renyi_one_bc}
\eea
Here we used $\beta_\pm$ as a shorthand notation to indicate that the quantity in parenthesis is shifted by an infinitesimal positive number $\delta$ (for instance, $3\beta_\pm/2=3\beta/2\pm\delta$).
The first line of equation (\ref{eq:renyi_one_bc}) is a consequence of the presence of the twist operator in the left system; the second and third lines are a consequence of the insertion of the measurement operator (\ref{eq:parity}) in the right side for $k=1,...,M/2$ (see Fig.~\ref{fig:contour_one} for an illustration).

Similarly, the denominator of equation (\ref{eq:renyi_one}) can also be represented by a path integral with the same action (\ref{eq:syk_action_one}).
The only difference is in the boundary conditions. In fact, the absence of a twist operator implies that the boundary conditions for the denominator read 
\bea
\begin{aligned}
    &
    \psi_i(0) = -\psi_i(\beta_-), \quad \psi_{i}(\beta_+) =  - \psi_{i}(2\beta), \quad 
    i=1,...,N,  \\
    & \psi_{2k-1}\left(\frac{\beta_-}2\right) = i \psi_{2k}\left(\frac{\beta_-}2\right), \quad \psi_{2k-1}\left(\frac{\beta_+}2\right) = -i \psi_{2k}\left(\frac{\beta_+}2\right),  \quad k = 1,..., M/2,  \\
    & \psi_{2k-1}\left(\frac{3\beta_-}2\right) = i \psi_{2k}\left(\frac{3\beta_-}2\right), \quad \psi_{2k-1}\left(\frac{3\beta_+}2\right) = -i \psi_{2k}\left(\frac{3\beta_+}2\right),  \quad k = 1,..., M/2. 
    \end{aligned}
    \label{eq:denbc}
\eea

The next step is to rewrite the large-$N$ action in terms of bilocal fields analogous to the ones defined in equations~(\ref{eq:bilocal}) and (\ref{eq:offdiagonal}), but with arguments in the range $\tau_{1,2}\in (0,2\beta)$. We can then derive and numerically solve the equations of motion in the saddle point approximation for the numerator and denominator of equation (\ref{eq:renyi_one}), and compute the corresponding on-shell actions $I_{num}(m)$ and $I_{den}(m)$. The details of this procedure are reported in Appendix \ref{append:renyi_SYK}. Finally, the Renyi-2 mutual information in the saddle point approximation is immediately given by $I_{LR}^{(2)}(\Psi(m))=2S_L^{(2)}(\Psi(m))=2(I_{num}(m)-I_{den}(m))$.

\begin{figure}
    \centering
    \includegraphics[width=0.5\textwidth]{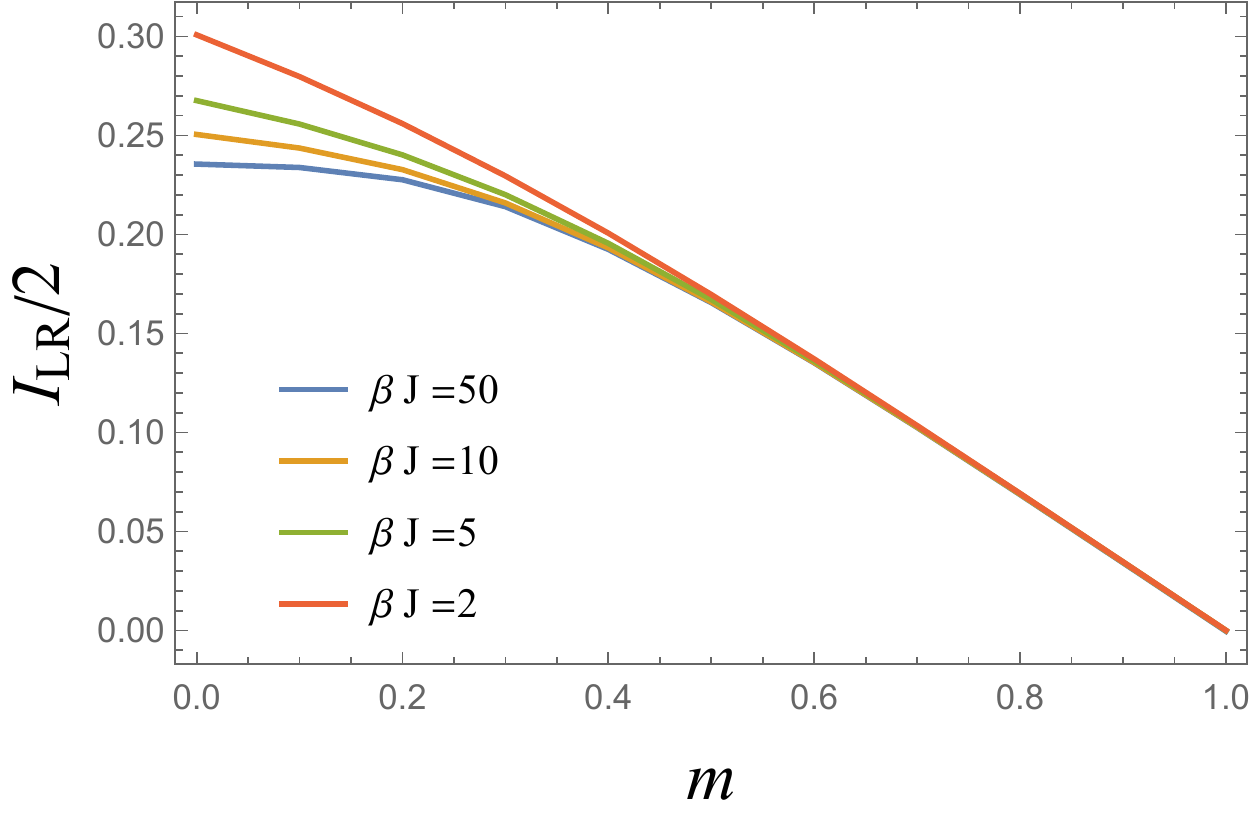}
    \caption{Renyi-2 mutual information $\mathcal{I}_{LR}^{(2)}(\Psi(m))$ between the left system and the unmeasured Majorana fermions on the right side. $m = M/N$ is the ratio of measured Majorana fermions on the right side. The mutual information is always non-vanishing for $m<1$, and vanishes only when the whole right side is measured ($m=1$).}
    \label{fig:syk-one-mi}
\end{figure}

The result of the numerical evaluation of the Renyi-2 mutual information $\mathcal I_{LR}^{(2)}(\Psi(m)) = I_{LR}^{(2)}(\Psi(m))/N$ is shown in Fig.~\ref{fig:syk-one-mi}. 
Note that the mutual information is finite for any value of $0< m < 1$, i.e. for any size of the subset $M$ of measured Majorana fermions. Only when all Majorana fermions are measured ($m=1$, corresponding to the analysis of \cite{kourkoulou2017pure}, see Appendix \ref{app:KM}) does the mutual information vanish. In other words, the left and right systems always remain entangled, no matter how large the measured subsystem in the right is. In the next section, we will find an analogous result by computing the holographic entanglement entropy in the bulk dual theory (given by JT gravity coupled to $N$ copies of a CFT). This will allow us to give in Section \ref{sec:teleportation} a holographic interpretation of our results in terms of bulk teleportation \cite{Antonini:2022sfm}: the effect of the partial measurement performed on the right system is to teleport bulk information previously contained in the entanglement wedge of the right system into the left system.
We also observe a phase transition in the behavior of the mutual information at large $\beta$ from nearly flat to linearly decaying. In the bulk dual picture, this corresponds to an entanglement wedge phase transition triggered by the measurement, as we will describe in Section \ref{sec:JT}.

\section{Partial measurement in JT gravity}
\label{sec:JT}

In this section, we explore the gravity dual of the SYK measurements considered above. In particular, our dual system will be Jackiw-Teitelboim (JT) gravity coupled to bulk matter, which we take to be described by $N$ copies of a CFT (capturing the $N$ Majorana fermions in the SYK model). In this and the following sections, we switch the side that is being measured; namely, we will measure the left side of the TFD. We can then calculate the mutual information in the post-measurement state between the right side and remaining fermions on the left side, this time using quantum extremal surface (QES) formula 
\cite{Engelhardt:2014gca}. In particular, the bulk description (which relies on the location of the QES) will make manifest the entanglement wedge of the right side and thus, in turn, demonstrate how information about the center of the bulk will shift around, becoming encoded in different boundary regions as a result of measurement. Specifically, when the number of measured fermions remains small ($m \lesssim 0.26$), a significant amount of the left asymptotic region will be encoded in the remaining (unmeasured) fermions of the left system. However, as the number of measured fermions becomes large ($m \gtrsim 0.26$), the quantum extremal surface will undergo a phase transition, and information about nearly all of the left side of the bulk will become accessible from the right side. As in \cite{Antonini:2022sfm}, we refer to this as teleportation of the bulk geometry, a notion which we will make more manifest in Section \ref{sec:teleportation} below.

Here, we start with a more in depth description of our bulk model in Section \ref{sec:jt_desc}, before moving to the holographic description of partial measurement and a computation of the mutual information in Section \ref{sec:jt_meas}.
Specifically, we propose for the QES prescription that the measured matter can be modeled by a boundary conformal field theory, as boundary measurements create an end-of-the-world brane in the bulk that serves as a boundary for the bulk conformal matter.
Along with a comparison with the SYK results above, we find strong agreement between the QES prescription and the SYK calculation.

\subsection{JT gravity with matter CFT} 
\label{sec:jt_desc}

The action for JT gravity coupled to matter CFTs is given by
\bea\label{eq:JT_action}
\begin{aligned}
    I =& I_{JT}(g,\phi) + I_M(g, \chi) \\
	I_{JT}(g, \phi) =& - \frac{\phi_0}{16\pi G_N} \Big[ \int d^2 x \sqrt{g}  R +  2\int_{\partial \mathcal M} K \Big] \\
	&- \frac{1}{16\pi G_N} \Big[ \int d^2 x \sqrt{g} \phi(R+2) + 2\int_{\partial \mathcal M} \phi|_\text{bdy} K\Big]\label{eq:jt_dyn}, 
\end{aligned}
\eea
where $\Phi = \phi_0 + \phi$ is a dilaton field, with $\phi$ a small fluctuation around a uniform background $\phi_0$. 
If we view this theory as coming from the dimensional reduction of a higher dimensional theory, $\phi_0$ would denote the area of the black hole horizon at extremality (with $\phi$ giving the area deviations for near extremal black holes), and is therefore related to ground state entropy of an extremal black hole.
The first line of the JT action is a purely topological term, so that the dynamical part of the action comes solely from the second line of equation (\ref{eq:jt_dyn}).  
We take the matter action $I_M(g,\chi)$ to be given by $N$ copies of a CFT with central charge $c$ (which we will take to be $c=1/2$ when comparing with the SYK results of the previous section, as each Majorana fermion can be thought of as half of a Dirac fermion).

The resulting equations of motion read
\bea
\begin{aligned}
	R &= -2, \\
	T_{\mu\nu}&= \frac1{8\pi G_N} (\nabla_\mu \nabla_\nu \phi - g_{\mu\nu} \nabla^2 \phi + g_{\mu\nu} \phi) ,
 \end{aligned}
\eea
where $T_{\mu\nu}$ is the stress tensor associated with the matter fields of $I_M$. 
The metric of the (Euclidean) eternal black hole is given by (see Appendix ~\ref{append:JT} for a derivation)
\bea
    ds^2 = \frac{4\pi^2}{\beta^2} \frac{d\sigma^2 + d\tau^2}{\sinh^2 \frac{2\pi}\beta \sigma},
\eea
where $\tau$ is the imaginary time.
In these coordinates, which cover one exterior portion of the two-sided black hole, the horizon is located at $\sigma \rightarrow \infty$,  and the asymptotic boundary is located at $\sigma = \epsilon$. At the boundary, the induced metric satisfies the condition $g = \frac1{\epsilon^2}$, and the dilaton field  satisfies $\phi(\epsilon) = \frac{\phi_r}{\epsilon}$. This leads to the dilaton profile, 
\bea
    \phi(\sigma) =  \phi_r \frac{2\pi}\beta \frac1{\tanh \frac{2\pi }{\beta}\sigma}.
\eea

\begin{figure}
    \centering
    \includegraphics[height=0.25\textwidth]{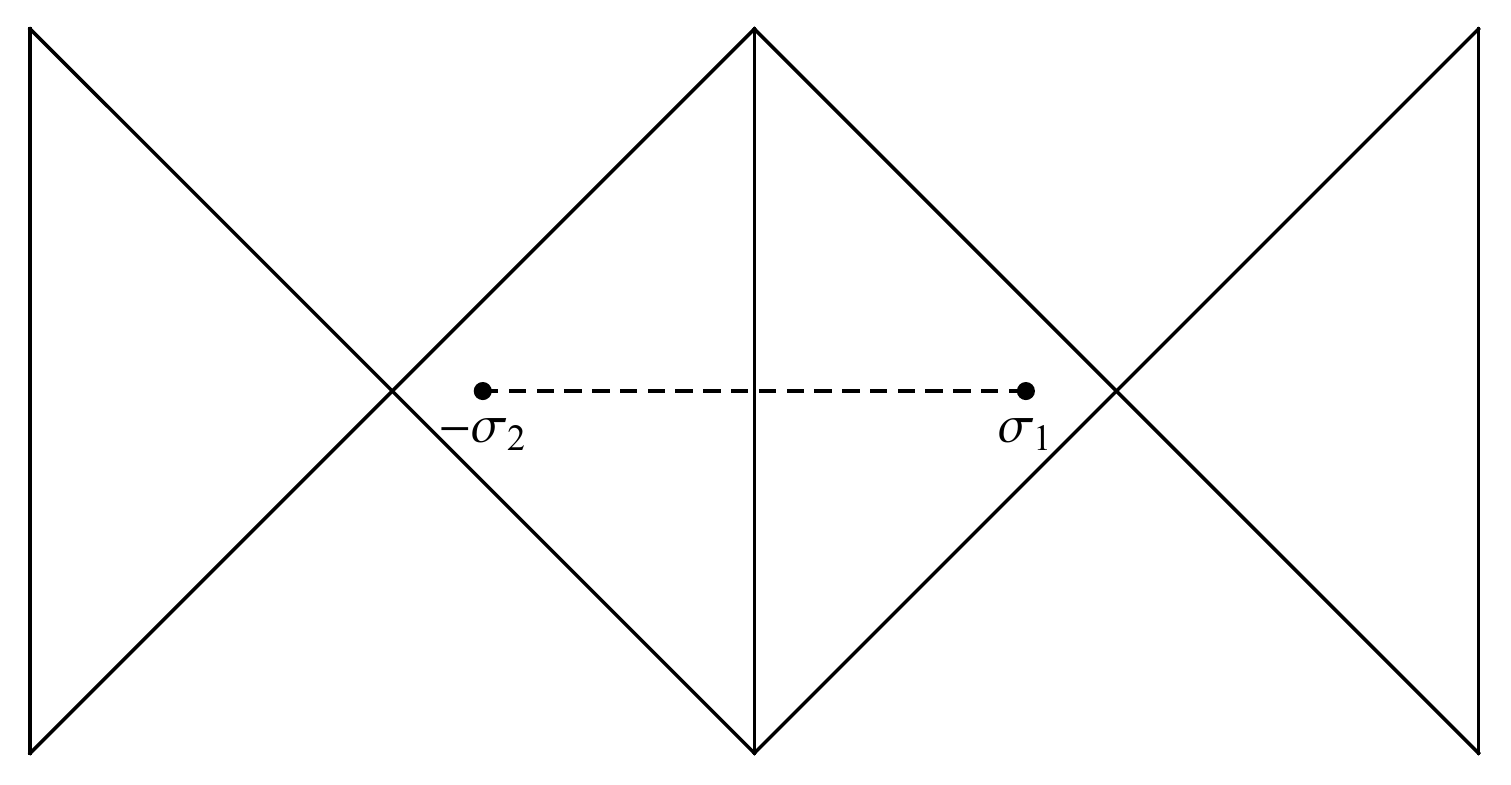}
    \caption{ 
    The doubled spacetime used to compute the bulk matter entropy. The boundary at the center of the figure is transparent. The left-most boundary and the right-most boundary are identified. 
    $\sigma_1$ and $-\sigma_2$ denote twist operators inserted in the first and second copies of the black hole.}
    \label{fig:JT_rindler}
\end{figure}

We will compute the entropy of a subset $R$ of the boundary system via the quantum extremal surface formula
\bea
    S(R) = \min\{\underset{\sigma}{\text{ext}} S_\text{gen}(\sigma)\},
\eea
i.e. the smallest, extremal value of $S_\text{gen}$, the generalized entropy, which is given by the sum of the area of $\sigma$ (divided by $4 G_N$) plus the bulk matter entropy in a region bound by $\sigma$.  More explicitly, we compute the entropy by $S_\text{gen} = \frac{\Phi(\sigma)}{4G_N} + S_\text{CFT}(\sigma)$, evaluated at an extremal value of $\sigma$, which we call the quantum extremal surface.  
In order to do this, we need to first calculate the bulk entanglement entropy of some region extending out to the boundary. We can do this by employing a ``doubling trick,'' as in \cite{nezami2021quantum}. The matter in the spacetime will obey reflecting boundary condition at the asypmtotic boundary $\sigma = \epsilon$. We can related the bulk entropy in this space to the entropy on a cylinder -- specifically, a doubled space constructed by gluing together two copies  of a black hole at their boundaries, this time with transparent boundary conditions (with a left moving mode in one copy joined to a right moving mode in the second copy). See Fig.~\ref{fig:JT_rindler} for a depiction.

In this doubled space, we can calculate the entropy of a bulk subregion (at $t=0$) by inserting twist operators at points $(\sigma_1,0)$ in the first copy and $(-\sigma_2,0)$ in the second copy. 
The entanglement entropy of the matter is then given by~(see Appendix~\ref{append:JT})
\bea\label{eq:s_double}
    S_\text{double}([-\sigma_2, \sigma_1]) =  \frac{c}6 \log \left[ \frac{4\sinh^2 \frac{\pi}\beta(\sigma_1 + \sigma_2)}{ \sinh \frac{2\pi}\beta \sigma_1 \sinh \frac{2\pi}\beta \sigma_2} \right],
\eea
where the subscript indicates the doubled theory. 
Therefore, in the original theory, a single twist operator at $(\sigma, 0)$ gives an entanglement entropy
\bea
    S_\text{CFT}(\sigma) = \frac12 S_\text{double}([-\sigma, \sigma ]) = \frac{c}{6} \log 2. 
\eea
Note that from (\ref{eq:s_double}), the entanglement entropy does not depend on the value of $\sigma$.

If we consider the entanglement entropy of the full right system, the quantum extremal surface will be at the bifurcation surface as determined by symmetry. The generalized entropy is then given by
\bea
\begin{aligned}
    S_\text{gen} 
    &= \frac{\phi_0}{4G_N} + \frac{\phi_r}{4G_N} \frac{2\pi}{\beta} + \frac{N c}{6} \log 2, \\
    &= \frac{\tilde\phi_0}{4G_N} + \frac{\phi_r}{4G_N} \frac{2\pi}{\beta}, \label{eq:right-side}
\end{aligned}
\eea
where in the first line, we have included $N$ copies of the matter CFT, and in the second line for simplicity, we have defined $ \frac{\tilde \phi_0}{4G_N} = \frac{\phi_0}{4G_N} + \frac{N c}{6} \log 2 $. 
To make contact with the SYK model, we note that the first term corresponds to the ground state entropy, and the second term is a correction linear in temperature.

\subsection{Measurement-induced entanglement wedge transition}
\label{sec:jt_meas}
\begin{figure}
    \centering
    \includegraphics[width=0.3\textwidth]{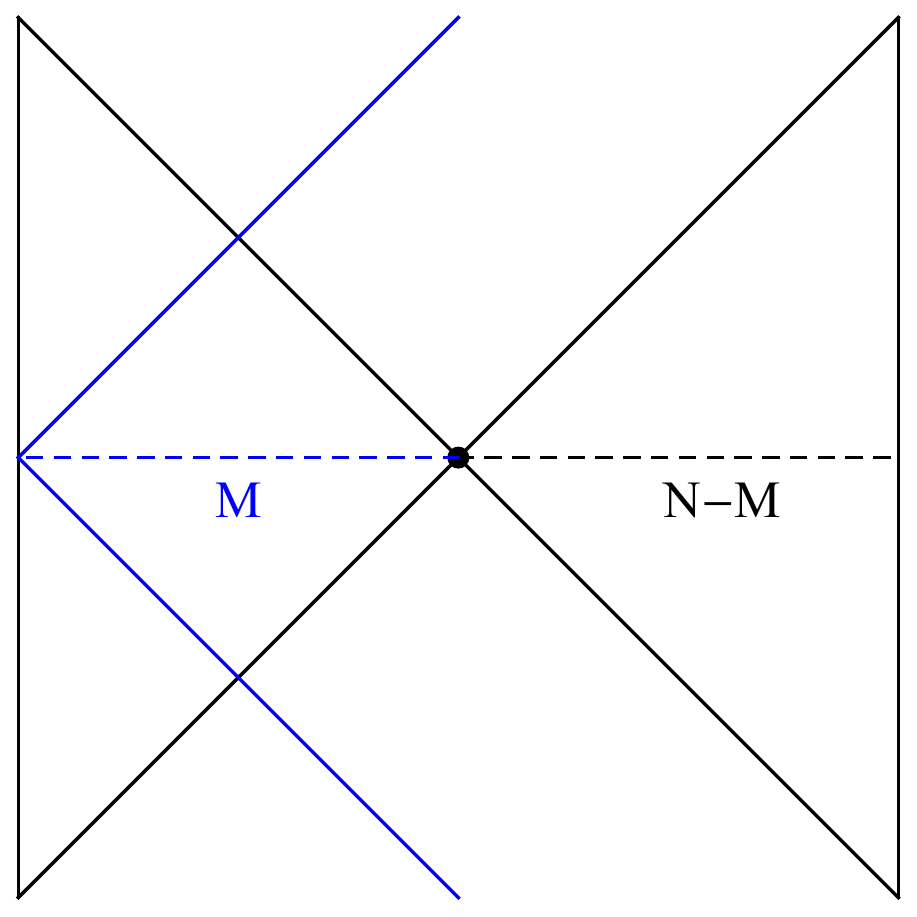}
    \caption{A boundary CFT in the $AdS_2$ background. 
    The right boundary is the conventional $AdS$ boundary.
    Boundary conditions associated with the projective measurement of part of the left Majorana fermions are imposed at the left boundary. An ETW brane visible to $M$ bulk fermions is depicted in blue, where $M$ is the number of fermions measured in the left side.}
    \label{fig:BCFT}
\end{figure}

With an expression for the bulk generalized entropy in hand, we can now investigate the effect of performing a measurement on one side, and in particular, calculate the mutual information between the two sides. We know that measuring the full left side will create an end-of-the-world brane in the left asymptotic region \cite{kourkoulou2017pure} (and, of course, measuring nothing will leave the bulk unchanged). 
We thus expect that a natural bulk dual to partial measurement is the creation of an end-of-the-world brane in the bulk, but one only visible to the bulk fields dual to the measured fermions. As such, we model the bulk matter for the measured fields by a BCFT living in the bulk, with their boundary at the end-of-the-world brane. Thus, if we measure $M$ Majorana fermions on the left side, we will have $N-M$ copies of a normal matter CFT corresponding to the unmeasured fermions, and $M$ copies of a BCFT corresponding to the measured fermions, as shown in Fig.~\ref{fig:BCFT}.

To calculate the bulk entropy arising from the measured fermions, we insert a twist operator for the BCFT at $(\sigma,0)$, which leads to the following entanglement entropy
\bea \label{eq:EE_BCFT}
    S_\text{BCFT}(\sigma) =  \frac{c}6 \log \left[ \frac{2\sinh \frac{\pi}\beta(2\sigma)}{ \sinh \frac{2\pi}\beta \sigma} \right] + \log g = \frac{c}6 \log 2 + \log g,
\eea
where $\log g$ is the boundary entropy of the BCFT~\cite{cardy1989boundary,affleck1991universal}.

For $M \ll N$, we do not expect the position of the quantum extremal surface to change much. We can verify this by calculating the entanglement entropy of the right side, 
\bea \label{eq:jt_sgen}
\begin{aligned}
    S_\text{gen}(M) &= \frac{\phi_0}{4G_N} + \frac{\phi_r}{4G_N} \frac{2\pi}\beta \frac1{\tanh \frac{2\pi }{\beta}\sigma} + (N-M) S_\text{CFT}(\sigma)  + M S_\text{BCFT}(\sigma), \\
    &=\frac{\tilde\phi_0}{4G_N} + \frac{\phi_r}{4G_N} \frac{2\pi}\beta + M \log g.
\end{aligned}
\eea
In the second line, because $S_\text{CFT}(\sigma)$ and $S_\text{BCFT}(\sigma)$ do not depend on the end point, we find $\sigma = \infty$ as the location for the quantum extremal surface, as expected.

Because $M \ll N$, there will still be a large mutual information between the right side and the unmeasured Majorana on the left after projection. Further, because the full state is pure, the entanglement entropy of the unmeasured  Majorana on the left side will be equal to that of the right. 
As a consequence, the entanglement wedge of the left side will also extend up to the bifurcation surface. As above, we include twist operators for all copies of matter fields at the QES, including $K=N-M$ copies of a CFT, and $M$ copies of a BCFT, yielding the same entropy as in (\ref{eq:jt_sgen}).

We can also consider the opposite limit, when we measure a large number of Majorana fermions, i.e., $M \sim N$ (such that the remaining number of unmeasured fermions is small $K \ll N$). 
To understand this limit, we invoke the bulk picture suggested in \cite{nezami2021quantum}: if we have access to the remaining $K\ll N$ unmeasured boundary Majorana fermions of the left system, we expect to be able to reconstruct their dual bulk fields in a small region near the boundary. In particular, the state of the fermions remains constant on time scales $\sim 1/J$. HKLL bulk reconstruction~\cite{hamilton2006holographic} then implies we can reconstruct the dual fields in a small wedge near the boundary (the bulk domain of dependence of this boundary timelike region). This wedge will have size on the order of the UV cutoff $\epsilon \sim 1/J$. As a result, these fermions will have a bulk entropy associated to this UV bulk region. Since the entanglement wedge is now at the UV cutoff, we take the QES to be empty (and hence take the area contribution from the dilaton to be zero). 

To compute the bulk entropy associated with the unmeasured fermions in the UV cutoff region, we note that the reduced density matrix of the unmeasured Majoranas can be found by projecting the thermal density matrix onto a fixed state. In the low temperature limit, the thermal density matrix reduces the effective available space to a low energy subsector. Therefore, the effect of measurement is constrained to that reduced low energy subspace as well. Because the SYK model is chaotic, we expect the entanglement entropy of the unmeasured fermions when $K \ll N$ will be dominated by subsystem entanglement entropy evaluated in the ground state, which is given by~\cite{huang2019eigenstate}
\bea\label{eq:SUV}
     \frac{S_\text{UV}}{N} =  \lambda \left( \frac{\log 2}{2} - \frac1{16} \arcsin \lambda^{3/2} \right), \quad \lambda = K/N. 
\eea
where the second term originates from the nontrivial spectral density near the spectrum edge in the SYK model~\cite{garcia2017analytical}.

\begin{figure}
    \centering
    \includegraphics[width=0.5\textwidth]{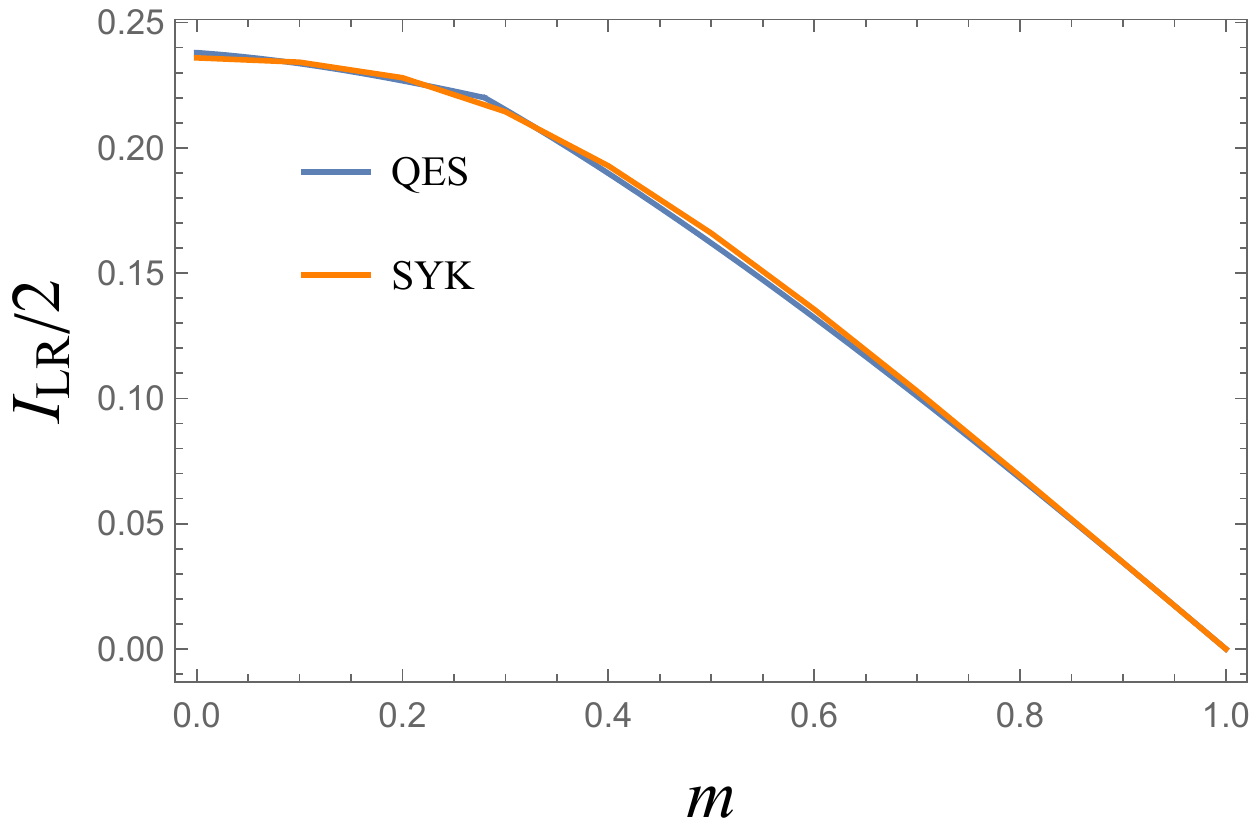}
    \caption{Comparison of the Von Neumann mutual information computed in this section using the QES prescription and the Renyi-2 mutual information compued in Section \ref{sec:SYKMI} for the SYK model.  We choose the parameters $S_0 = 0.2324 N$, $C = 0.28 $, $c= 0.5$, and $\beta \mathcal J = 50$. The two results are in good agreement. In the bulk calculation, the phase transition is sharp, while in the SYK calculation it is smoothed out.}
    \label{fig:syk_partial-measure}
\end{figure}

With these two limits fairly well understood, in the intermediate regime ($K\sim M$), we take the entropy to be given by $S_\text{QES} = \min\{S_\text{gen}, S_\text{UV}\}$. 
While this bulk protocol may potentially seem ad hoc, we can justify the prescription 
by comparing it to the SYK calculations done in Section \ref{sec:measurement}. As demonstrated in Fig.~\ref{fig:syk_partial-measure}, we find a good match between the two approaches. To make this comparison, we take
\bea
    \frac{\tilde\phi_0}{4G_N} = S_0, \quad  \frac{\phi_r}{4G_N} = \frac{N C}{2\pi \mathcal J}.
\eea
and $\log g= - \frac{\log2}{12}$~\footnote{This value of the boundary entropy is determined by matching the SYK data. However, it is interesting to observe that if we imagine the bulk matter is dual to $N$ decoupled EPR states, then after one partner in each EPR state is measured, the entanglement entropy for the other partner simply vanishes. This means that the boundary entropy should cancel the bulk contribution to the entropy for the BCFT, which would predict $\log g = - \frac{c}6 \log2$ from (\ref{eq:EE_BCFT}). In the $c=1/2$ case of our interest, this gives exactly the value of the boundary entropy obtained by matching the SYK result.}. 
Notice that since the full state is pure, the mutual information is given by $I_{LR} = 2 S_\text{QES}$.

\begin{figure}
    \centering
    \subfigure[]{
    \includegraphics[width=0.25 \textwidth]{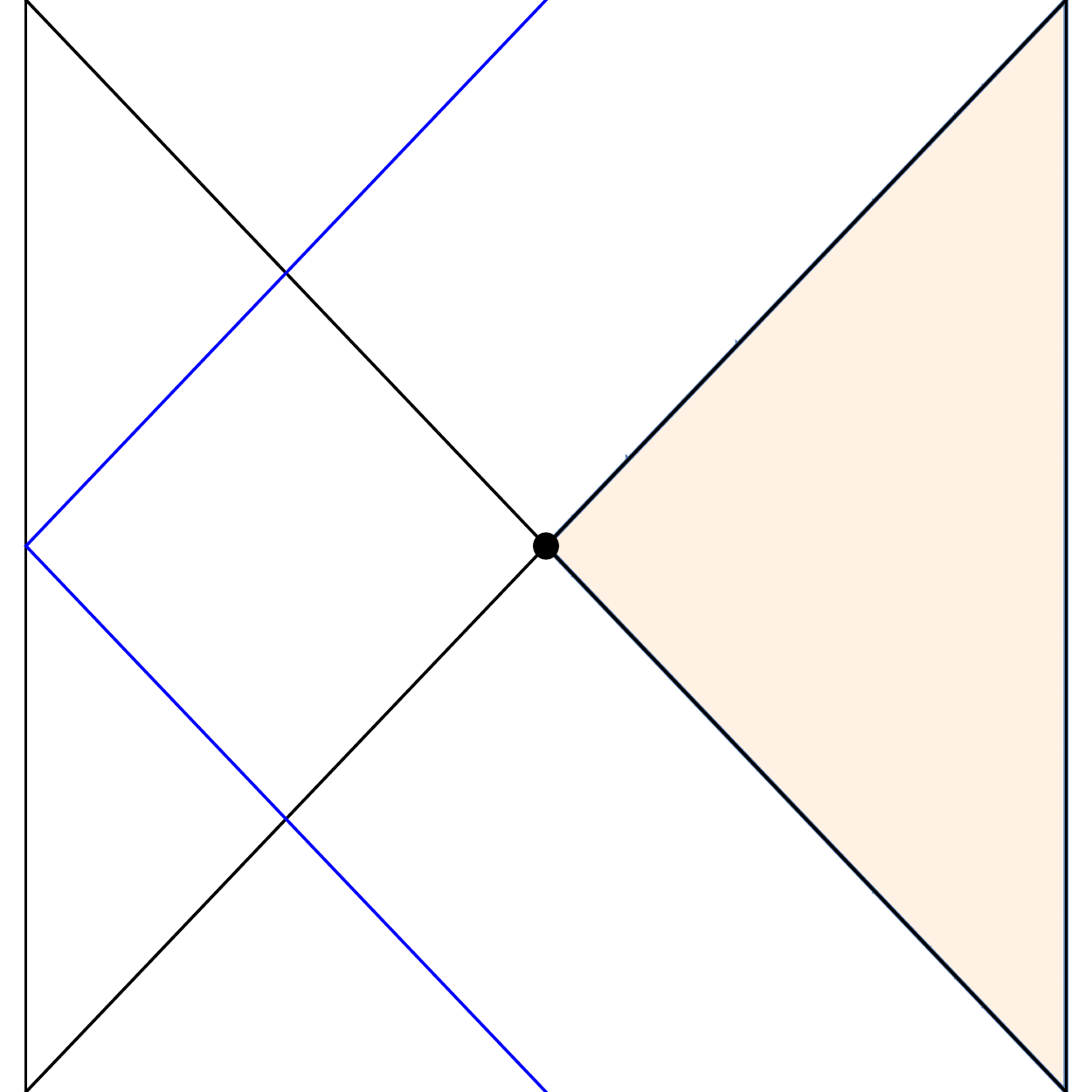}} \quad \quad \quad 
    \subfigure[]{
    \includegraphics[width=0.25 \textwidth]{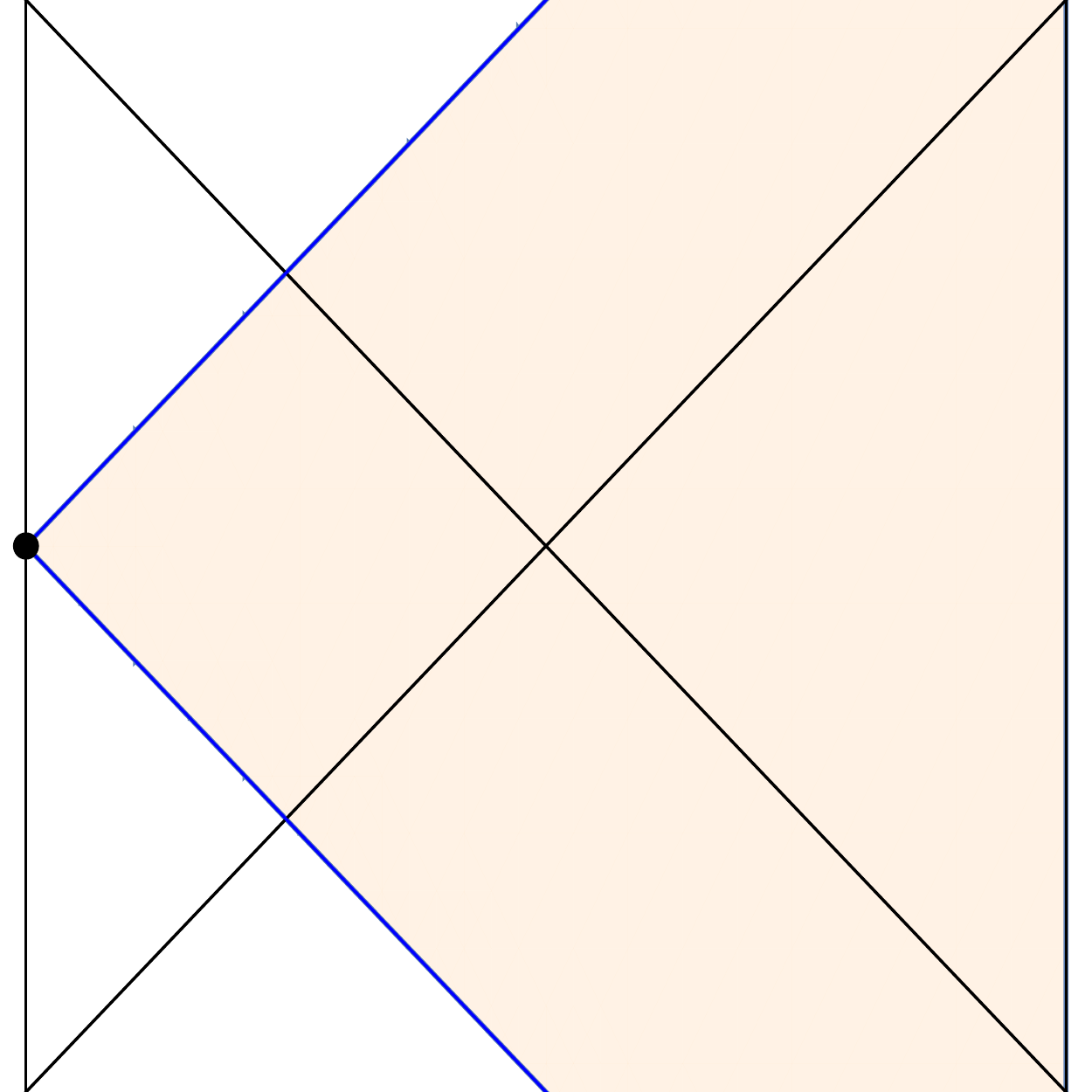}}
    \caption{Measurement-induced entanglement wedge transition. 
    The tan color denotes the entanglement wedge of the right side. 
    (a) When a small subset of Majorana fermions is measured in the left side, i.e. $m<m_*$, the entanglement wedge of the right side is the right Rindler patch.
    (b) When a large subset of Majorana fermions is measured in the left side, i.e. $m>m_*$, the entanglement wedge of the right side includes part of the interior and of the left Rindler patch. 
    }
    \label{fig:EW_transition}
\end{figure}

For $S_0 = 0.2324 N$, $C = 0.28 $, $c= 0.5$, and $\beta \mathcal J = 50$, the phase transition in the quantum extremal surface  occurs at $\lambda_* \approx 0.72 $, see Fig. \ref{fig:syk_partial-measure}. Therefore, when $M >m_* N \approx 0.26 N $ (with $m_*=1-\lambda_*$)  Majorana fermions are  measured in the left system, the information about the bulk that would have been encoded in those fermions becomes available to the right system, similar to \cite{Antonini:2022sfm}. In other words, we find that performing a projective measurement on about a third of the left system is sufficient to allow us to teleport the information about the left side of the bulk into the right boundary, except for a small wedge at the cutoff of the left boundary. On the other hand, when a small number of fermions are measured, i.e. $M<m_*N\approx 0.26 N$, the bulk information contained in the $M$ measured fermions is teleported into the unmeasured $K=N-M$ fermions in the same side. Therefore, the quantum extremal surface still sits at the bifurcation surface and the information about the left side of the bulk can still be accessed from the unmeasured fermions in the left side.
An illustration of the entanglement wedge transition is shown in Fig.~\ref{fig:EW_transition}.  

We will make efforts to better understand this transfer of bulk information by exploring quantum teleportation and traversable wormholes in the next section.

\section{Quantum teleportation}
\label{sec:teleportation}

\begin{figure}
    \centering
    \subfigure[]{
    \includegraphics[width=0.25\textwidth]{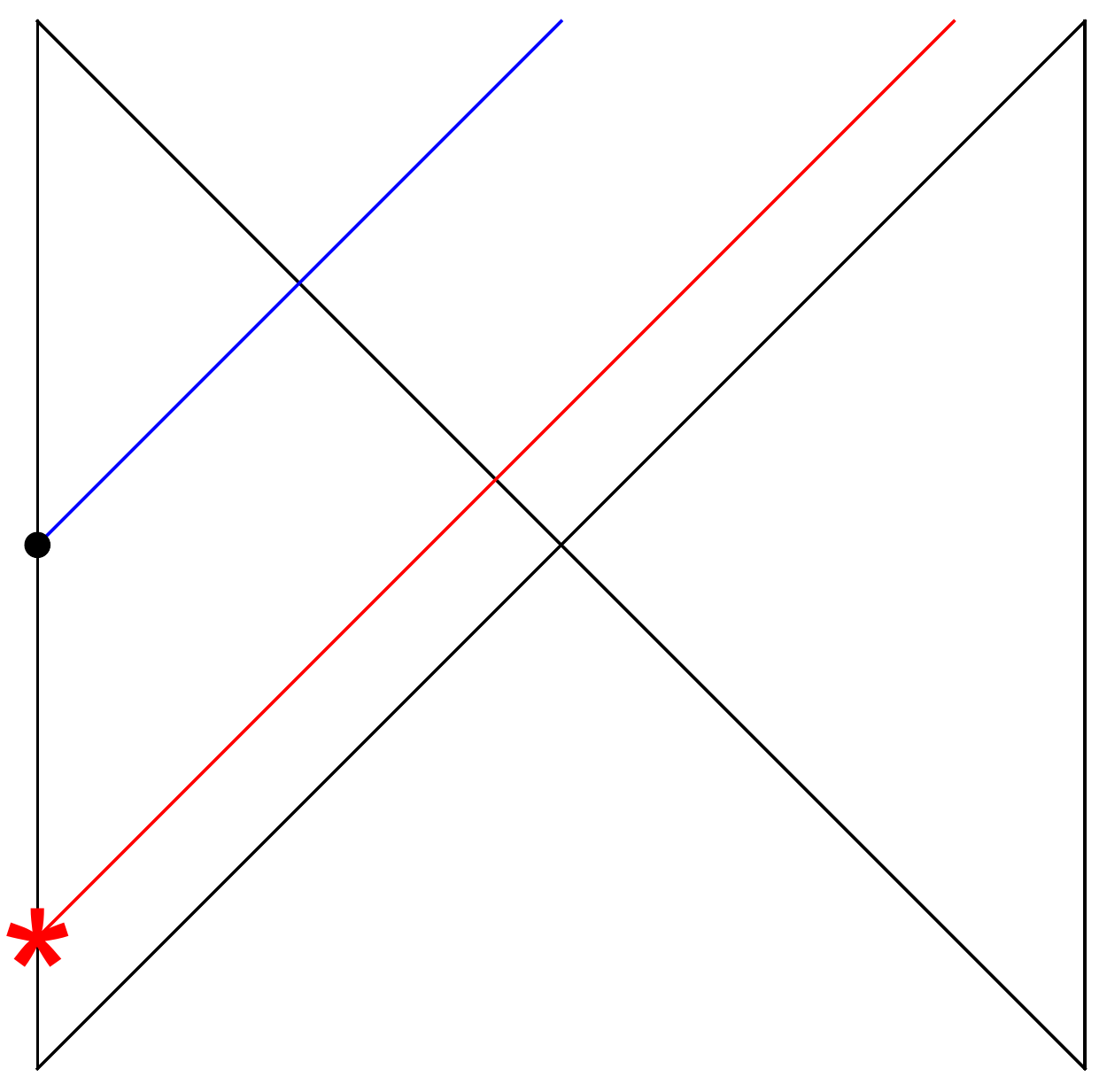}} \quad \quad \quad \quad 
    \subfigure[]{
    \includegraphics[width=0.25\textwidth]{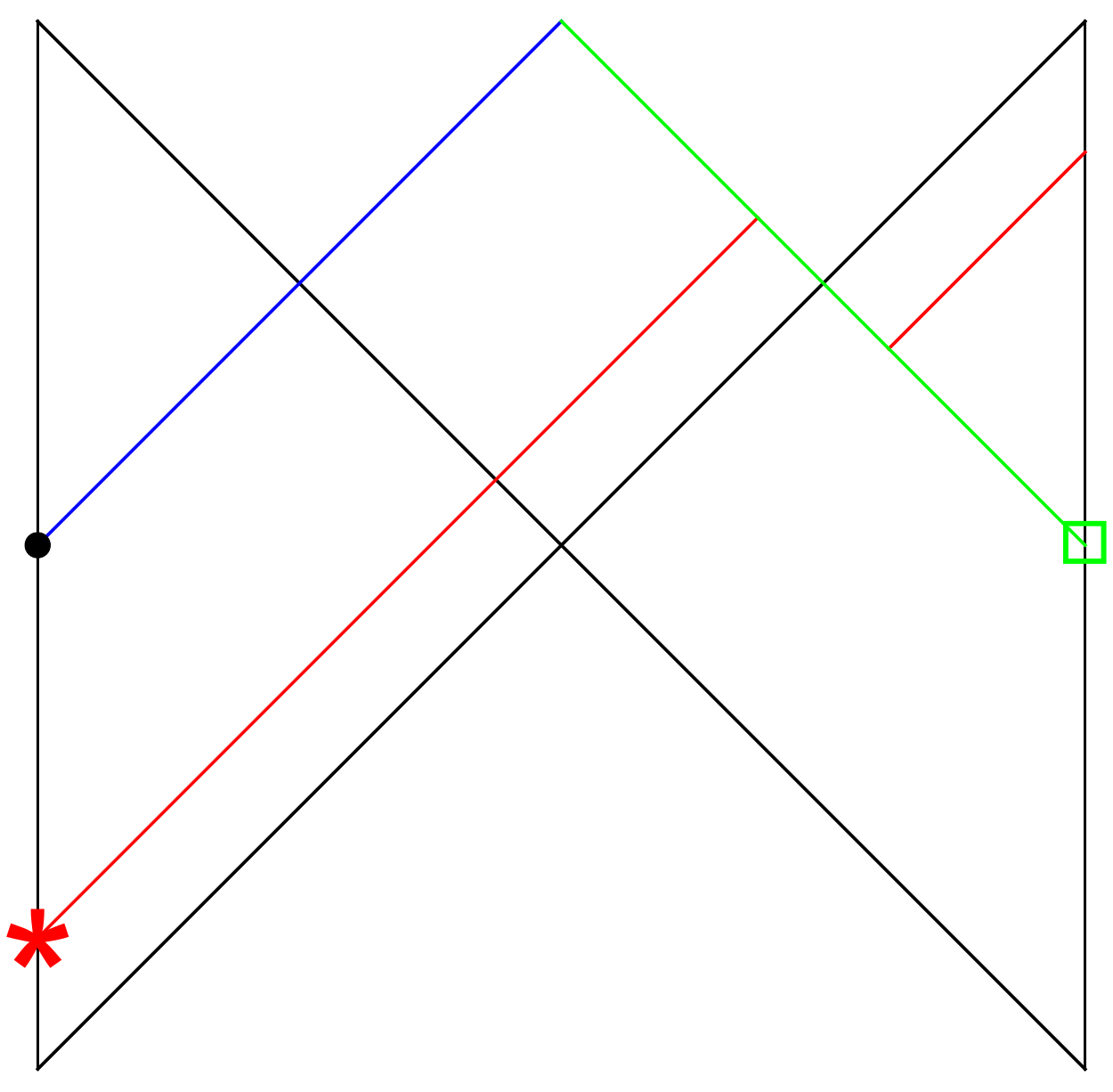}}
    \caption{The red line denotes the qubit that is inserted in the left side at time $-t$. 
    (a) Measurement is implemented at time $t=0$ in the left side, and it creates an end-of-the-world brane. 
    The qubit cannot escaped from the horizon to the right side.
    (b) Measurement is implemented at time $t=0$ in the left side, in the mean time, a decoding operator according to the measurement outcome is implemented in the right side. 
    This measurement-decoding protocol can extract the qubit out from horizon to the right side at time $t$. }
    \label{fig:traverse}
\end{figure}

As noted in the previous section, when a sufficient number of boundary fermions are measured ($m > m^*$), the entanglement wedge of the right boundary jumps to contain nearly the entire bulk (see Fig.~\ref{fig:EW_transition} (b)).  Consequently, almost all of the bulk (including most of the left wedge) can be reconstructed solely from the right system. 
To make this concrete, we can imagine releasing a particle on the left side at time $t=-t_0$, then measuring the left side at time $t=0$.
The measurement will create an end-of-the world brane in the bulk. However,  this will not effect the particle, which will continue to fall into the black hole, and in particular will not escape out to the right boundary, as illustrated in Fig.~\ref{fig:traverse} (a).
To extract the information from the right side, we must construct a decoding operator using information from the measurement.
In the following section, we will see that by applying such a decoding operator to the right side, we will be able to extract the particle at $t = t_0$ in the right side. 
This is simply a quantum teleportation protocol, which can be used to explain the quantum information origin of the entanglement wedge transition explored above. 
Unsurprisingly, this boundary teleportation protocol can be related to a traversable wormhole in the bulk~\footnote{Note that the wormhole will be traversable from only one side, similar to the analysis of  \cite{maldacena2017diving}.}.
In the following, we will first describe the teleportation protocol in the SYK model, which is realized as a quantum channel, and then calculate a left-right correlation function, (\ref{eq:lr_measurement}), that is closely related to not only the teleportation fidelity under such a quantum channel, but also traversability in the dual picture. 
We will then show that the dual of this protocol provides a realization of traversable wormhole from the left side. 
Finally, we consider an encoding method to reach nearly perfect teleportation fidelity of an arbitrary qubit state.

\subsection{Teleportation protocol}
Motivated by the measurement-induced entanglement wedge phase transition studied above, we want to construct a protocol to recover the information initially stored in the left side from the right side after measurement, when the entanglement wedge extends up to the left boundary.
Before describing our protocol, we should note that similar teleportation protocols have been discussed in Refs.~\cite{brown2019quantum, nezami2021quantum, schuster2022many}. 
Nevertheless, in the previous literature, instead of the measurement-decoding protocol normally associated with quantum teleportation, a two-sided operator is implemented in the SYK model to render the wormhole traversable.   
While it was argued that the two-sided coupling could be written in terms of a measurement and decoding ~\cite{maldacena2017diving,gao2021traversable,Milekhin:2022bzx}, the precise decoding protocol was not studied in detail. 
For instance, in the SYK model, the two-sided operator has the form of $\exp(i \theta \sum_i \psi_{L,i} \psi_{R,i})$, which consists of a fermionic operator in the left and right side. Thus, it is not possible to measure the fermionic operator and get an outcome.

Here, we investigate the usual measurement-decoding quantum teleportation protocol using a measurement operator of the form $i \psi_{L,2k-1} \psi_{L,2k}$ acting  on the left side. 
Since it is a bosonic operator, we can get a measurement outcome and accordingly implement the decoding operator in the right side.

\begin{figure}
    \centering
    \includegraphics[width=0.37\textwidth]{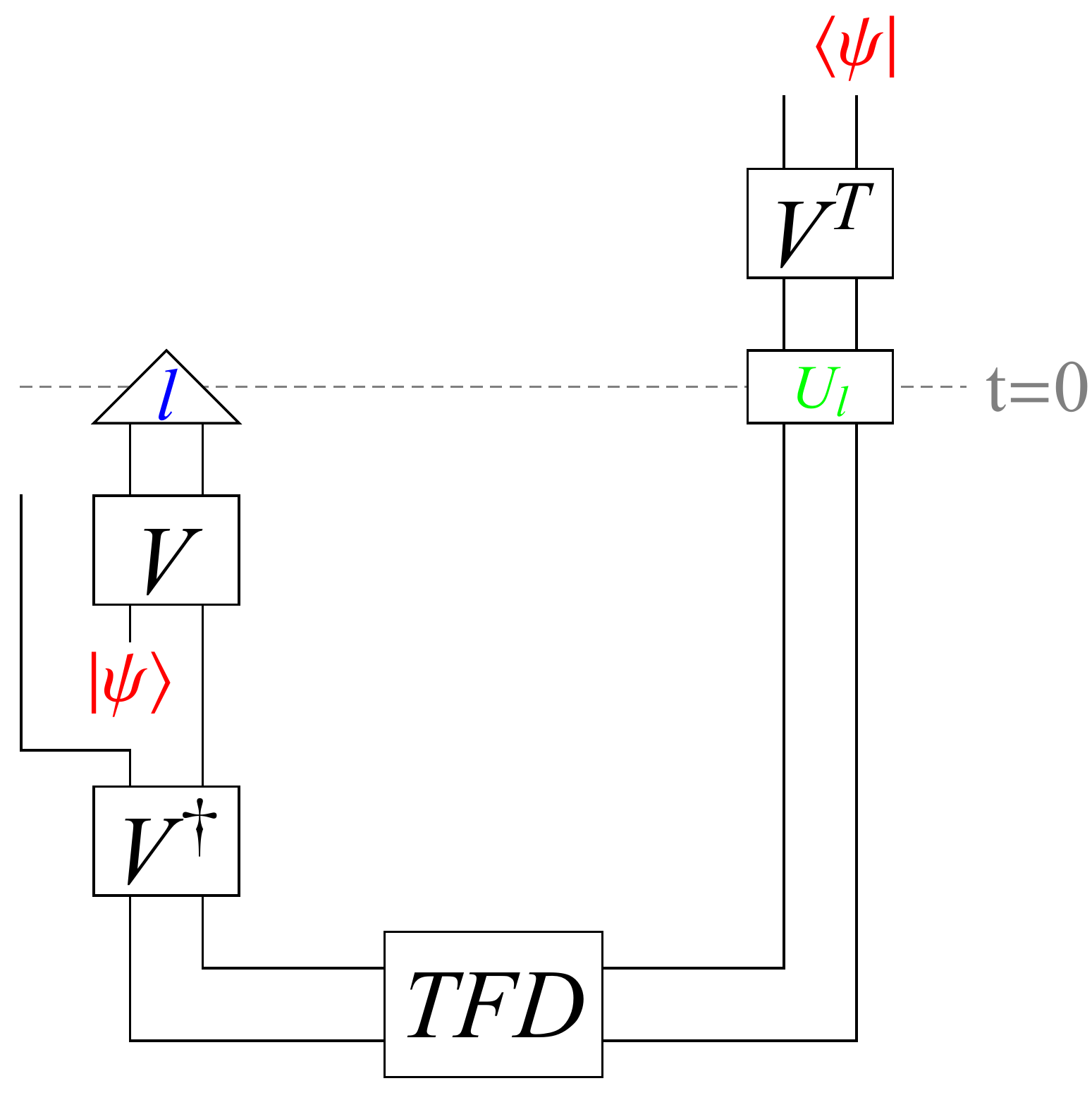}
    \caption{The circuit of the teleportation protocol. 
    In the left side, an unknown state $|\psi \rangle$ is inserted in the past $-t_0$, and a measurement is applied at $t=0$. 
    The projection is indicated by the triangle with the measurement outcome $\textbf l$.
    According to the measurement outcome, a decoding operator, denoted by $U_\textbf{l}$, is implemented in the right side at the same time $t=0$. 
    After a time evolution $t_0$, the state $|\psi \rangle$ is teleported to the right side. 
    $V$ denotes the time evolution operator.}
    \label{fig:circuit}
\end{figure}

The circuit version of Fig.~\ref{fig:traverse} is shown in Fig.~\ref{fig:circuit}. 
The teleportation protocol consists of three steps:
\begin{enumerate}
    \item At time $t = -t_0$, an unknown state $|\psi \rangle$ is inserted in the left side;
    \item At time $t = 0$, a projective measurement is performed in the left side and a decoding operator is implemented in the right side according to the measurement outcome~\footnote{Here, we consider measuring the whole left side for simplicity. In general, measuring any subset of the Majorana fermions exceeding the critical value $m^*$ (as suggested by the entanglement wedge transition) will be enough. It would be interesting to determine the critical number of measured fermions needed for a successful teleportation. We leave this to future investigation.};
    \item At time $t = t_0$, the state $|\psi \rangle$ is teleported to the right side.
\end{enumerate}
At time $t=0$, the measurement-decoding protocol should be understood as a quantum channel, which can be defined by Kraus operator,
\bea \label{eq:kraus}
    K_{\textbf l} = ( | L_\textbf{l} \rangle  \langle  L_\textbf{l} |) \otimes U_{\textbf l}, 
\eea
where $|L_\textbf{l} \rangle$ is the measurement outcome on the $k=1,...,N/2$ qubits, i.e.,
\bea \label{eq:measurement_operator_l}
    -i 2 \psi_{L,2k-1} \psi_{L, 2k} | L_\textbf{l} \rangle =  l_k |L_\textbf{l}\rangle, \quad k=1,...,N/2.
\eea
This means that the decoding operator we perform on the right side explicitly depends on the measurement outcome in the left side.
Conditioned on this measurement outcome, the full protocol is shown by the circuit in Fig.~\ref{fig:circuit}.

To quantify the teleportation fidelity, we consider the {\it left-right} correlation function $C_{\textbf{l},Q}$ conditioned on the measurement outcome~\footnote{This correlation function is closely related to the state fidelity, see Appendix~\ref{append:fidelity}.},
\bea \label{eq:lr_measurement}
\begin{aligned}
     C_{\textbf{l},Q} &= \langle TFD |  Q_R(t_0) \left(   |L_{\textbf l} \rangle \langle L_{\textbf l} |  \otimes U_{\textbf l} \right) Q_L(-t_0) |TFD \rangle, \\
    Q_L( -t_0) &= (V_L Q V^\dag_L) \otimes \mathds 1, \quad  Q_R(t_0) = \mathds 1 \otimes (V_R^\dag  Q V_R),
\end{aligned}
\eea
where $V_{L,R}$ denotes the time evolution operator in the left and right side. 
$Q$ is the operator being teleported. 
We assume the operator is properly normalized, $Q^\dag Q = \mathds 1$, so that the magnitude of the correlation function through the measurement-decoding channel (summing over all Kraus operators) is at most one.

As discussed in Section~\ref{sec:measurement}, in the SYK model with measurement given by~(\ref{eq:measurement_operator_l}), the Born probability does not depend on the measurement outcome. 
The decoding operator associated with the measurement is 
\bea \label{eq:decode}
    U_{\textbf l} =   \exp\left( -i \frac{\theta}q \sum_{k=1}^{N/2} i l_k \psi_{R,2k-1} \psi_{R,2k}\right).
\eea
where $\theta$ is a tuning parameter. 
Later, we will tune $\theta$ to maximize left-right correlation. 
Moreover, we will see in the next subsection that the left-right correlation function does not depend on the measurement outcome. 
Therefore, we define the following left-right correlation function for the SYK model
\bea \label{eq:lr_correlation}
    C_Q = \frac{\langle TFD | Q_R^\dag(t_0) \left( | L_{\textbf l}\rangle \langle L_{\textbf l} | \otimes U_{\textbf l} \right) Q_L(-t_0) |TFD \rangle}{\langle TFD |  \left( | L_{\textbf l}\rangle \langle L_{\textbf l} | \otimes \mathds 1 \right) |TFD \rangle },
\eea
where we have omitted the subindex denoting measurement outcome because the result is independent of it. 
The denominator is the probability of having the measurement outcome $\textbf{l}$. 
Further, because the probability is independent of the outcome, we have $\langle TFD |  \left( | L_{\textbf l}\rangle \langle L_{\textbf l} | \otimes \mathds 1 \right) |TFD \rangle = 1/2^{N/2}$, due to the $2^{N/2}$ different outcomes. 
Thus, $C_Q = \sum_{\textbf l} C_{\textbf l, Q}$, and since we have properly normalized the operator $Q$, this correlation function has maximal magnitude one.
In the following, we will use large-$q$ techniques to calculate this left-right correlation function~(\ref{eq:lr_correlation}), and show that by properly tuning the parameter $\theta$, the fidelity can be made close to one.

\subsection{Left-right correlation function in the large $q$ limit}

Motivated by the left-right correlation function~(\ref{eq:lr_correlation}), we define the following {\it twisted} correlation function
\bea
    \mathcal G_Q = \frac{\langle TFD | Q_R^\dag(t_0) \left( | L_{\textbf l}\rangle \langle L_{\textbf l} | \otimes U_{\textbf l} \right) Q_L(-t_0) |TFD \rangle}{\langle TFD |  \left( | L_{\textbf l}\rangle \langle L_{\textbf l} | \otimes U_{\textbf l} \right) |TFD \rangle }.
\eea
It is related to the left-right correlation function~(\ref{eq:lr_correlation}) by
\bea \label{eq:relation}
    C_Q = \frac{\langle TFD |  \left( | L_{\textbf l}\rangle \langle L_{\textbf l} | \otimes U_{\textbf l} \right) |TFD \rangle }{\langle TFD |  \left( | L_{\textbf l}\rangle \langle L_{\textbf l} | \otimes \mathds 1 \right) |TFD \rangle } \mathcal G_Q. 
\eea
In this subsection, we will calculate both the twisted correlation function and the prefactor in the above relation.

In the large-$N$ limit, the basic twisted correlation function is given by $Q_L = \sqrt{2} \psi_{L,i}$ and $Q_R = - i \sqrt{2} \psi_{R,i}$, 
where the factor of $\sqrt 2$ is to have a properly normalized $Q$. 
Teleportation of an arbitrary Majorana string, i.e. an operator consisting of an arbitrary product of SYK Majorana operators, can be obtained using the basic correlation function. In fact, the disconnected diagram built from the basic correlation function will dominate the diagrammatic expansion of correlation functions for Majorana strings in the large-$N$ limit. 
The basic twisted correlation function reads
\bea
\begin{aligned}
    \frac{\mathcal G_\psi}2 &= \frac{\langle TFD | (-i\psi_{R,i}(t_0)) \left( | L_{\textbf l}\rangle \langle L_{\textbf l} | \otimes U_{\textbf l} \right) \psi_{L,i}(-t_0) |TFD \rangle}{\langle TFD |  \left( | L_{\textbf l}\rangle \langle L_{\textbf l} | \otimes U_{\textbf l} \right) |TFD \rangle } \\
    &= \frac{\langle L_\textbf{l} | \psi_{L,i}(-t_0) e^{-\beta H_L/2} e^{-i \frac{\theta}{q} \sum_k i l_k \psi_{L,2k-1} \psi_{L,2k}} \psi_{L,i}(t_0) e^{-\beta H_L/2} | L_\textbf{l} \rangle  }{\langle L_\textbf{l} |  e^{-\beta H_L/2} e^{-i \frac{\theta}{q} \sum_k i l_k \psi_{L,2k-1} \psi_{L,2k}}  e^{-\beta H_L/2} | L_\textbf{l} \rangle },
\end{aligned}
\eea
where we have used the properties of the thermofield double~(\ref{eq:tfd}) to bring operators from the right side to the left side.
When it is clear we will drop the subindex $L$ for simplicity. 
We define the following imaginary time ordered correlation function,
\bea
    G_\psi(\tau_1, \tau_2) &=& \frac{\langle L_\textbf{l} |e^{- \beta H} \mathcal T \left[ e^{-i \frac{\theta}q \sum_k l_k i \psi_{2k-1}(\frac\beta2) \psi_{2k}(\frac\beta2)} \psi_{i}(\tau_1) \psi_{i}(\tau_2) \right] | L_\textbf{l} \rangle}{\langle L_{\textbf l} |  e^{- \beta H} \mathcal T \left[ e^{-i \frac{\theta}q \sum_k l_k i \psi_{2k-1}(\frac\beta2) \psi_{2k}(\frac\beta2)} \right]| L_\textbf{l} \rangle },
\eea
where $\mathcal T$ denotes imaginary time ordering, and as above $|L_{\textbf l} \rangle$ is the projected state with measurement outcomes given by $\textbf l = (l_1, ..., l_{N/2})$.
Owing to the large-$N$ structure, $G_\psi$ does not depend on the index $i$.
Note that the imaginary time correlation function and the basic twisted correlation function differ by a factor of two.

\begin{figure}
    \centering
    \subfigure[]{
    \includegraphics[width=0.4\textwidth]{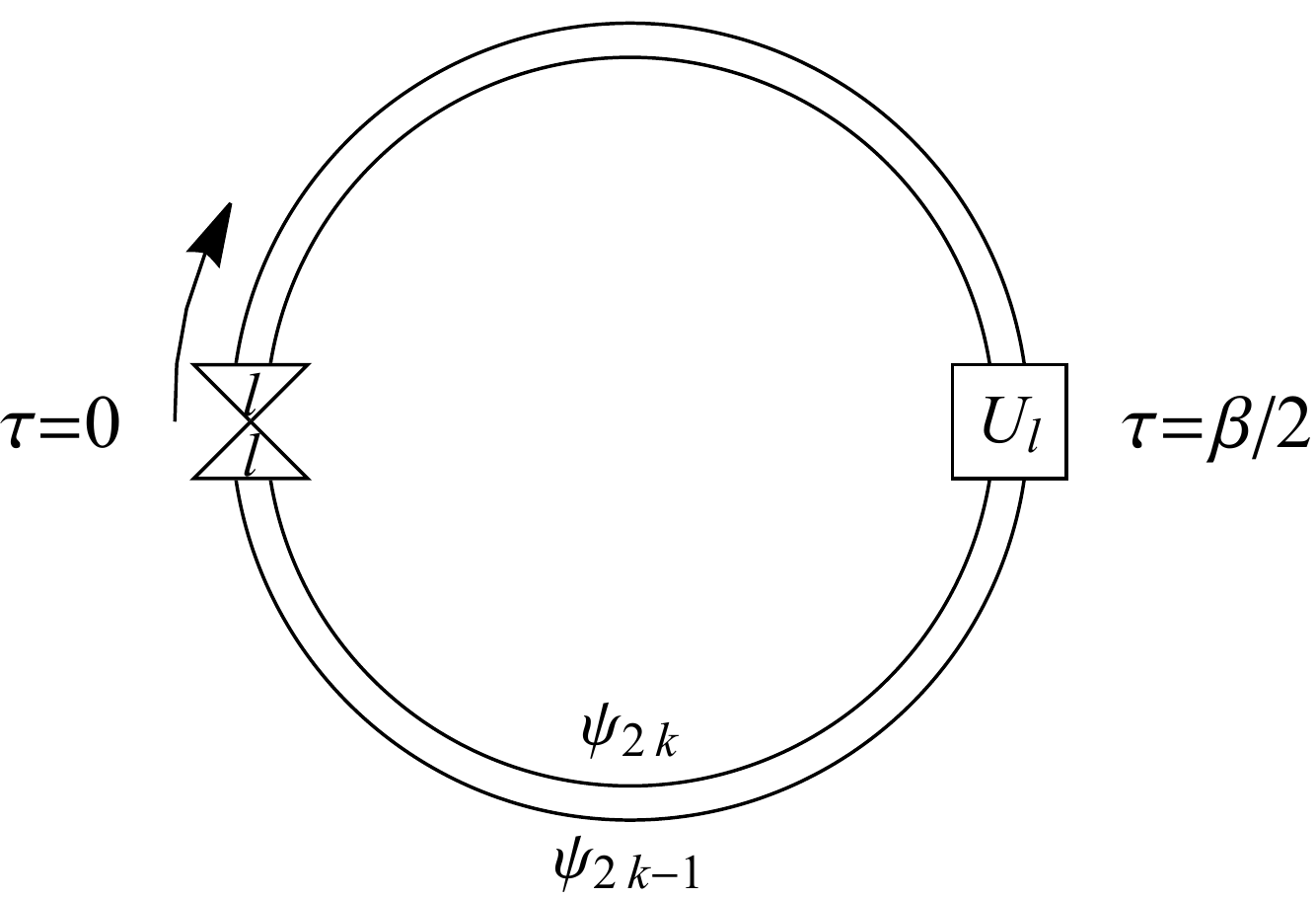}} \quad \quad 
    \subfigure[]{\includegraphics[width=0.36\textwidth]{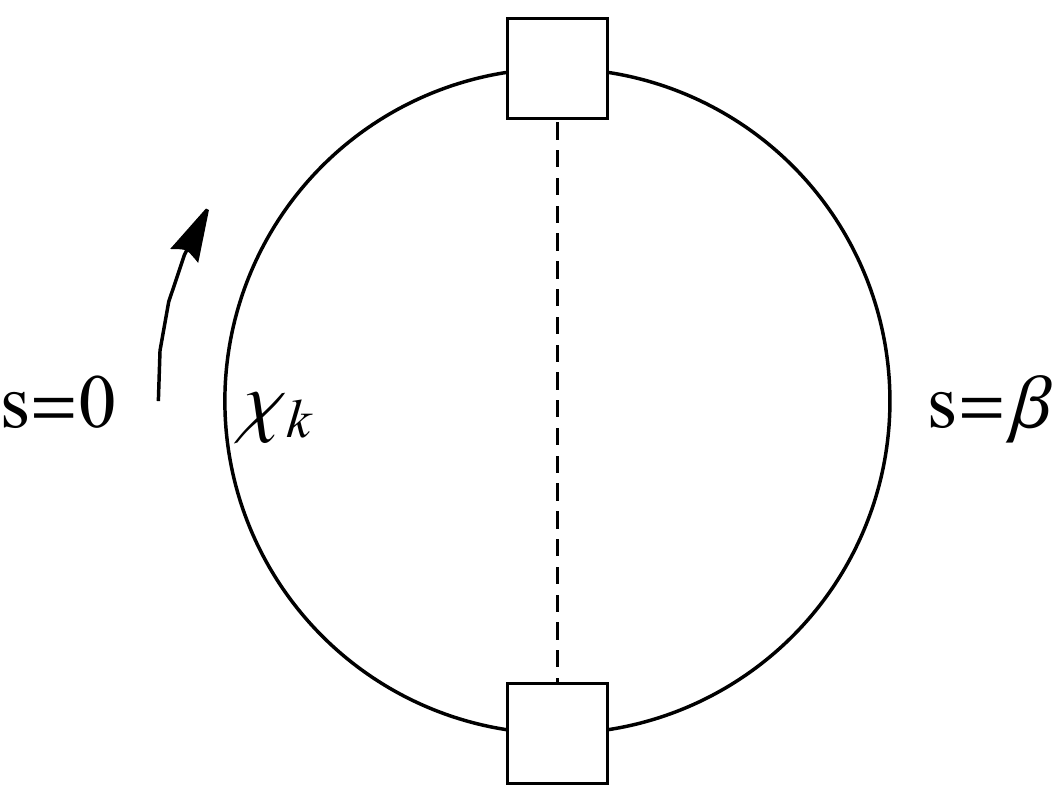}}
    \caption{(a) Imaginary time contour for the twisted correlation function. 
    In the left side ($\tau = 0, \beta$), a projective measurement $| L_{\textbf l} \rangle \langle L_{\textbf l}| $ with outcome $\textbf{l}$ (denoted by the two triangles) is performed, and in the right side ($\tau = \frac\beta2$), an associated decoding operator $U_{\textbf l}$ is applied.
    (b) Imaginary time contour for the Majorana field $\chi_k$. 
    The boundary condition from the decoding operator becomes a rotation between $s= \frac\beta2$ and $s= \frac{3\beta}2$, indicated by the dashed line. }
    \label{fig:imaginary_time_contour}
\end{figure}

In the imaginary time contour (see Fig.~\ref{fig:imaginary_time_contour} (a)), the measurement operator leads to boundary condition at $\tau = 0$ and $\tau = \beta$ given by,
\bea \label{eq:bc_left}
    \psi_{2k-1}(0) = -i l_k \psi_{2k}(0), \quad \psi_{2k-1}(\beta) = i l_k \psi_{2k}(\beta),
\eea
and the decoding operator~(\ref{eq:decode}) leads to the following boundary condition at $\tau = \beta/2$,
\bea \label{eq:bc_right}
    \left( \ba{cccc} \psi_{2k-1}(\frac{\beta^+}{2}) \\ i l_k \psi_{2k}(\frac{\beta^+}{2})\ea \right) =\left( \ba{cccc} \cos \frac{\theta}{q} & - i \sin \frac{\theta}{q} \\ - i \sin \frac{\theta}q & \cos \frac{\theta}{q} \ea \right) \left( \ba{cccc} \psi_{2k-1}(\frac{\beta^-}{2}) \\ i l_k \psi_{2k}(\frac{\beta^-}{2})\ea \right).
\eea

\begin{figure}
    \centering
    \includegraphics[width=0.35\textwidth]{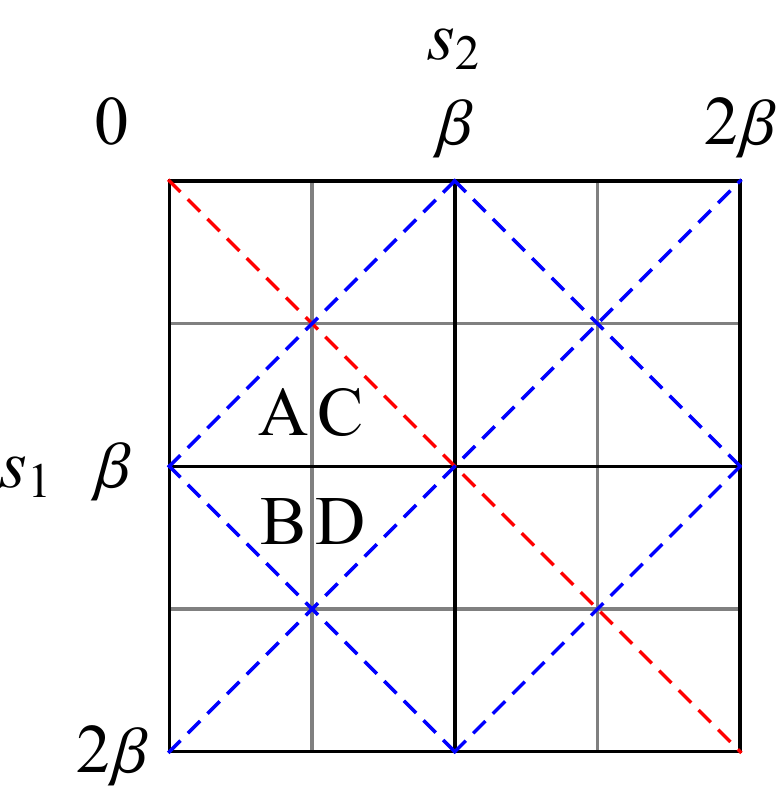}
    \caption{The full domain of twisted correlation function. The blue dashed lines indicate the reflection boundary condition, and the red dashed line indicates the anti-reflection boundary condition.
    With the help of these boundary conditions, we can reduce the calculation to the fundamental domains $A$, $B$, $C$ and $D$.
    The gray solid lines indicate the boundary condition at $s_{1,2} = \beta/2, 3\beta/2$. }
    \label{fig:fundamental_domain}
\end{figure}

In order to take care of the boundary condition at $\tau = 0, \beta$, we introduce a new field $\chi_k$ as in~(\ref{eq:chi}), 
\bea
    \chi_k(s) = 
    \begin{cases}
    \psi_{2k-1}(s), \quad & 0<s<\beta, \\
    i l_k \psi_{2k}(2\beta -s ), \quad & \beta< s < 2\beta,
    \end{cases}
\eea
and accordingly the bilocal field,
\bea
    G_\chi(s_1, s_2) = \frac2{N} \sum_{k=1}^{N/2} \chi_k (s_1) \chi_k(s_2).
\eea
Here, $G_\chi$ denotes the correlation function in the presence of the twisted boundary condition. Note that the arguments of this bilocal field $s$, run from $s=0$ to $s=2\beta$.
As above, $G_\chi$ will satisfy the Schwinger-Dyson equation,
\bea
    G_\chi = (\partial - \Sigma)^{-1} , \quad \Sigma_\chi(s_1, s_2) = \frac{\mathcal J^2}{q} P(s_1, s_2) [2 G_\chi(s_1, s_2)]^{q-1}.
\eea
Now, however, we will have a different boundary condition at $\tau = \beta/2$ (\ref{eq:bc_right}). In terms the correlation function, this boundary condition becomes
\bea \label{eq:bc_s}
    \left( \ba{cccc} \lim_{s_1/s_2 \rightarrow \frac{\beta^+}{2}} G_\chi(s_1, s_2) \\ 
    \lim_{s_1/s_2 \rightarrow \frac{3\beta^-}{2}} G_\chi(s_1, s_2)\ea \right) = \left( \ba{cccc} \cos \frac{\theta}q & -i \sin \frac{\theta}{q} \\ -i \sin \frac{\theta}{q} & \cos\frac{\theta}{q} \ea \right) \left( \ba{cccc} \lim_{s_1/s_2 \rightarrow \frac{\beta^-}{2}} G_\chi(s_1, s_2) \\ 
    \lim_{s_1/s_2 \rightarrow \frac{3\beta^+}{2}} G_\chi(s_1, s_2) \ea \right). 
\eea
There are various symmetries that will further simplify the calculation~(see Appendix \ref{append:correlation_function}).
In particular, the twisted correlation function satisfies various reflection conditions, which are illustrated in Fig.~\ref{fig:fundamental_domain}.
The blue (red) dashed lines indicate (anti-)reflection conditions. 
Then the calculation on the full domain can be reduced to the fundamental domain, denoted by $A$, $B$, $C$, and $D$. The boundary condition~(\ref{eq:bc_s}) is illustrated by the gray lines.

In the following, we will use a large-$q$ expansion to calculate the twisted correlation function in the fundamental domain. 
In the large $q$ limit, and assuming $G_\chi(s_1, s_2) = G_0(s_1, s_2) \left( 1 + \frac1q g(s_1, s_2) \right)$, where $G_0(s_1, s_2) = \frac12 \sgn(s_1 - s_2)$ is the bare propagator, we arrive at a Liouville equation similar to (\ref{eq:liou}),
\bea
    \partial_{s_1} \partial_{s_2} [ G_0(s_1, s_2) g(s_1, s_2)] = \mathcal J^2 P(s_1, s_2) [2 G_0(s_1, s_2)]^{q-1} e^{g(s_1, s_2)}.
\eea

The remaining problem boils down to calculating the Liouville equation in the fundamental domains $A$, $B$, $C$, and $D$ with the boundary condition~(\ref{eq:bc_s}). 
In the large-$q$ limit, this boundary condition can be simplified (as in~\cite{qi2019quantum,gao2021traversable}), to give
\bea \label{eq:bc_large_q}
    G_{A}(s_1, \frac\beta2) \approx e^{i \frac{\theta}q} G_C(s_1, \frac\beta2), \quad G_{B}(s_1, \frac\beta2) \approx e^{i \frac{\theta}q} G_D(s_1, \frac\beta2). 
\eea
Because the portions of the fundamental domain $C$ and $D$ do not cross the twist boundary condition,
we assume that the solution in $C$ and $D$ has time translational symmetry~\cite{qi2019quantum}. 
On the other hand, for the fundamental domain regions $A$ and $B$, we need to solve the Liouville equation supplemented with the boundary condition~(\ref{eq:bc_large_q}).
Via a tedious calculation, we can obtain the solution in the fundamental domains,
\bea
\begin{aligned}
   & G_A(s_1, s_2) =   \hat{G}_{11}(s_1, s_2)  \frac{e^{i \theta/q}}{\left(1 + \frac{e^{i\theta}-1}{\sin \gamma} \frac{\sin(\alpha(s_1 - \beta/2))\sin(\alpha(s_2 - \beta/2))}{\sin(\alpha |s_{12}| + \gamma)} \right)^{2/q} }, \\
   & G_B(s_1, s_2) =  \hat{G}_{21}(s_1,s_2)  \frac{e^{i \theta/q}}{\left(1 + \frac{e^{i\theta}-1}{\sin \gamma} \frac{\sin(\alpha \beta/2)\sin(\alpha(s_2 - \beta/2))}{\sin(\alpha (\beta - s_{2}) + \gamma)} \right)^{2/q} } , \\
   & G_C(s_1, s_2) = \hat{G}_{11}(s_1,s_2), \quad G_D(s_1, s_2) =  \hat G_{21}(s_1,s_2),
   \end{aligned}
\eea
where $\hat{G}_{11}$ and $\hat{G}_{21}$ are given in~(\ref{eq:solution_I}) and~(\ref{eq:solution_II}), and the constant parameters are determined by~(\ref{eq:constant}). 
Solutions in other regions can be obtained by the symmetries described above.

With these results, the prefactor in~(\ref{eq:relation}) is now straightforward to calculate. 
We can start with the following equation,
\bea
\begin{aligned}
     &\frac{\partial}{\partial \theta} \log \left[ \frac{\langle TFD |  \left( | L_{\textbf l}\rangle \langle L_{\textbf l} | \otimes U_{\textbf l} \right) |TFD \rangle }{\langle TFD |  \left( | L_{\textbf l}\rangle \langle L_{\textbf l} | \otimes \mathds 1 \right) |TFD \rangle } \right] \\
    & = \frac{\langle L_\textbf{l} | e^{-\beta H} \mathcal T \left[ \frac1q \sum_k l_k \psi_{2k-1}(\frac\beta2) \psi_{2k}(\frac\beta2)  e^{-i \frac{\theta}{q} \sum_k i l_k \psi_{2k-1}(\frac\beta2)\psi_{2k}(\frac\beta2) } \right]| L_\textbf{l} \rangle }{\langle L_\textbf{l} | e^{-\beta H} \mathcal T \left[ e^{-i \frac{\theta}{q} \sum_k i l_k \psi_{2k-1}(\frac\beta2)\psi_{2k}(\frac\beta2) } \right]| L_\textbf{l} \rangle} \\
    & = \frac{i}q \frac{N}2 G_\chi \left(\frac{3\beta}2, \frac\beta2 \right)  = \frac{i}q \frac{N}4 (\sin \gamma )^{4/q},
\end{aligned}
\eea
which we can integrate to get 
\bea
    \frac{\langle TFD |  \left( | L_{\textbf l}\rangle \langle L_{\textbf l} | \otimes U_{\textbf l} \right) |TFD \rangle }{\langle TFD |  \left( | L_{\textbf l}\rangle \langle L_{\textbf l} | \otimes \mathds 1 \right) |TFD \rangle } = \exp \left[ \frac{i N \theta}{4q} (\sin\gamma)^{4/q} \right].
\eea
Thus, the difference~(\ref{eq:relation}) between the left-right correlation function and the twisted correlation function is only a phase.
Further note that this phase does not depend on the operator $Q$, and therefore the magnitude of these two correlation function is the same.

\begin{figure}
    \centering
    \includegraphics[width=0.4\textwidth]{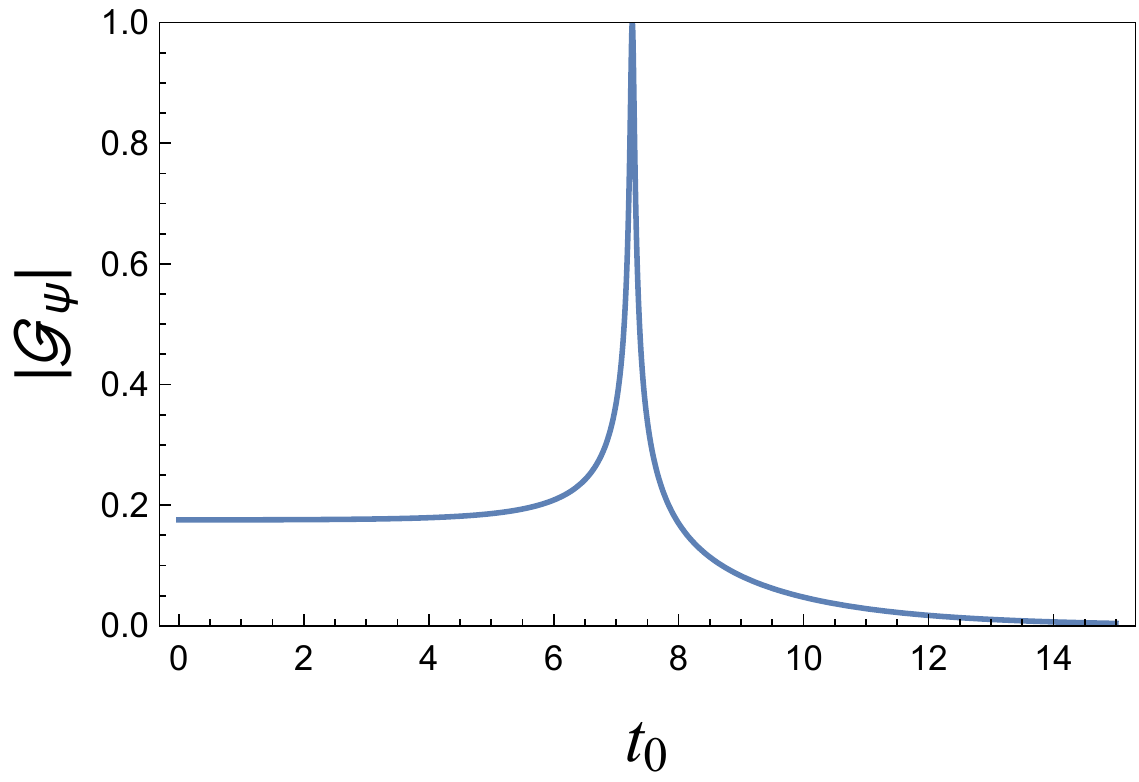}
    \caption{The magnitude of the twisted correlation function as a function of $t_0$. Here the parameter is chosen to be $\beta \mathcal =100$, $\theta = 10^{-4}$ and $q=8$. The magnitude is close to one at a time $t_0$ that is related to the parameter $\theta$.}
    \label{fig:norm}
\end{figure}

To get the teleportation fidelity, we take $s_1 = \beta^- + i t_0$ and $s_2 = \frac{\beta^-}2 + i t_0$, which is in region $A$, and find
\bea
    \mathcal G_\psi = 2 G_A \left(\beta^- + i t_0, \frac{\beta^-}2 + i t_0 \right) \approx e^{i\theta/q} \left( \frac{\sin \gamma}{1 - \frac{\theta e^{2\alpha t_0} }{4 \sin \gamma} e^{i\gamma} } \right)^{2/q},
\eea
where we have expanded the equation using $e^{ - \alpha t_0} \ll 1$, and $\theta \ll 1$. 
When the tuning parameter is chosen to be $ \theta = 4 e^{-2\alpha t_0} \sin \gamma $, the twisted correlation reaches its maximal magnitude,
\bea \label{eq:twist_max}
    \mathcal G_\psi \approx \exp\left( i  \frac{\pi}q \right) \left( \cos \frac{\gamma}2\right)^{2/q}. 
\eea
In the low energy limit, $\gamma = \frac\pi{\beta \mathcal J} + \mathcal O\left(\frac1{(\beta \mathcal J)^2} \right)$, the magnitude of the twisted correlation function is close to one.
This means that the measurement-decoding channel can successfully teleport the Majorana operator from the left side to the right side. 
A plot of the magnitude of the left-right correlation function is shown in Fig.~\ref{fig:norm}.

\subsection{Traversable wormhole}

To relate our results to a traversable wormhole, it will be simpler to start with~(\ref{eq:lr_measurement}). 
We consider the sum over measurement outcome, i.e., $\sum_{\textbf l} C_{\textbf l, Q}$.
The sum over measurement outcome can be rewritten   as
\bea
      \sum_\textbf{l} |L_\textbf{l} \rangle \langle L_\textbf{l} | \otimes U_\textbf{l} = \sum_\textbf{l} \left( \otimes_k |l_k \rangle \langle l_k | \right) \otimes e^{i \frac{\theta}{2q} \sum_k l_k M_{R,k}},
\eea
where we have introduced $M_{R,k} = -i 2 \psi_{R,2k-1} \psi_{R,2k}$, $M_{L,k} = -i 2 \psi_{L,2k-1} \psi_{L,2k}$, and $M_{L,k} | l_k \rangle = l_k |s_k \rangle$. 
Note that the projection operator is an operator-valued delta function, i.e.,
\bea
    | l_k \rangle \langle l_k | = \delta(M_{L,k} - l_k ).
\eea
Using this delta function, we arrive at a two-sided operator~\cite{gao2021traversable}
\bea
    \sum_{\textbf l} |L_\textbf{l} \rangle \langle L_\textbf{l} | \otimes U_\textbf{l} = \sum_\textbf{l} \otimes_k \delta(M_{L,k} - l_k) \otimes e^{i \frac{\theta}{2q} \sum_k l_k M_{R,k}} = e^{i \frac{\theta}{2q} \sum_k M_{L,k} M_{R,k}}.
\eea
After summing over all outcomes, the left-right correlation function is given by
\bea
\begin{aligned}
    \sum_{\textbf l} C_{\textbf{l},Q} &= \langle TFD | Q_R(t_0) \sum_{\textbf l} \left(   |L_{\textbf l} \rangle \langle L_{\textbf l} |  \otimes U_{\textbf l} \right) Q_L(-t_0) |TFD \rangle \\
    &= \langle TFD | Q_R(t_0) \left(  e^{i \frac{\theta}{2q} \sum_k M_{L,k} M_{R,k}}  \right) Q_L(-t_0) |TFD \rangle.
    \end{aligned}
\eea
This is the two-sided operator that makes the wormhole traversable~\cite{gao2017traversable,maldacena2017diving,brown2019quantum,nezami2021quantum,gao2021traversable}. 

We can compare our results with the two-sided correlation function from an $AdS_2$ gravity calculation. 
As depicted in Fig.~\ref{fig:traverse}, we consider the following physical setup in an $AdS_2$ black hole:
a particle created by the operator $Q$ is released in the past $t=-t_0$ on the left side; 
at time $t_0 = 0$, a projective measurement is performed on the left side, and simultaneously, according to the measurement outcome, the decoding operator is implemented in the right side with a properly chosen tuning parameter;
the particle is received in the right side at time $t=t_0$ with some probability. 
This process is effectively quantified by the two-sided correlation function~\cite{maldacena2017diving}
\bea
    C_\text{grav} = \langle Q_R(t_0) e^{i \frac{g}{N/2} \sum_{k} M_{L,k} M_{R,k}} Q_L(-t_0) \rangle 
\eea
where the expectation value is evaluated in the $AdS_2$ black hole state. 
Here, to make the comparison, we denote the two-sided deformation as $M$. 
In the probe limit, where the back reaction of the particle created by $Q$ on the geometry is neglected, the two-sided correlation function reads~\cite{maldacena2017diving}
\bea
    C_\text{probe} \propto \frac1{\left( 2 - g \frac{\Delta_M}{2^{2\Delta_M + 2}} G_N e^{ \frac{2\pi}\beta t_0} \right)^{2\Delta_Q}}
\eea
where $\Delta_M$ and $\Delta_Q$ are the scaling dimension of operator $M$ and $Q$, respectively, and $G_N$ is Newton's constant. 
It is easy to see that when $g \sim \frac{\exp[-\frac{2\pi}\beta t_0]}{G_N}$, the correlation function will diverge.
Of course, this divergence is not physical and can be cured by smearing out the particle, 
but, in essence, it shows the particle traverses from the left side to the right side~(see \cite{maldacena2017diving} for a discussion). 

Closely related to this phenomena, we find that the left-right correlation function of an arbitrary Majorana string $Q$ is given by
\bea
    C_Q \propto \left(\frac{\pi/(\beta \mathcal J)}{ 1 - g \frac{\beta \mathcal J}{4 \pi N} e^{2\alpha t_0} e^{i\gamma}} \right)^{2\Delta_Q}, 
\eea
where $\Delta_Q = p/q$, $p$ is the number of Majorana operators contained in the Majorana string $Q$, and $g = \frac{N}4 \theta $. 
In the low energy limit, $\gamma \rightarrow 0$, we encounter a seemly divergent expression for the left-right correlation at $g \sim N \exp[-2 \alpha t_0]$, which is identical to the gravity calculation with $G_N = 1/N$. 
Nevertheless, it is regularized by $\gamma \approx \frac{\pi}{\beta \mathcal J} $ from the UV complete SYK model. 
This shows that the teleportation protocol realizes a traversable wormhole.

\subsection{Fidelity bound}

In the previous sections, we studied the teleportation of an arbitrary Majorana string $Q$, and found that our protocol yields a left-right correlation that is close to one. 
In this subsection, we show that the left-right correlation function gives a lower bound for state teleportation, and we construct an encoding method that can achieve a nearly perfect teleportation fidelity~\cite{gao2021traversable}.
Note that the state teleportation fidelity is related to the fidelity of the distillation of EPR pairs~\cite{yoshida2019disentangling},
\bea
    \langle F_\psi \rangle = \frac{d}{d+1} F_{EPR} + \frac1{d+1},
\eea
where $d$ is the dimension of Hilbert space of the teleported state $|\psi\rangle$, and $\langle \cdot \rangle$ implies an average over all states. Therefore, we will consider $F_{EPR}$ below. As shown in Appendix~\ref{append:fidelity}, the distillation fidelity is lower bounded by $F_{EPR} \ge \left| \frac1{d^2} \sum_{\textbf{l},Q} C_{\textbf{l},Q} \right|^2$,
where $C_{\textbf{l},Q}$ is given by~(\ref{eq:lr_measurement}), and the sum of operators $Q$ is over a complete basis for the $d$-dimensional Hilbert space. 
In the large-$N$ SYK model, as the left-right correlation function does not depend on the measurement outcome, the bound of the distillation fidelity simplifies to
\bea \label{eq:bound_syk}
    F_{EPR} \ge \left| \frac1{d^2} \sum_{Q} C_{Q} \right|^2.
\eea
where $C_Q$ is given in (\ref{eq:lr_correlation}).

Now for simplicity, we consider the teleportation of an arbitrary qubit state, for which the Hilbert space consists of two Majorana operators,
$\Psi_1, \Psi_2$. 
In order to achieve a high fidelity, we need to embed these two operators into $p$ SYK Majorana operators in the the following way,
\bea
\begin{aligned}
    &\Psi_1 = 2^{\frac{p}2} i^{\frac{p(p+1)}p} \psi^1 \psi^5 ... \psi^{4p-3}, \\
    &\Psi_2 = 2^{\frac{p}2} i^{\frac{p(p+1)}p} \psi^3 \psi^7 ... \psi^{4p-1}.
    \end{aligned}
\eea
where $p$ is an odd number and the prefactor is chosen such that these two operators satisfy $\{\Psi_i, \Psi_j\} = 2\delta_{ij}$, $\Psi_i^\dag = \Psi_i$. 

The complete basis operators for the qubit state are $\mathds 1, \Psi_1, \Psi_2, i \Psi_1 \Psi_2$.
The fidelity bound then becomes
\bea \label{eq:bound_SYK}
    F_{EPR} = \left| \frac14 \left( C_{\mathds 1} + C_{\Psi_1} + C_{\Psi_2} + C_{i \Psi_1 \Psi_2} \right) \right|^2 = \left| \frac14 \left( 1 + \mathcal G_{\Psi_1} + \mathcal G_{\Psi_2} + \mathcal G_{i \Psi_1 \Psi_2} \right) \right|^2,
\eea
where in the second equality, we have used the fact that the twisted correlation function is equal to the left-right correlation function up to a constant phase. 
Here, we can see an interesting distinction between the teleportation of a Marjorana string such as $\Psi$ and the teleportation of an arbitrary qubit state.
Apart from the requirement that the magnitude of the twisted correlation function be maximal, there is also a necessary condition that the phases of these correlation functions should align to zero modulo $2\pi$. 
This provides good reason to embed the qubit state into Majorana strings $\Psi_i$ with $p$ SYK Majorana operators.

In the large-$N$ limit, the leading contribution to the twisted correlation function is disconnected. 
Thus the twisted correlation function for $\Psi_i$ is given by the basis twisted correlation function,
\bea
    \mathcal G_{\Psi_1} = \mathcal G_{\Psi_2} \approx  i^p (\mathcal G_\psi)^p, \quad \mathcal G_{i\Psi_1 \Psi_2} \approx - (\mathcal G_{\Psi})^{2p}.
\eea
We observe that though the tuning parameter $\theta$ is fixed by the time of state insertion, we still have freedom to tune $p$ to align the phase of the correlation function. 
In particular, from~(\ref{eq:twist_max}), the fidelity is bounded by
\bea
    F_{EPR} =  \left| \frac14 \left( 1+ 2 e^{i2\pi(\frac{p}4 + \frac{p}{2q}) } (\cos \gamma)^{2p/q} + e^{i 2\pi (p + \frac12 +\frac{p}{q}) }  (\cos \gamma)^{4p/q}  \right) \right|^2,
\eea
To align the phases, we require $\frac{p}4 + \frac{p}{2q}$ and $p + \frac12 +\frac{p}{q}$ to be close to integers. 
For simplicity, we consider $q \in 4 \mathds Z$. We can then see that 
\bea
    p= \begin{cases} \frac{q}{2} + 1 , \quad q = 4 (2k+1) \\
    \frac{q}2 - 1, \quad q = 8 k
    \end{cases} \quad k \in \mathds Z,
\eea
which is consistent with the fact that $p$ is an odd number. 
In the zero temperature limit, we take $\gamma \rightarrow 0$. The fidelity bound in the large-$q$ limit is then simply given by
\bea
    F_{EPR} \ge \left|\frac14 \left( 1+ 2 e ^{i\frac{\pi}{q} }  + e^{i  \frac{2\pi}{q}} \right) \right|^2 = \left(\cos \frac{\pi}{2q} \right)^4 \approx 1 - \frac{\pi^2}{2q^2}.
\eea
We can see that the embedding into $p$ Majorana operators can lead to a fidelity for the teleportation of an arbitrary qubit state that is close to one in the large-$q$ limit.

\section{Discussion}

\label{sec:discussion}

In this work, we studied the effect of projective measurements on the thermofield double state dual to an eternal black hole; in particular, we focused on the SYK model and its dual, JT gravity coupled to matter. 
To characterize the post-measurement state, we calculated the Renyi-2 mutual information between the two sides of the SYK thermofield double state, which showed a transition phenomenon at low temperature.
We then reproduced this mutual information calculation in the holographic JT plus matter dual. In this gravitational dual, the measurement creates an end-of-the-world brane in the bulk, and in particular, the matter fields dual to the measured fermions are then modeled by a boundary CFT. 
The gravitational picture makes manifest that the mutual information transition can be associated to an entanglement wedge transition, see Fig.~\ref{fig:EW_transition}. Upon measurement of a sufficiently large fraction of fermions, the entanglement wedge of the unmeasured side can jump to contain large regions behind the horizon of the black hole. On the boundary side, this change in the entanglement wedge can be understood as post-selection teleporting information about the bulk into the unmeasured boundary. 
We then explored this teleportation interpretation further, using the SYK model to construct an explicit protocol realizing such a teleportation. Finally, we related this telportation protocol to the physics of traversable wormholes.

In general, measurement is a nonlocal operation on a many-body quantum state, which radically changes the entanglement structure.
Thus, it is not clear when a postselected state from an arbitrary measurement will give rise to a nice holographic dual (e.g. a smooth, semiclassical geometry).
One exception is local projection onto a product state\footnote{In the jargon of conformal field theory, a projection onto a Cardy state~\cite{cardy1989boundary,miyaji2014boundary}.}, which can be described by a boundary conformal field theory~\cite{takayanagi2011holographic,fujita2011aspects,numasawa2016epr}, and has been recently studied in the context of holography~\cite{numasawa2016epr,Antonini:2022sfm}. 
Continuing in the same vein, the particular measurement operator we study is the fermion parity operator for each qubit formed by two adjacent even and odd Majorana operators~(\ref{eq:measurement_operator}); this is also a local projection. 
This measurement procedure preserves many of the nice properties of the original correlation function in the absence of measurement.
Specifically, the ``diagonal'' correlation function remains intact after measurement, indicating the dual background $AdS_2$ geometry remains valid. 
This is consistent with the expectation that local projective measurement leaves the dual spacetime intact \cite{Antonini:2022sfm}.

While a dual bulk geometry still exists, it is not left unchanged by this local projective measurement. It is now clear that these measurements create an end-of-the-world brane in bulk \cite{numasawa2016epr,Antonini:2022sfm}.
Whereas the Ryu-Takayanagi formula receives only minor modifications in the presence of bulk end-of-the-world branes in the bulk, it is not well understood how the matter entropy contribution from bulk fields---a crucial contribution in e.g. black hole evaporation and the formation of  islands~\cite{penington2020entanglement,penington2022replica,Almheiri_2020}---is modified in the presence of a projection on the boundary.
In a broader context, what the quantum extremal surface formula~\cite{Engelhardt:2014gca} is in a system with non-unitarity in the boundary---like that introduced by measurement---is an important open question. 
Our model above suggests that the bulk matter dual to a measured boundary can be effectively captured by a boundary CFT in two dimensional gravity. 
It remains to be seen whether such a proposal could be extended to more general, higher dimensional gravitational systems.

Apart from a scalar-type quantity to characterize entanglement, like the entanglement entropy, the dynamics of how information ``flows'' in the many-body system is often a key to understanding the physics from an information theory perspective.  
For instance, studying the information flow of an evaporating black hole has inspired tremendous progress in the black hole information paradox.
It was proposed in Ref.~\cite{Antonini:2022sfm} that the information flow induced by local projective measurement on the boundary can be largely understood as information teleportation from the measured region to the unmeasured region, thanks to the (pre-measurement) entanglement resource between the two boundary subregions. 
Here, we have provided an explicit protocol using the UV complete SYK model to realize this expectation.

As a consequence of this information flow, an entanglement wedge transition shown in Fig.~\ref{fig:EW_transition} naturally arises. 
It will be interesting to draw an analogy between this measurement-induced entanglement wedge transition and the transition between the empty set and the island in an evaporating black hole~\cite{penington2020entanglement,penington2022replica,Almheiri_2020}. 
A relation of this sort may be natural in light of the Horowitz-Maldacena resolution of the black hole information problem, which invokes a post-selection at the horizon to explain how information might escape from a black hole \cite{Horowitz:2003he}. In more modern language, the excess number of black hole states late in the evaporation process (in a semi-classical description) can be understood via a non-isometric map~\cite{akers2022quantum, akers2022black}. Of course, projective measurements like the ones we have studied above  can also give rise to non-isometries \cite{Antonini:2022sfm}. We leave exploration of this potential connection to future work.

As stated above, local projection is a special kind of measurement that can preserve large portions of the dual bulk geometry. 
There are many unknown questions regarding ``holographic measurement''. What are the neccessary properties a measurement operation should satisfy to still give rise to a dual bulk geometry? 
If a measurement can retain a portion of the dual spactime, what kind of geometric changes might arise from the effect of measurement (apart from the creation of an ETW brane)?
An interesting extension of our results might be to to consider a soft measurement that interpolates between an identity operator and a projection. 
Since a dual description exists at both ends of this interpolation, namely, the pre-measurement spacetime on one end and the one including an ETW brane on the other, it seems likely that an effective geometric description exists for this soft measurement.

The attentive reader may have realized that state preparation through Euclidean time evolution can be regarded as a soft measurement of the Hamiltonian operator.
An interesting question that has been recently studied in the quantum information science and many-body physics community is that of hybrid circuit evolution, where the quantum circuits consist of not only unitary operations but also non-unitary ones.
In particular, when the non-unitary gates are realized as a measurement interspersing the full circuit, it also induces a transition of the entanglement structure of the final state, the so-called measurement-induced entanglement phase transition~\cite{li2019measurement,skinner2019measurement,chan2019unitary,jian2021measurement,Bentsen:2021ukm}. 
Our measurement-induced entanglement wedge transition can be viewed as the effect of ``one time'' measurement, in contrast to the hybrid evolution. 
This provides a first step in understanding the dynamical measurement-induced entanglement transition in a context with holographic duality.
When the system can be effectively described by an evolution of a generic non-Hermitian Hamiltonian, several attempts have been made to explore the dual spacetime~\cite{Milekhin:2022bzx,kawabata2022dynamical,garcia2022keldysh,goto2022entanglement}. 
Due to the non-Hermiticity in the boundary degrees of freedom, an analysis involving complex metrics is inevitable; nevertheless, given the larger analytical control it provides, we believe lower dimensional gravity is a promising platform to further explore the dynamical effects arising in the presence of measurements.

\acknowledgments
We would like to thank Gregory Bentsen, Charles Cao, Raphael Bousso, Yasunori Nomura, Ping Gao, Zhenbin Yang, Zhuo-Yu Xian for helpful discussions. We acknowledge support from the Simons Foundation via It From Qubit (S-K.J.), from the U.S. Department of Energy grant DE-SC0009986 (S-K.J. and B.G.S.), from the AFOSR under FA9550-19-1-0360 (B.G-W.), and from the U.S. Department of Energy, Office of Science, Office of Advanced Scientific Computing Research, Accelerated Research for Quantum Computing program ``FAR-QC'' (S.A.) .

\appendix

\section{One-sided measurement in the SYK model: Renyi-2 entropy} 
\label{append:renyi_SYK}

In this appendix, we derive the Schwinger-Dyson equations and the formula for the on-shell actions $I_{num}(m)$ and $I_{den}(m)$ introduced in Section \ref{sec:SYKMI}. These are needed ingredients to compute the SYK Renyi-2 mutual information between the left side and the unmeasured Majorana fermions in the right side after a projective measurement on a subset of Majorana fermions is performed in the right side.
We also provide some more detail about our numerical analysis.

\subsection{On-shell action for the numerator of equation (\ref{eq:renyi_one})}

Let us start by computing $I_{num}(m)$. As a first step, the boundary conditions (\ref{eq:renyi_one_bc}) suggest introducing bilocal fields analogous to the ones defined in equations~(\ref{eq:bilocal})-(\ref{eq:offdiagonal}) and the corresponding Lagrange multiplier fields $\tilde{\Sigma}_{ii}(\tau_1,\tau_2)$, $i=1,2,3$. The arguments now have range $\tau_{1,2}\in (0,2\beta)$.
In terms of the bilocal fields, the action (\ref{eq:syk_action_one}) takes the form
\bea \label{eq:action_renyi_one_numer}
    && -I = \int d\tau_1 d \tau_2 \Big[ -  \frac12 \Big( \sum_{k=1}^{M/2} \psi_{2k-1}(\tau_1) (\partial - \tilde{\Sigma}_{11}) \psi_{2k-1}(\tau_2) + \sum_{k=1}^{M/2} \psi_{2k}(\tau_1) (\partial - \tilde{\Sigma}_{22}) \psi_{2k}(\tau_2) \nn \\ 
    && + \sum_{j=M+1}^N \psi_{j}(\tau_1) (\partial - \tilde{\Sigma}_{33}) \psi_{j}(\tau_2) \Big)  
    -  \frac12 \left( \frac{M}2 \tilde{\Sigma}_{11} \tilde{G}_{11} + \frac{M}2 \tilde{\Sigma}_{22} \tilde{G}_{22} + (N-M) \tilde{\Sigma}_{33} \tilde{G}_{33} \right) \nn \\
    && + \frac{J^2}{2q N^{q-1}} \left( \frac{M}2 \tilde{G}_{11}(\tau_1, \tau_2) + \frac{M}2 \tilde{G}_{22}(\tau_1, \tau_2) + (N-M) \tilde{G}_{33}(\tau_1, \tau_2) \right)^q \Big].
\eea

Similar to what we did in Section \ref{sec:pathint}, we can now introduce a new Majorana field defined as
\bea \label{aeq:chi}
    \chi_k(s) = \begin{cases}
    \psi_{2k-1}(s), & \quad 0< s < \beta/2, \\
    i \psi_{2k}(\beta-s), & \quad \beta/2 < s < \beta, \\
    -i \psi_{2k}(2\beta - s), & \quad \beta < s < 3\beta/2, \\
    \psi_{2k-1}(s-\beta), & \quad 3\beta/2 < s < 2\beta, \\
    \psi_{2k-1}(s-\beta) , & \quad 2 \beta < s < 5\beta/2, \\
    i \psi_{2k}(4\beta-s), & \quad 5\beta/2 < s < 3\beta, \\
    - i \psi_{2k}(5\beta -s), & \quad 3\beta < s < 7\beta/2, \\
    \psi_{2k-1}(s-2\beta), & \quad 7\beta/2< s< 4\beta.
    \end{cases}
\eea
The new Majorana satisfies the conventional anti-periodic boundary conditions for fermionic fields, i.e.  $\chi_k (4\beta + s) = - \chi_k(s)$. We can then integrate out the Majorana fermions in the path integral, obtaining the following action for the path integral over the bilocal fields:
\bea
\begin{aligned}
    - \frac{I}N =& \frac{m}4  \log \det(\hat{G}_{t}^{-1} - \hat{\tilde{\Sigma}}) + \frac{1-m}2 \log \det (G_{t}^{-1} - \tilde{\Sigma}_{33}) \\
    & - \frac12  \int d\tau_1 d\tau_2 \left(\frac{m}{2} \tilde{\Sigma}_{11} \tilde{G}_{11} + \frac{m}{2} \tilde{\Sigma}_{22} \tilde{G}_{22} + (1-m) \tilde{G}_{33} \right) \\
    & + \frac{J^2}{2q} \left( \frac{m}2 ( \tilde{G}_{11}(\tau_1, \tau_2) +  \tilde{G}_{22}(\tau_1, \tau_2) + (1-m) \tilde{G}_{33}(\tau_1, \tau_2) \right)^q,
    \end{aligned}
    \label{eq:collaction}
\eea
where we introduced the free fermion propagator $G_t=\partial^{-1}$ and a compact notation for the self energy
\bea \label{aeq:compact}
    \hat{\tilde{\Sigma}}(s_1,s_2) = \left[ \ba{cccccccc} 
    \tilde{\Sigma}_{11}(11) & 0 & 0 & \tilde{\Sigma}_{11}(12) & \tilde{\Sigma}_{11}(13) & 0 & 0 & \tilde{\Sigma}_{11}(14) \\
    0 & -\tilde{\Sigma}_{22}(11) & \tilde{\Sigma}_{22}(12) & 0 & 0 & -\tilde{\Sigma}_{22}(13) & \tilde{\Sigma}_{22}(14) & 0 \\ 
    0 & \tilde{\Sigma}_{22}(21) & -\tilde{\Sigma}_{22}(22) & 0 & 0 & \tilde{\Sigma}_{22}(23) & -\tilde{\Sigma}_{22}(24) & 0 \\ 
    \tilde{\Sigma}_{11}(21) & 0 & 0 & \tilde{\Sigma}_{11}(22) & \tilde{\Sigma}_{11}(23) & 0 & 0 & \tilde{\Sigma}_{11}(24) \\
    \tilde{\Sigma}_{11}(31) & 0 & 0 & \tilde{\Sigma}_{11}(32) & \tilde{\Sigma}_{11}(33) & 0 & 0 & \tilde{\Sigma}_{11}(34) \\
    0 & -\tilde{\Sigma}_{22}(31) & \tilde{\Sigma}_{22}(32) & 0 & 0 & -\tilde{\Sigma}_{22}(33) & \tilde{\Sigma}_{22}(34) & 0 \\ 
    0 & \tilde{\Sigma}_{22}(41) & -\tilde{\Sigma}_{22}(42) & 0 & 0 & \tilde{\Sigma}_{22}(43) & -\tilde{\Sigma}_{22}(44) & 0 \\ 
    \tilde{\Sigma}_{11}(41) & 0 & 0 & \tilde{\Sigma}_{11}(42) & \tilde{\Sigma}_{11}(43) & 0 & 0 & \tilde{\Sigma}_{11}(44)
    \ea \right].  \nn \\
\eea
In \ref{aeq:compact}, we have used the short-hand notation $\tilde{\Sigma}_{11}(ij) =\tilde{\Sigma}_{11} \left((i-1)\beta/2 + s_1, (j-1)\beta/2 + s_2 \right)$ and $\tilde{\Sigma}_{22}(ij) = \tilde{\Sigma}_{22}\left((i-1)\beta/2 - s_1, (j-1)\beta/2 - s_2 \right)$ with $0< s_{1,2} < \beta/2$.
The boundary condition resulting from the twist operator is given by
\bea
    && G_{t}(s_1, s_2) = \frac12 \sgn(s_1 - s_2), \quad s_{1,2} \in (0, 2\beta), \\
    && \hat G_{t}(s_1, s_2) = \frac12 \sgn(s_1 - s_2), \quad s_{1,2} \in (\beta, 3\beta) \text{ or } s_{1,2} (0,\beta) \cup (3\beta, 4\beta).
\eea
and $G_t(s_1, s_2) = \hat G_t (s_1, s_2) = 0$ otherwise. 

We can now derive the Schwinger-Dyson equations by varying the action (\ref{eq:collaction}) for the bilocal fields:
\bea
\begin{aligned}
    & \hat G = (\hat G_{t}^{-1} - \hat \Sigma)^{-1}, \quad G = (G_{t}^{-1} - \Sigma)^{-1},\\
    & \Sigma_{ii} =  J^2 \left( \frac{m}2 (G_{11} + G_{22}) + (1- m ) G_{33} \right), \quad i=1,2,3.
    \end{aligned}
    \label{aeq:renyi_one_eom}
\eea
Here we omitted the ~$\tilde{}$~ to indicate that the bilocal fields are on shell. $G_{11}$ and $G_{22}$ are related to $\hat G$ by 
{\small
\bea \label{aeq:compact_G}
\begin{aligned}
    &G_{11}(s_1,s_2) = \\
    & 
    \left[ \ba{cccc} \hat G(s_1,s_2) & \hat G(s_1, 3\beta/2+s_2) & \hat G(s_1, 2\beta+s_2) & \hat G(s_1, 7\beta/2+s_2) \\ 
    \hat G(3\beta/2+ s_1,s_2) & \hat G(3\beta/2+ s_1, 3\beta/2+s_2) & \hat G(3\beta/2 + s_1, 2\beta+s_2) & \hat G(3\beta/2 + s_1, 7\beta/2+s_2) \\
    \hat G(2\beta+ s_1,s_2) & \hat G(2\beta + s_1, 3\beta/2 +s_2) & \hat G(2\beta + s_1, 2\beta+s_2) & \hat G(2\beta + s_1, 7\beta/2+s_2) \\
    \hat G(7\beta/2+ s_1,s_2) & \hat G(7\beta/2 + s_1, 3\beta/2+s_2) & \hat G(7\beta/2 + s_1, 2\beta+s_2) & \hat G(7\beta/2 + s_1, 7\beta/2+s_2) 
    \ea
    \right],  \\
    & G_{22}(s_1,s_2) =
    \\ 
    & \left[ \ba{cccc} -\hat G(\beta - s_1, \beta - s_2) & \hat G(\beta - s_1, 3\beta/2-s_2) & - \hat G(\beta - s_1, 3\beta - s_2) & \hat G(\beta - s_1, 7\beta/2-s_2) \\ 
    \hat G(3\beta/2 - s_1, \beta - s_2) & -\hat G(3\beta/2 - s_1, 3\beta/2-s_2) & \hat G(3\beta/2 - s_1, 3\beta - s_2) & -\hat G(3\beta/2 - s_1, 7\beta/2 - s_2) \\
    -\hat G(3\beta - s_1, \beta - s_2) & \hat G(3\beta - s_1, 3\beta/2 -s_2) & -\hat G(3\beta - s_1, 3\beta-s_2) & \hat G(3\beta - s_1, 7\beta/2-s_2) \\
    \hat G(7\beta/2- s_1, \beta - s_2) & -\hat G(7\beta/2-, 3\beta/2-s_2) & \hat G(7\beta/2-  s_1, 3\beta+s_2) & -\hat G(7\beta/2-  s_1, 7\beta/2-s_2) 
    \ea
    \right].
    \end{aligned}
\eea}

We can then discretize the imaginary time contour and numerically solve the Schwinger-Dyson equations (\ref{aeq:renyi_one_eom}) combined with equations (\ref{aeq:compact}) and (\ref{aeq:compact_G}) by iteration.
Using the Schwinger-Dyson equations (\ref{aeq:renyi_one_eom}), the on-shell action $I_{num}(m)$ can be written as
\bea \label{aeq:onshell_one_numer}
    - \frac{I_{num}(m)}N &=& \frac{m}4  \left(\log\det[ \hat G^{-1} \hat G_{t}] + 2 \log 2\right) +  \frac{1-m}2 \left(\log \det[ G^{-1} G_{t}] + \log 2 \right) \\
    && - \frac{J^2}{2} \left(1- \frac1q\right)  \int d\tau_1 d\tau_2 \left( \frac{m}2 (G_{11} + G_{22}) + (1-m) G_{33}  \right)^q,
\eea
where we have used a normalization for the bare propagators
\bea
    \log (G_{t}^{-1}) - \log 2 = 0, \quad \log (\hat G_{t}^{-1}) - 2 \log 2 = 0.
\eea
This is readily understood if we consider that $G_{t}$ is the same as a propagator for a free fermion on a contour $s \in (0,2\beta)$ and $\hat G_{t}$ is the same as a propagator for two free fermions on two separately contours $s \in (\beta, 3\beta)$ and $s \in (0,\beta) \cup (3\beta, 4\beta)$.
The partition function of a free fermionic mode is simply $2$ (accounting for the occupied and unoccupied states).

Once the Schwinger-Dysons equations (\ref{aeq:renyi_one_eom}) are numerically solved, the on-shell action (\ref{aeq:onshell_one_numer}) can be evaluated. In order to minimize the error due to the discretization of the imaginary time contour, we obtain results using different finite discretizations and then extrapolate the result in the continuum limit (i.e. infinitely fine discretization). 
More explicitly, we divide each contour $(0,\beta)$ into $L=40,60,80,100$ segments, and then extrapolate the results in the $L \rightarrow \infty$ limit. 

\subsection{On-shell action for the denominator of equation (\ref{eq:renyi_one})}

In a similar fashion, the denominator of equation (\ref{eq:renyi_one}) can be written in terms of a path integral
\bea
    \overline{(\Tr[\Psi(m)])^2} = \int D\psi e^{-I},
\eea
where the action is again given by equation (\ref{eq:syk_action_one}). We can then rewrite the action in terms of bilocal fields:
\bea
\begin{aligned}
    - \frac{I}N =& \frac{m}4  \log \det(\hat{G}_{i}^{-1} - \hat{\tilde{\Sigma}}) + \frac{1-m}2 \log \det (G_{i}^{-1} - \tilde{\Sigma}_{33}) \\
    & - \frac12  \int d\tau_1 d\tau_2 \left(\frac{m}{2} \tilde{\Sigma}_{11} \tilde{G}_{11} + \frac{m}{2} \tilde{\Sigma}_{22} \tilde{G}_{22} + (1-m) \tilde{G}_{33} \right) \\
    & + \frac{J^2}{2q} \left( \frac{m}2 ( \tilde{G}_{11}(\tau_1, \tau_2) +  \tilde{G}_{22}(\tau_1, \tau_2)) + (1-m) \tilde{G}_{33}(\tau_1, \tau_2) \right)^q
    \end{aligned}
\eea
where again $G_i=\partial^{-1}$. The boundary conditions (\ref{eq:denbc}) now lead to 
\bea
    && G_{i}(s_1, s_2) = \frac12 \sgn(s_1 - s_2), \quad 0< s_{1,2} < \beta \text{ or } \beta< s_{1,2} < 2\beta, \\
    && \hat G_{i}(s_1, s_2) = \frac12 \sgn(s_1 - s_2), \quad 0< s_{1,2} < 2\beta \text{ or } 2\beta < s_{1,2} < 4\beta.
\eea
and $G_i(s_1, s_2) = \hat G_i (s_1, s_2) = 0$ otherwise.
The Schwinger-Dyson equations are given by
\bea
    && \hat G = (\hat G_{i}^{-1} - \hat \Sigma)^{-1}, \quad G = (G_{i}^{-1} - \Sigma)^{-1},  \\
    && \Sigma_{ii} =  J^2 \left( \frac{m}2 (G_{11} + G_{22}) + (1- m ) G_{33} \right), \quad i=1,2,3. 
\eea
with $\hat \Sigma$ ($\hat G$) given by~(\ref{aeq:compact}) and~(\ref{aeq:compact_G}). 
The on-shell action can then be written as
\bea \label{aeq:onshell_one_deno}
    - \frac{I_{den}(m)}N &=& \frac{m}4  \left(\log\det[ \hat G^{-1} \hat G_{i}] + 2 \log 2\right) +  \frac{1-m}2 \left(\log \det[ G^{-1} G_{i}] + 2 \log 2 \right) \nn \\
    && - \frac{J^2}{2} (1- \frac1q)  \int d\tau_1 d\tau_2 \left( \frac{m}2 (G_{11} + G_{22}) + (1-m) G_{33}  \right)^q,
\eea
where the normalization is 
\bea
    \log (G_{i}^{-1}) - 2\log 2 = 0, \quad \log (\hat G_{i}^{-1}) - 2 \log 2 = 0.
\eea

We can then numerically solve the Schwinger-Dyson equations and evaluate the on-shell action $I_{den}(m)$ like we did for $I_{num}(m)$. Once we have the numerically-obtained value of both on-shell actions~(\ref{aeq:onshell_one_numer}) and~(\ref{aeq:onshell_one_deno}), we can finally compute the Renyi-2 entropy---given by $S_L^{(2)}(\Psi(m)) = I_{num}(m) - I_{den}(m)$ in the saddle point approximation---and the Renyi-2 mutual information---given by $I_{LR}^{2}(\Psi(m))=2S_L^{(2)}(\Psi(m))$. The numerical results obtained following the procedure described in this appendix are reported in Fig. \ref{fig:syk-one-mi}. 

\section{The $m=1$ case: Kourkoulou-Maldacena state}
\label{app:KM}

The setup studied by Kourkoulou and Maldacena~\cite{kourkoulou2017pure} can be seen as a special case of our setup involving a one-sided measurement on a TFD state. In particular, when all the Majorana fermions on one side are projected using the measurement operator~(\ref{eq:parity}) (namely, $m=1$), the resulting post-measurement state on the other side is pure and given by a Kourkoulou-Maldacena state.
For completeness, in this appendix we calculate the Renyi-2 entropy of a subset of Majorana fermions in the Kourkoulou-Maldacena state~\cite{zhang2020entanglement} obtained in the $m=1$ case of the analysis of Section \ref{sec:measurement} and Appendix \ref{append:renyi_SYK}. 
Without loss of generality, we assume the Majorana fermions on the right side are projected.

We can then divide the Majorana fermions on the left side into two subsets
$A=\{\psi_i|i=1,...,N_A\}$ and $A^c=\{\psi_i|i=N_A + 1,..., N\}$ and define $\lambda = N_A/N$. 
The Renyi-2 entropy of the subset $A$ is given by 
\bea 
    e^{-S_A^{(2)}(\lambda)} =  \frac{\overline{\Tr_A \left[ \left( \Tr_{\bar A} [\rho_\text{KM}] \right) ^2 \right]}}{ \overline{\left( \Tr [\rho_\text{KM}] \right)^2}},  \quad \rho_\text{KM} = \Tr_R| \Psi \rangle \langle \Psi |,
    \label{eq:renyiKM}
\eea
where $|\Psi \rangle = |\Psi(m=1) \rangle$ is defined in equation~(\ref{eq:state_one}) and we traced over the right side (which after the measurement is in a product state with the left side). 
Like in our previous analysis, the numerator and denominator of equation (\ref{eq:renyiKM}) can be represented using a Euclidean path integral. The corresponding action for the numerator is the same as (\ref{eq:syk_action_one}) on a contour $\tau \in (0, 2\beta)$, but the boundary conditions are modified to take into account the insertion of a twist operator only for the first $N_A$ Majorana fermions on the left and the fact that all Majorana fermions are measured on the right: 
\bea
\begin{aligned}
    &
    \psi_i(0) = -\psi_i(2\beta), \quad \psi_i(\beta_-) = \psi_i(\beta_+), \quad
    i=1,...,N_A, \\
    & \psi_{N_A+j}(0) = - \psi_{N_A+j}(\beta_-), \quad \psi_{N_A+j}(\beta_+) = - \psi_{N_A+j}(2\beta), \quad j=1,...,N-N_A, \\
    & \psi_{2k-1}(\frac{\beta_-}2) = i \psi_{2k}(\frac{\beta_-}2), \quad \psi_{2k-1}(\frac{\beta_+}2) = -i \psi_{2k}(\frac{\beta_+}2),  \quad k = 1,..., N/2, \\
    & \psi_{2k-1}(\frac{3\beta_-}2) = i \psi_{2k}(\frac{3\beta_-}2), \quad \psi_{2k-1}(\frac{3\beta_+}2) = -i \psi_{2k}(\frac{3\beta_+}2),  \quad k = 1,..., N/2.
    \end{aligned}
\eea
The first line gives the anti-periodic boundary conditions for Majorana fermions in $A$ (for which a twist operator is inserted in the left); the second line gives the anti-periodic boundary conditions for Majorana fermions in $A^c$ (for which no twist operator is inserted); the last two lines are a consequence of the measurement of all Majorana fermions in the right side.

On the other hand, the action for the denominator is again given by equation~(\ref{eq:syk_action_one}), while the boundary conditions now reflect the fact that no twist operators are inserted in the left side: 
\bea
\begin{aligned}
    & \psi_{j}(0) = - \psi_{j}(\beta_-), \quad \psi_{j}(\beta_+) = - \psi_{j}(2\beta), \quad j=1,...,N, \\
    & \psi_{2k-1}(\frac{\beta_-}2) = i \psi_{2k}(\frac{\beta_-}2), \quad \psi_{2k-1}(\frac{\beta_+}2) = -i \psi_{2k}(\frac{\beta_+}2),  \quad k = 1,..., N/2, \\
    & \psi_{2k-1}(\frac{3\beta_-}2) = i \psi_{2k}(\frac{3\beta_-}2), \quad \psi_{2k-1}(\frac{3\beta_+}2) = -i \psi_{2k}(\frac{3\beta_+}2),  \quad k = 1,..., N/2.
    \end{aligned}
\eea

\begin{figure}
    \centering
    \includegraphics[width=0.5\textwidth]{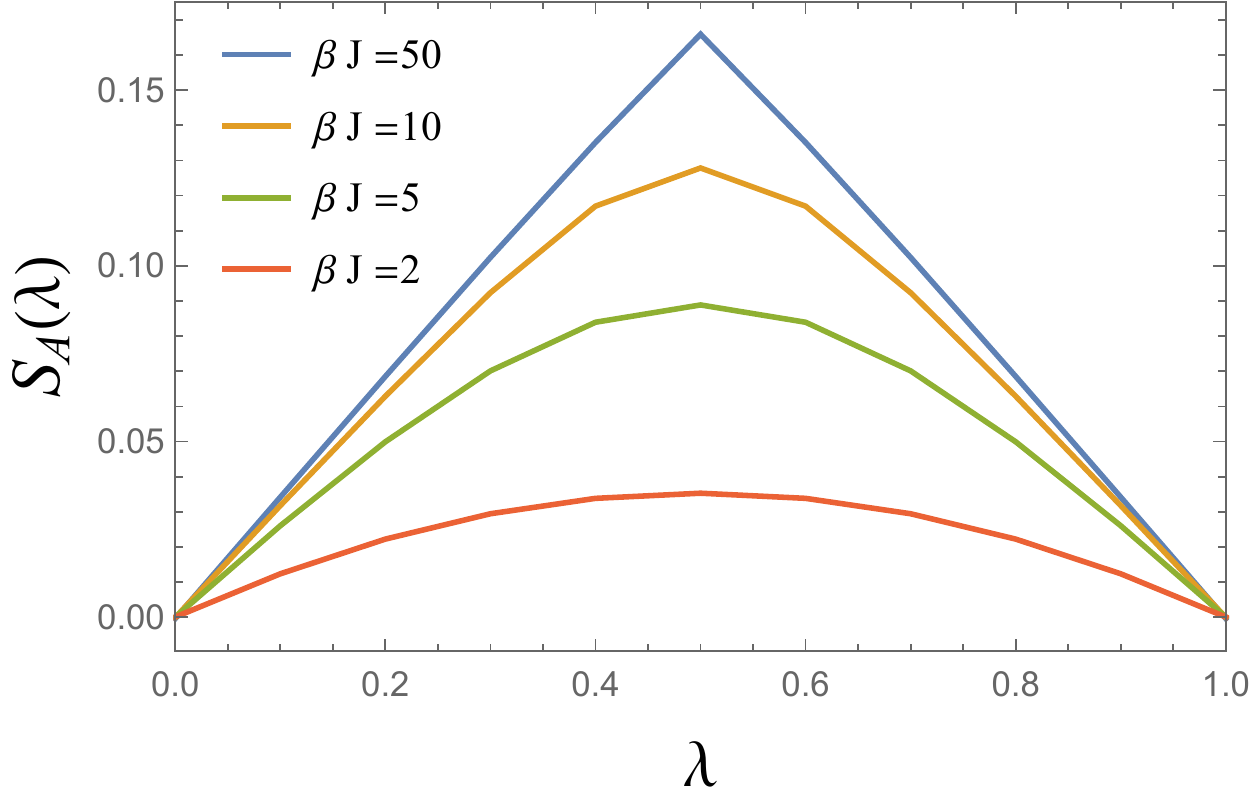}
    \caption{Renyi-2 entropy per Majorana fermion of a subset $A$ of fermions in the left side after all fermions in the right side have been measured. The left system is in the Kourkoulou-Maldacena state.
    $\lambda$ denotes the fraction of left fermions included in subsystem $A$. 
    Different colors denote different values of $\beta$. The Renyi-2 entropy is symmetric around $\lambda=0.5$ due to the purity of the Kourkoulou-Maldacena state and the $O(N)$ symmetry of the large-$N$ SYK model. Increasing $\beta$ we increase the amount of Euclidean time evolution preparing the Kourkoulou-Maldacena state, and therefore the entanglement among fermions.}
    \label{fig:syk-one-temperature}
\end{figure}

We can then introduce again the bilocal fields, derive and solve numerically the Schwinger-Dysons equations, evaluate the on-shell action for the numerator and denominator, and finally compute the Renyi-2 entropy in the saddle point approximation. This procedure is completely analogous to the one carried out in Section \ref{sec:measurement} and  Appendix~\ref{append:renyi_SYK}, to which we refer for technical details.
The results of our numerical evaluation of the Renyi-2 entropy as a function of the fraction $\lambda_A$ of fermions in subsystem $A$ are shown in Fig.~\ref{fig:syk-one-temperature} for different values of $\beta$.

A general feature for all curves is that the entropy increases with $\lambda$ for $\lambda < 1/2$, and decreases for $\lambda > 1/2$.
This is a consequence of the fact that the Kourkoulou-Maldacena state---obtained in the left side by measuring all fermions in the right side---is a pure state. 
For this reason and due to the $O(N)$ symmetry of the SYK model at large $N$, the Renyi-2 entropy is symmetric with respect to $\lambda = 1/2$. 
Notice also that the Renyi-2 entropy is lower for smaller values of $\beta$. 
This behavior can be easily understood. For $\beta=0$, the Kourkoulou-Maldacena state is simply given by the eigenstate of the measurement operator, which is a product state of the Majorana fermions (and therefore any subsystem of Majorana fermions has vanishing entropy). Increasing $\beta$, the Kourkoulou-Maldacena state is given by the same product state evolved by an amount $\beta$ of Euclidean time. Such Euclidean time evolution introduces entanglement among different Majorana fermions, increasing the Renyi-2 entropy \cite{kourkoulou2017pure,Antonini:2021xar}.

\section{Two-sided measurement and mutual information in the SYK model}
 \label{append:syk_two-side}

In this appendix we consider the case where the projective measurement is performed on a subset of fermions in both sides of a SYK thermofield double.
Similar to the analysis of Section \ref{sec:measurement}, we measure the fermionic parity on the left and right side. The measurement operators are given by
\bea \label{eq:measurement-flavor}
\begin{aligned}
    & M_{R,k} = -i 2 \psi_{R,2k-1} \psi_{R, 2k}, \quad k=1,..., M/2 \\ 
    & M_{L,k} = -i 2 \psi_{L,2k-1} \psi_{L,2k}, \quad k= (N-M)/2 + 1,..., N/2,
    \end{aligned}
\eea
where again we assume $M$ and $N$ are both even, and we take the large-$N$ limit with fixed $m = M/N$. 
The eigenvalues of the measurement operators are $r_k, l_k = \pm 1$.
Assume we measure the TFD state and observe the measurement outcomes $\textbf{l}=(l_{(N-M)/2+1},...,l_{N/2})$ and $\textbf{r}=(r_1,...,r_{M/2})$ for the left side and right side, respectively. Then the post-measurement states for the measured Majorana fermions on the left and right side are given by  $|L_\textbf{l}(m)\rangle$ and $|R_\textbf{r}(m)\rangle$, where
\bea
\begin{aligned}
    & M_{R,k} | R_\textbf{r}(m) \rangle =  r_k |R_\textbf{r}(m)\rangle, \quad k=1,...,M/2,  \label{eq:measurement_two1} \\
    & M_{L,k} | L_\textbf{l} (m) \rangle =  l_k |L_\textbf{l} (m)\rangle, \quad k=(N-M)/2+1,...,N/2. \label{eq:measurement_two2}
\end{aligned}
\eea
The resulting full (unnormalized) post-measurement state for the two-sided system is given by
\bea \label{eq:state_two}
    |\Phi_{\textbf{l,r}}(m) \rangle =  \left( \mathds 1^{(1-m)}_R \otimes |R_\textbf{r}(m) \rangle \langle R_\textbf{r}(m) | \right) \otimes \left( \mathds 1^{(1-m)}_L \otimes |L_\textbf{l}(m) \rangle \langle L_\textbf{l}(m) | \right) |TFD \rangle, 
\eea
where $\mathds 1^{(1-m)}_{L,R}$ denotes the identity operator acting on the unmeasured Majorana in the left and right sides. 

We are interested in computing the mutual information between the unmeasured Majorana fermions in the two sides. 
Because the state $|\Phi_{\textbf{l,r}}(m) \rangle$ is pure, the mutual information satisfies
\bea
    I_{LR}(\Phi_\textbf{l,r}(m)) = S_L(\Phi_\textbf{l,r}(m)) + S_R(\Phi_\textbf{l,r}(m)) - S_{LR}(\Phi_\textbf{l,r}(m)) = 2 S_L(\Phi_\textbf{l,r}(m)),
\eea
where $\Phi_\textbf{l,r}(m) = |\Phi_\textbf{l,r}(m) \rangle \langle \Phi_\textbf{l,r}(m) | $ is the associated unnormalized density matrix. Like in Section \ref{sec:SYKMI}, we focus on the Renyi-2 entropy 
\bea
    e^{-S_L^{(2)}(\Phi(m))} = \frac{ \overline{\Tr_L \left[ ( \Tr_R[\Phi(m)^2]) \right]}}{ \overline{ \left(\Tr[\Phi(m)] \right)^2}}.
    \label{eq:apprenyi2}
\eea
Since our analysis does not depend on the measurement outcomes, we omitted the subscripts in equation (\ref{eq:apprenyi2}) and in the following calculation we set for simplicity $l_k = r_k = 1$ for $k=1,...,M/2,(N-M)/2+1,...,N/2$.

\begin{figure}
    \centering
    \subfigure[$1 \le i \le M$]{\includegraphics[height=0.4\textwidth]{Figures/fig_contour_one_numer1.pdf}} \quad 
    \subfigure[$M<i \le N-M$]{\includegraphics[height=0.4\textwidth]{Figures/fig_contour_one_numer2.pdf}}
    \quad
    \quad
    \subfigure[$N-M <i \le N$]{\includegraphics[height=0.4\textwidth]{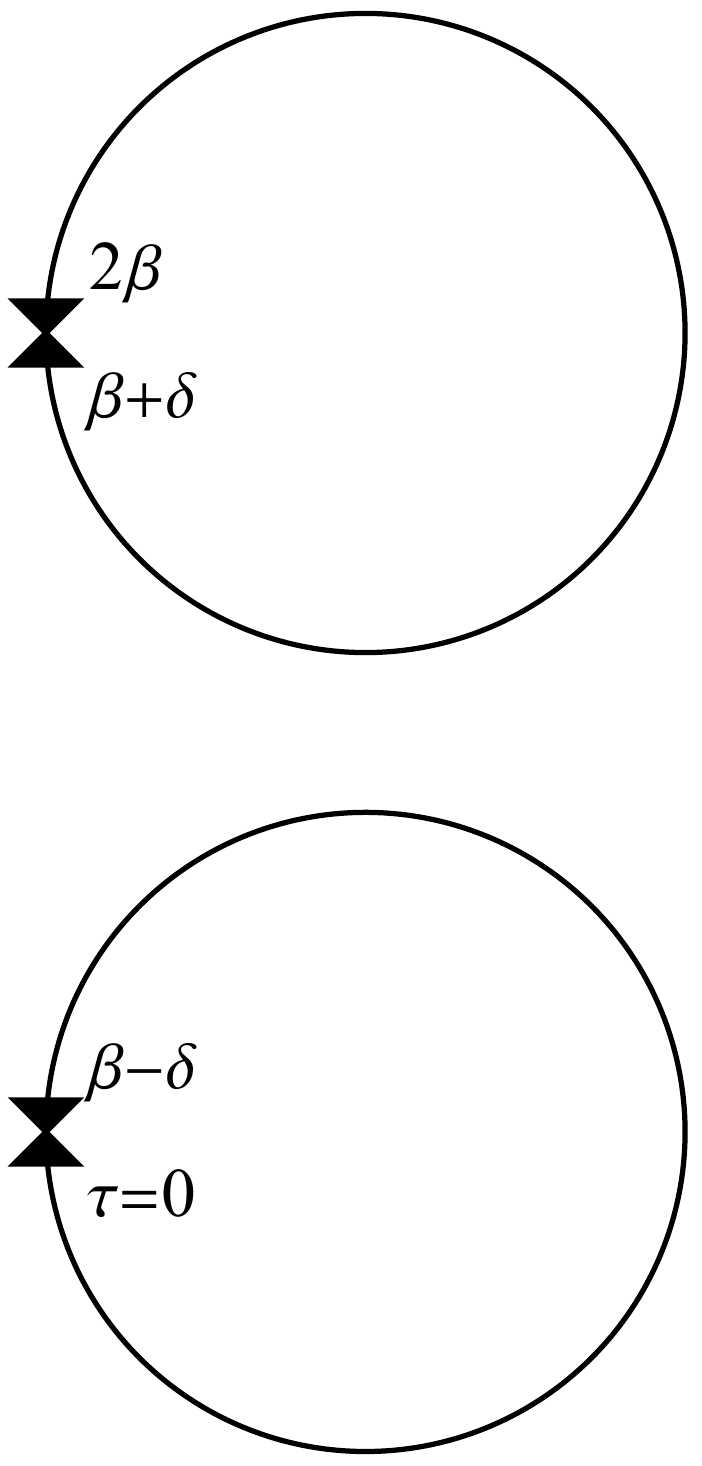}}
    \caption{Imaginary time contour $\tau\in (0,2\beta)$ for the Euclidean path integral computing the numerator of equation (\ref{eq:apprenyi2}) when $M<N/2$.
    The twist operator inserted in the left system is indicated by the dashed line. The insertion of a measurement operator in the left and right sides is indicated by the triangles. When the projective measurement operator is inserted in the left side, the effect of the twist operator is trivial and we therefore omit the dashed line.  (a) Boundary conditions (\ref{eq:renyi_bc_two1}) for Majorana fermions $\psi_i$, $1\le i\le M$. (b) Boundary conditions (\ref{eq:renyi_bc_two2}) for Majorana fermions $\psi_i$, $M < i \le N-M$. (c) Boundary conditions (\ref{eq:renyi_bc_two3}) for Majorana fermions $\psi_i$, $N-M < i \le N$. }
    \label{fig:syk_bc_two1}
\end{figure}

The path integral representation  for the numerator of equation (\ref{eq:apprenyi2}) is
\bea
    \overline{\Tr_L \left[ ( \Tr_R[\Phi(m)^2]) \right]} = \int D\psi e^{-I},
\eea
with the action
\bea \label{eq:syk_action_two}
 -I = -\int d\tau \sum_j \frac12 \psi_j(\tau) \partial_\tau \psi_j(\tau) +  \int d\tau_1 d\tau_2 \frac{J^2}{2q N^{q-1}} \left( \sum_j \psi_j(\tau_1) \psi_j(\tau_2) \right)^q.
\eea
The imaginary time runs from $\tau=0$ to $\tau=2\beta$. 
We now need to specify the correct boundary conditions taking into account both the measurement operators and the insertion of a twist operator in the left side. To this end, we need to discuss two separate cases, i.e. $M<N/2$ and $M>N/2$.

\subsubsection*{\quad $\textbf{M<N/2}$}

For $M<N/2$, the measurement on the right side is performed on Majorana fermions $\psi_i$ with $i\le M$, while on the left side it is performed on Majorana fermions $\psi_i$ with $i>N-M$. 
For the Majorana fermions $\psi_i$ with $M < i \le N-M$, there is no measurement in either side. A graphical representation of the boundary conditions is given in Fig.~\ref{fig:syk_bc_two1}.

For $1 \le i \le M$, the boundary conditions must take into account the measurement operator inserted at $\tau = \beta/2, 3 \beta/2 $ as well as the twist operator inserted at $\tau =0, \beta,2\beta$ (see Fig.~\ref{fig:syk_bc_two1} (a)):
\bea
\begin{aligned}
    & 
    k = 1,..., M/2  \\
    & \psi_{2k-1}(\frac{\beta_-}2) = i \psi_{2k}(\frac{\beta_-}2), \quad  \psi_{2k-1}(\frac{\beta_+}2) = -i \psi_{2k}(\frac{\beta_+}2), \\
    & \psi_{2k-1}(\frac{3\beta_-}2) = i \psi_{2k}(\frac{3\beta_-}2), \quad  \psi_{2k-1}(\frac{3\beta_+}2) = -i \psi_{2k}(\frac{3\beta_+}2),\\ 
    & \psi_{2k-1}(0) = -\psi_{2k-1}(2\beta), \quad  \psi_{2k}(0) = - \psi_{2k}(2\beta), 
    \end{aligned}
    \label{eq:renyi_bc_two1} 
\eea
where we used the shorthand notation $\beta_\pm$ defined in Section \ref{sec:SYKMI}.

For $ M< i \le N-M$, there is no measurement operator and the twist operator is again inserted at $\tau =0, \beta,2\beta$ (see Fig.~\ref{fig:syk_bc_two1} (b)):
\bea 
    i=M+1,...,N-M, \quad \psi_i(0) = -\psi_i(2\beta). \label{eq:renyi_bc_two2}
\eea
For $N-M< i \le N$, the boundary conditions must take into account the insertion of the measurement operator at $\tau = 0,\beta, 2\beta$.
The twist operator at the same point is now trivial because $\langle L_\textbf{l}(m) | L_\textbf{l}(m) \rangle = 1$ (see Fig.~\ref{fig:syk_bc_two1} (c)):
\bea
\begin{aligned}
    & k = (N-M)/2+1,..., N/2  \label{eq:renyi_bc_two3} \\
    & \psi_{2k-1}(0) = -i \psi_{2k}(0), \quad \psi_{2k-1}(\beta_-) = i \psi_{2k}(\beta_-),\\
    & \psi_{2k-1}(\beta_+) = -i \psi_{2k}(\beta_+), \quad \psi_{2k-1}(2\beta) = i \psi_{2k}(2\beta).
    \end{aligned}
\eea

\begin{figure}
    \centering
    \subfigure[$1 \le i \le N-M$]{\includegraphics[height=0.4\textwidth]{Figures/fig_contour_one_numer1.pdf}} \quad 
    \subfigure[$N-M < i \le M $]{\includegraphics[height=0.4\textwidth]{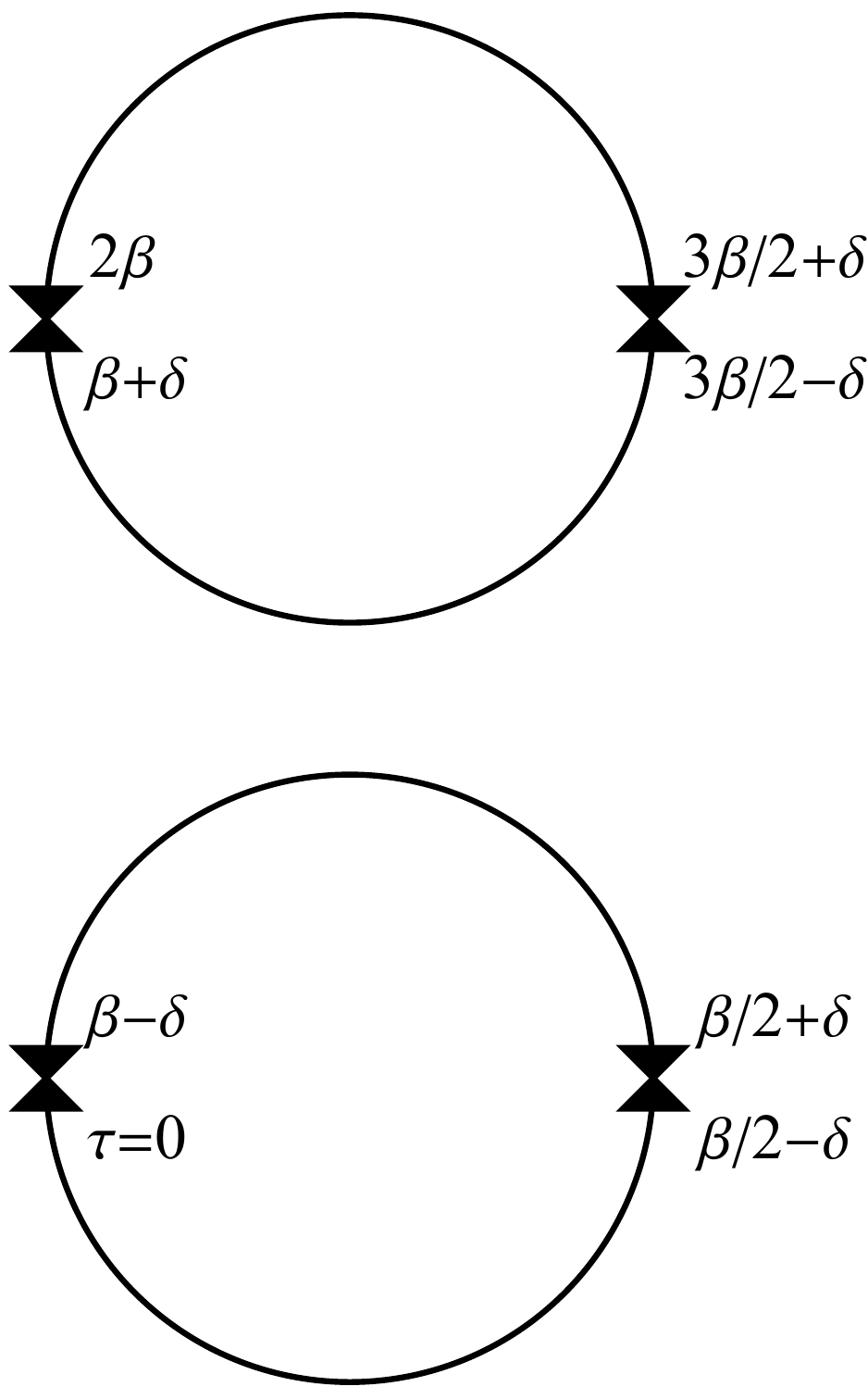}}
    \quad
    \subfigure[$M <i \le N$]{\includegraphics[height=0.4\textwidth]{Figures/fig_contour_two_numer3.pdf}}
    \caption{Imaginary time contour $\tau\in (0,2\beta)$ for the Euclidean path integral computing the numerator of equation (\ref{eq:apprenyi2}) when $M>N/2$.
    The twist operator inserted in the left system is indicated by the dashed line. The insertion of a measurement operator in the left and right sides is indicated by the triangles. When the projective measurement operator is inserted in the left side, the effect of the twist operator is trivial and we therefore omit the dashed line.  (a) Boundary conditions (\ref{eq:renyi_bc_two4}) for Majorana fermions $\psi_i$, $1\le i\le N-M$. (b) Boundary conditions (\ref{eq:renyi_bc_two5}) for Majorana fermions $\psi_i$, $N-M < i \le M$. (c) Boundary conditions (\ref{eq:renyi_bc_two6}) for Majorana fermions $\psi_i$, $M < i \le N$. }
    \label{fig:syk_bc_two2}
\end{figure}

\subsubsection*{\quad $\textbf{M>N/2}$}

Increasing the number of measured Majorana fermions such that $M>N/2$, the measurement prescription in the right side and left side is the same as before, but now the measurement occurs in both sides for the Majorana fermions $\psi_i$ with $N-M< i \le M$.\footnote{For simplicity, we postselect the outcomes of the right-side measurement and the left-side measurement to be the same so that the method we developed in Section \ref{sec:measurement} and Appendix \ref{append:renyi_SYK} can be directly applied. If the measurement outcomes are different, we cannot simply introduce a new Majorana fermion $\chi_k$, because it would not satisfy the conventional anti-periodic boundary conditions for fermionic fields.}
A graphical representation of the resulting boundary conditions is given in Fig.~\ref{fig:syk_bc_two2}. 

For $1\le i \le N-M$, the boundary conditions must account for the measurement operator inserted at $\tau = \beta/2, 3 \beta/2 $ as well as the twist operator inserted in $\tau = 0,\beta,2\beta$ (see Fig.~\ref{fig:syk_bc_two2} (a)):
\bea
\begin{aligned}
    & k = 1,..., (N-M)/2  \label{eq:renyi_bc_two4} \\
    & \psi_{2k-1}(\frac{\beta_-}2) = i \psi_{2k}(\frac{\beta_-}2), \quad  \psi_{2k-1}(\frac{\beta_+}2) = -i \psi_{2k}(\frac{\beta_+}2), \\
    & \psi_{2k-1}(\frac{3\beta_-}2) = i \psi_{2k}(\frac{3\beta_-}2), \quad  \psi_{2k-1}(\frac{3\beta_+}2) = -i \psi_{2k}(\frac{3\beta_+}2),  \\ 
    & \psi_{2k-1}(0) = -\psi_{2k-1}(2\beta), \quad  \psi_{2k}(0) = - \psi_{2k}(2\beta). 
    \end{aligned}
\eea
For $ N-M < i \le M$, there are measurement operators inserted in both sides---namely at $\tau =0, \beta/2,\beta, 3\beta/2, 2\beta$---and the action of the twist operator is trivial. This leads to the boundary conditions (see Fig.~\ref{fig:syk_bc_two2} (b))
\bea
\begin{aligned}
    & k = (N-M)/2 +1,..., M/2 \label{eq:renyi_bc_two5} \\
    & \psi_{2k-1}(0) = -i \psi_{2k}(0), \quad
    \psi_{2k-1}(\beta_-) = i \psi_{2k}(\beta_-), 
    \\
    & \psi_{2k-1}(\beta_+) = -i \psi_{2k}(\beta_+), \quad
    \psi_{2k-1}(2\beta) = i \psi_{2k}(2\beta),
    \\
    & \psi_{2k-1}(\frac{\beta_-}2) = i \psi_{2k}(\frac{\beta_-}2),  \quad
    \psi_{2k-1}(\frac{\beta_+}2) = -i \psi_{2k}(\frac{\beta_+}2),  \\
    & \psi_{2k-1}(\frac{3\beta_-}2) = i \psi_{2k}(\frac{3\beta_-}2), \quad
    \psi_{2k-1}(\frac{3\beta_+}2) = -i \psi_{2k}(\frac{3\beta_+}2). 
\end{aligned}
\eea
Finally, for $M< i \le N$, the boundary conditions must take into account the measurement operator inserted at $\tau = 0,\beta, 2\beta$, while the twist operator at the same point is trivial (see Fig.~\ref{fig:syk_bc_two2} (c)):
\bea
    \begin{aligned}
    & k = M/2+1,..., N/2  \label{eq:renyi_bc_two6} \\
    & \psi_{2k-1}(0) = -i \psi_{2k}(0), \quad \psi_{2k-1}(\beta_-) = i \psi_{2k}(\beta_-), \\
    & \psi_{2k-1}(\beta_+) = -i \psi_{2k}(\beta_+), \quad \psi_{2k-1}(2\beta) = i \psi_{2k}(2\beta).
    \end{aligned}
\eea

The denominator of equation (\ref{eq:apprenyi2}) can also represented by a path integral with the same action~(\ref{eq:syk_action_two}).
The difference is in the boundary conditions, because there is no twist operator. 
More explicitly, the boundary conditions for the denominator are given by equations (\ref{eq:renyi_bc_two1})-(\ref{eq:renyi_bc_two6}) after replacing the last line in equation (\ref{eq:renyi_bc_two1}), equation (\ref{eq:renyi_bc_two2}), and the last line in equation (\ref{eq:renyi_bc_two4}) by
\bea
    \psi_{i} (0) = - \psi_i(\beta_-), \quad \psi_{i} (\beta_+) = - \psi_i(2\beta), \quad i=1,...,N-M.
\eea

Having identifying the boundary conditions, we can now proceed to introduce bilocal fields, derive and numerically solve the Schwinger-Dyson equations, evaluate the numerator and denominator on-shell actions $I_{num}(m)$ and $I_{den}(m)$, and finally compute the Renyi-2 mutual information---which is given in the saddle point approximation by $I_{LR}^{(2)}(\Phi(m))=2(I_{num}(m)-I_{den}(m))$. This analysis is completely analogous to the one carried out for the one-sided measurement in Section \ref{sec:SYKMI} and Appendix \ref{append:renyi_SYK}, and we will therefore omit it here.

\begin{figure}
    \centering
    \includegraphics[width=0.5 \textwidth]{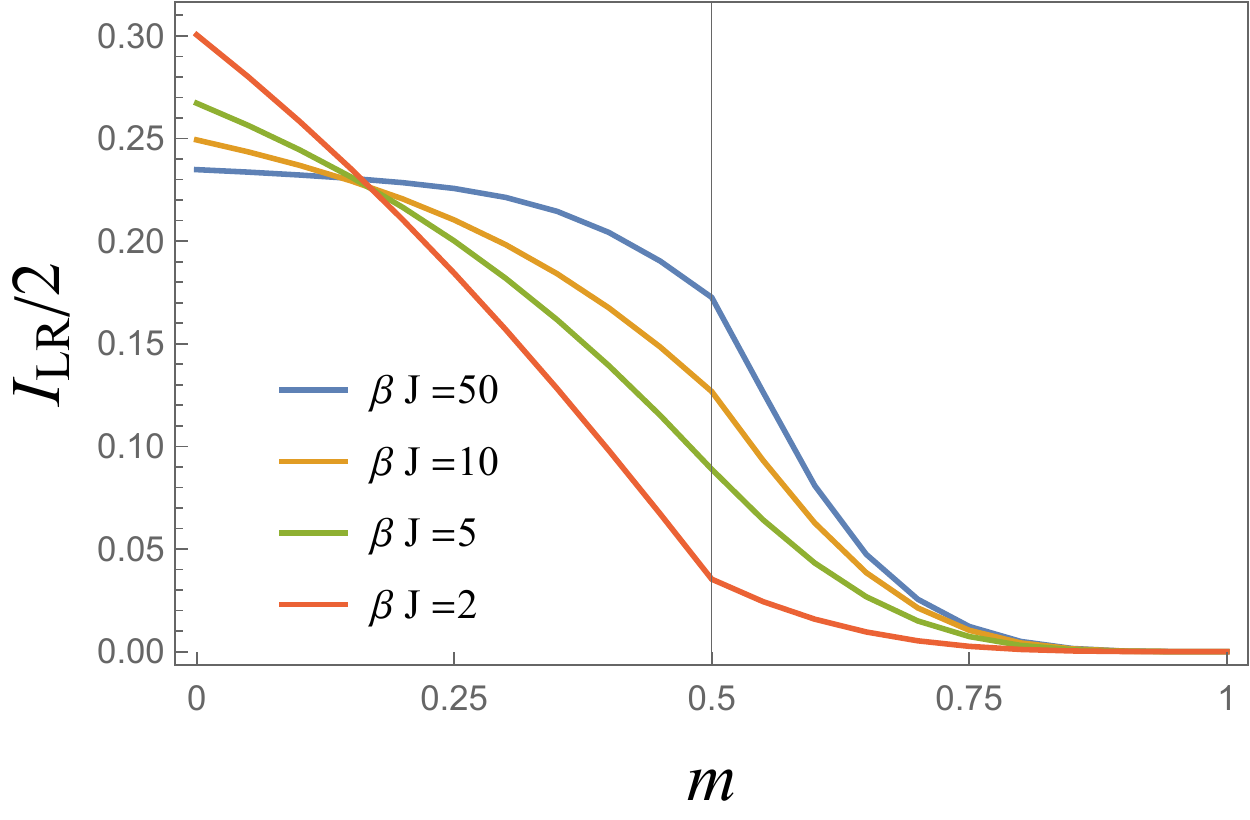}
    \caption{Renyi-2 mutual information $\mathcal{I}_{LR}^{(2)}(\Phi(m))$ between unmeasured Majorana fermions on the left and right sides for the two-sided measurement case. $m = M/N$ is the ratio of measured Majorana fermions on the left and right sides. The mutual information is always non-vanishing for $m<1$, and vanishes only when all the Majorana fermions are measured ($m=1$).}
    \label{fig:syk-two-mi}
\end{figure}

Our numerical results for the Renyi-2 mutual information $\mathcal{I}^{(2)}_{LR}(\Phi(m)) = I_{LR}^{(2)}(\Phi(m))/N$ are shown in Fig.~\ref{fig:syk-two-mi}. 
As in the one-sided measurement case, the mutual information remains finite for any $m<1$ and only vanishes for $m=1$. In other words, the left and right systems are completely disentangled only when all Majorana fermions are measured. 
One way to understand this result is to remember that at infinite temperature the TFD state reduces to EPR pairs between the left and right Majorana fermions.
In this case the mutual information is
\bea
    \mathcal I^\Phi_{LR}(m) = \begin{cases} (1-2m) \log 2, & m< 1/2,  \\
    0, & m \ge 1/2.
    \end{cases}
\eea
The transition between the entangled phase and the completely disentangled phase occurs at $m=1/2$. This can be readily understood if we recall that for $m=1/2$ (which is $M=N/2$) one fermion is measured for each one of the $N$ EPR pairs. In particular, the right fermion of each EPR pair is measured for $i=1,...,N/2$, while the left fermion of each EPR pair is measured for $i=N/2+1,...,N$.
Now imagine we turn on a small imaginary time evolution, i.e. we consider the TFD state at high but finite temperature. Then the all-to-all interaction in the Hamiltonian will generate entanglement between the two sides even for $m>1/2$. 
This is particularly evident for example in the $\beta \mathcal J = 2$ numerical result in Fig.~\ref{fig:syk-two-mi}. 

Finally, we would like to remark that the holographic dual analysis of the two-sided measurement setup described in the present Appendix is complicated by the fact that when large subsets of fermions are measured on both sides, it is not clear whether a geometric bulk dual description still exists. 
Therefore, when $m$ is small we expect the bulk dual analysis to remain substantially similar to the one described in Section \ref{sec:JT} for the one-sided measurement case. But when $m$ is large, it is not clear what bulk geometry is left behind by the measurement, if any. 
For intermediate values of $m$, the transition from a semiclassical geometry to a non-geometric description could be radical and it is not well-understood. We thus leave the bulk dual analysis of the two-sided measurement case for future work.

\section{Entanglement entropy in JT gravity with matter CFT} 
\label{append:JT}

Varying the dilaton and metric in the JT plus matter action (equation ~(\ref{eq:JT_action})), respectively, yield the equations of motion
\bea
    && R = -2, \\
	&& \frac1{8\pi G_N} (\nabla_\mu \nabla_\nu \phi - g_{\mu\nu} \nabla^2 \phi + g_{\mu\nu} \phi) = T_{\mu\nu},
\eea
where $T_{\mu\nu}$ is the stress tensor of the matter CFT. 
We first consider the solution at zero temperature, setting $T_{\mu\nu}$ to zero.
The metric in Euclidean Poincar\'e coordinates is
\bea
	ds^2 = \frac{dt^2 + dz^2}{z^2},
\eea
and we implement the following boundary conditions at $z = \epsilon$~\cite{maldacena2016conformal},
\bea
    g|_\text{bdy} = \frac1{\epsilon^2}, \quad \phi(\epsilon) = \frac1{\epsilon}.
\eea 
The solution for the dilaton is then given by
\bea
	\phi = \frac{\alpha + \gamma t + \delta (t^2 + z^2)}{z},
\eea
where $\alpha, \gamma, \delta$ are constants determined by the boundary conditions, $\alpha = 1$, $\gamma = \delta = 0$. 

To describe a thermal state with temperature $1/\beta$, we can make a series of coordinate transformations. 
To this end, we first introduce $x = z + i t$, $\bar x = z - it$ with
$ ds^2 = \frac{4dx d \bar x}{(x+ \bar x)^2}$.
Now then take
\bea\label{eq:thermal}
    x =  \frac{\beta}{2\pi} \tanh \frac{\pi}{\beta} y , \quad \bar x = \frac{\beta}{2\pi}  \tanh \frac{\pi}\beta \bar y.  
\eea
Finally, with $y = \sigma + i \tau$, $\bar y = \sigma - i \tau$, the metric becomes 
\bea
    ds^2 = \frac{4\pi^2}{\beta^2} \frac{d\sigma^2 + d\tau^2}{\sinh^2 \frac{2\pi}\beta \sigma}.
\eea
Implementing the boundary condition at $\sigma = \epsilon$,
$g_{\tau\tau} = \frac1{\epsilon^2}$, 
$\phi(\epsilon) = \frac{\phi_r}{\epsilon}$, 
yields the dilaton solution
\bea
    \phi(\sigma) =  \phi_r \frac{2\pi}\beta \frac1{\tanh \frac{2\pi }{\beta}\sigma}.
\eea

\begin{figure}
    \centering
    \includegraphics[width=0.4\textwidth]{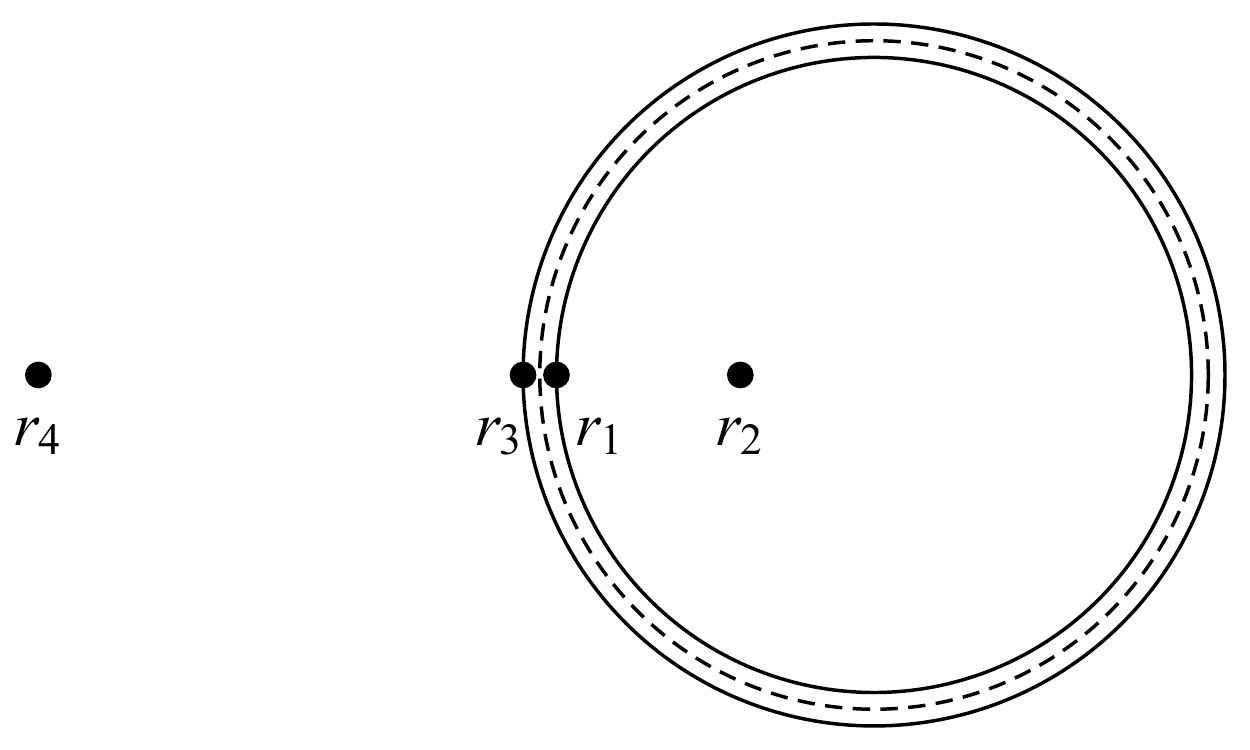}
    \caption{Two copies of Poincare disk with one of them filling outside $|z|>1$. }
    \label{fig:JT_euclidean}
\end{figure}

We can now turn to understanding the bulk matter entropy, computed via a ``doubling trick.'' We introduce a second Euclidean copy of the spacetime, glued together as in Fig.~\ref{fig:JT_euclidean}. To this end, we employ a (final) coordinate transformation,
\bea
\begin{aligned}
    & \text{first copy} \\ 
    & z = e^{\frac{2\pi}\beta(\sigma + i \tau)}, \quad \bar z = e^{\frac{2\pi}\beta(\sigma - i\tau)}, \quad ds^2 =  \frac{4 dz d\bar z}{(1- |z|^2)^2}, \quad |z| < 1, \\
    & \text{second copy} \\
    & z = e^{-\frac{2\pi}\beta(\sigma + i \tau)}, \quad \bar z = e^{-\frac{2\pi}\beta(\sigma - i\tau)}, \quad ds^2 =  \frac{4 dz d\bar z}{(1- |z|^2)^2}, \quad |z|>1.
\end{aligned}
\eea
The metric is identical in the two copies,  except one copy is mapped to $|z|<1$, and the other one to $|z|>1$. In other words,
the two copies are related by the transformation $z \rightarrow \frac1z$.
The free Dirac fermion can propagate in the entire plane, as shown in Fig.~\ref{fig:JT_euclidean}. 
Note that the UV cutoff of the two copies is located at $|z| = 1 \pm \frac{2\pi \epsilon}{\beta}$, respectively.

Since the metric differs from a flat space by only a Weyl factor $ds^2 = \frac{\sum_i dx^i dx^i}{\Omega(x)^2}$---and as noted above, the fermions propagate in the entire plane---we know the entanglement entropy between two twist operators at $\bf x_1, x_2$ near the vacuum state is given by
\bea
	S_\text{CFT} ({\bf x}_1, {\bf x}_2) = \frac{c}6 \log \frac{{\bf x_{12}^2 }}{\Omega({\bf x_1}) \Omega({\bf x_2})},
\eea
where ${\bf x}_{12} = {\bf x}_1 - {\bf x}_2 $. 
Finally, the dilaton field in this coordinate is given by
\bea
    \phi(z, \bar z) =   \phi_r \frac{2\pi}\beta \frac{1+|z|^2}{|1-|z|^2|}.
\eea

\section{Haar random unitary model for TFDs with measurement}
\label{append:random_tfd}

\begin{figure}
    \centering
    \includegraphics[width=0.3\textwidth]{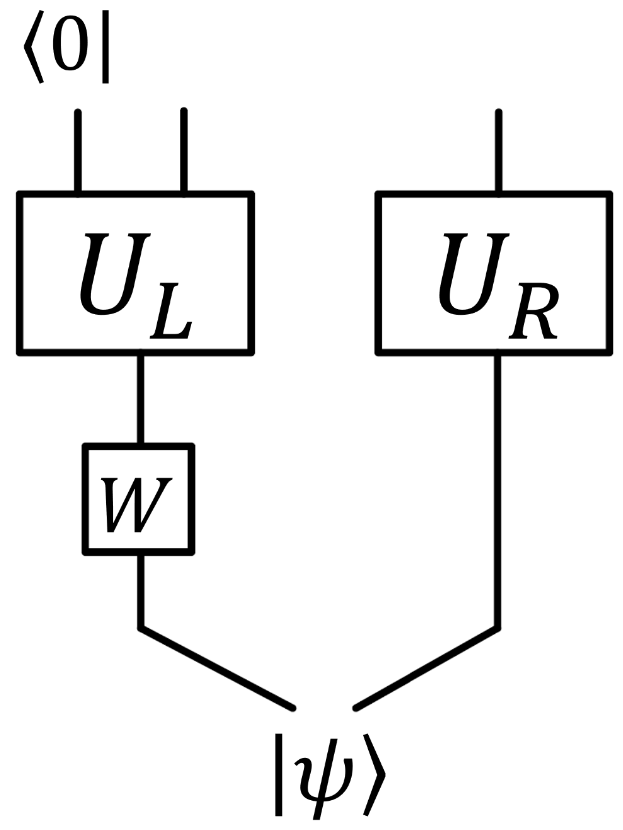}
    \caption{A model with two random unitaries, $U_L$ and $U_R$, representing the two sides each with $N$ qubits.
    $|\psi \rangle$ denotes a bulk state. 
    $M$ out of $N$ qubits in the left side are projected to state $|0\rangle$.
    $W$ denotes an operator acting on the left bulk Hilbert space. 
    }
    \label{fig:random_tfd}
\end{figure}

In this appendix, we construct a simple model that shows the essential physics of measurement-induced entanglement wedge phase transitions in the context of bulk reconstruction. 
Our simple model consists of two random unitaries $U_L$ and $U_R$ with equal dimension $2^N$, drawn from a Haar distribution. 
To allow for a more general setup, we take the bulk state of the system to be $|\psi \rangle$.
As shown in Fig.~\ref{fig:random_tfd}, these random unitaries map the state $|\psi \rangle $ from bulk to boundary (we take the incoming legs at the bottom of each unitary to be the bulk Hilbert space, and the outgoing legs on top of each unitary to be the boundary Hilbert space, which we refer to as the left and right sides for simplicity).

In analogy with the measurements considered in the main text, we measure $M$ qubits in the left boundary Hilbert space. 
In the main text, we found a phase transition occurs when the fraction $m=M/N$ of measured fermions exceeds a critical value $m^*$ (see e.g. Fig.~\ref{fig:summary}). When $m>m^*$, the information that was stored in the left side before measurement is teleported to the right side. 
Here, in this simple model, we ask a similar question. Suppose we have an operator $W$ in the left bulk Hilbert space acting on the state $|\psi \rangle$. How many qubits $M$ need to be measured in the left side such that this operator can be successfully reconstructed in the right boundary Hilbert space? 
In Fig.~\ref{fig:random_tfd}, $M$ out of $N$ qubits in the left side are projected to state $|0\rangle$. 
The measurement outcome is not important here since we are interested in the typical behavior of random unitaries drawn from a Haar distribution.

To answer this question, we employ the technique introduced in Ref.~\cite{akers2022black, Antonini:2022sfm} to give a bound on the reconstruction error.
This calculation is straightforward, and we refer to Ref.~\cite{Antonini:2022sfm} for details. 
Here we just present the final result. 
Let $W_R$ be an operator acting on the right boundary Hilbert space, and $V = \frac1{2^{M/2}} U_L \otimes U_R$ be the linear map from the bulk to the boundary for both sides (here the prefactor comes from measurements). The reconstruction error is defined by
\bea
    || V W |\psi \rangle  -  W_R V |\psi \rangle || \le \epsilon_1,
\eea
where $||.||$ denotes the norm of the state.
The reconstruction error for a typical error modeled by a random unitary drawn from the Haar distribution is bounded by $\epsilon_1 \le \sqrt{\epsilon_2}$, where
\bea
     \epsilon_2 \le 2 \sqrt{\frac{2^{N-M}}{e^{ S^{(2)}_L(\Psi) }}}.
\eea
$S^{(2)}_L(\Psi) = \Tr_R \left[ \Tr_L(\Psi)^2 \right]$ is the pre-measurement Renyi-2 entanglement entropy between left and right side evaluated for the state $\Psi = |\psi \rangle \langle \psi |$. 
This bound intuitively states that the reconstruction would be successful if the measurement is implemented on a subset of fermions large enough that the entanglement of the remaining $N-M$ unmeasured qubits is much smaller than the pre-measurement entanglement between the two sides, i.e. the entanglement resource.
With this understanding, we can give an estimate of the critical value of $m$ by solving $\frac{2^{N-M}}{e^{ S^{(2)}_L(\Psi) }} = 1$. The transition point is then given by 
\bea
    m^* = 1 - \frac{S^{(2)}_L(\Psi)}{N \log 2}.
\eea
Roughly speaking, this means that the critical point is determined by the pre-measurement entanglement between the two sides: the greater the entanglement, the fewer number of qubits needs to be measured to achieve the transition.

\section{Bound on teleportation fidelity}
\label{append:fidelity}

In this appendix, we are going to relate the bound on teleportation fidelity to the left-right correlation functions studied in Section~\ref{sec:teleportation}. 
Our discussion is similar to that of Ref.~\cite{schuster2022many}.
We will use diagrams to represent states and amplitudes. Assuming two copies of the same Hilbert space, 
each  consisting of $N/2$ qubits, we define some relevant states and projection operators as follows,
\bea
\begin{aligned}
    \figbox{0.4}{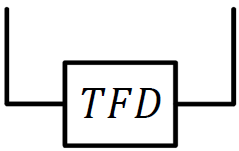} &= \frac1{(\Tr[e^{-\beta H}])^{1/2}}\sum_{n=1}^{N/2} e^{-\beta E_n} | E_n \rangle \otimes |E_n \rangle, \\ \figbox{0.4}{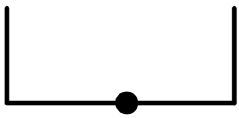} &= \frac1{d^{1/2}}\sum_{i=1}^{d} | i \rangle \otimes |i \rangle, \\
    \figbox{0.35}{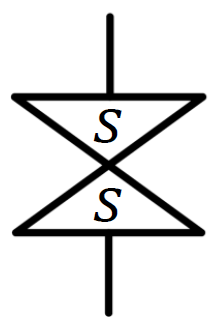} &= \bigotimes_{k=1}^{N/2} | s_k \rangle \langle s_k |,
\end{aligned}
\eea
where $H$ denotes the Hamiltonian and $|E_n \rangle$ denotes the energy eigenstates. 
$| s_k \rangle, s_k = \pm $ are two possible states of a qubit, and $ |s_k \rangle \langle s_k |$ denotes a projection operator. 
In this appendix, we use $s_k$ to denote the measurement outcome (note the change in notation from Section~\ref{sec:teleportation}, where we used $l_k$ to denote the measurement outcome).

The state teleportation protocol conditioned on measurement outcome $s_k$ is given by
\bea \label{eq:teleport_diagram}
    \figbox{0.4}{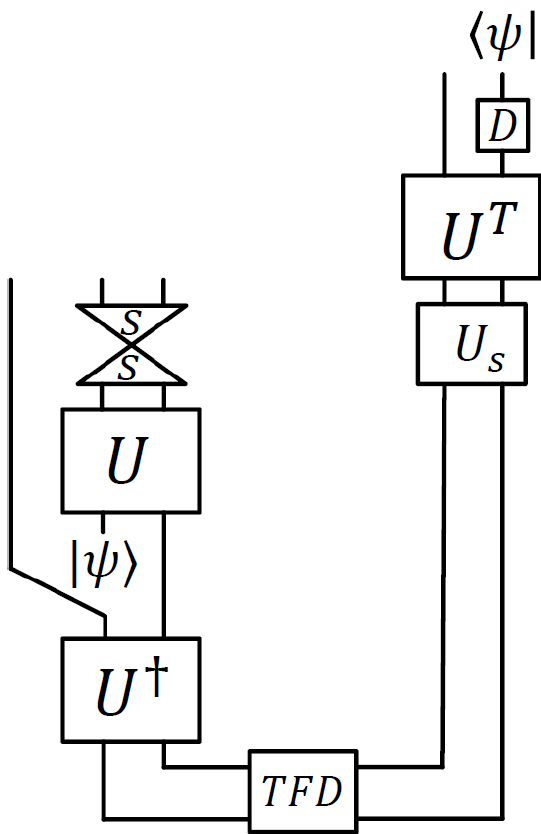}  
\eea
where $U$ is a time evolution operator and $U_s$ is a unitary operator dependent on the measurement outcome, which we refer to as a decoding operator.
$D$ is another decoding operator at the end of the teleportation.
Thus, the teleportation protocol consists of three parts:
\begin{enumerate}
    \item At time $t = -t_0$, a state $|\psi \rangle$ is inserted in the left side;
    \item At time $t = 0$, a projective measurement is performed in the left side and a decoding operator is implemented in the right side according to the measurement outcome;
    \item At time $t = t_0$, another decoding operator $D$ is performed in the right side, and the state is teleported to the right side.
\end{enumerate}
At time $t=0$, the measurement-decoding protocol should be understood as a quantum channel, which can be defined by a Kraus operator,
\bea
    K_{\textbf s} = \left( \otimes_k |s_k \rangle \langle s_k | \right) \otimes U_{\textbf s}, \quad \sum_{\textbf s} K_{\textbf s}^\dag K_{\textbf s}=1, 
\eea
where the decoding operator we perform on the right side depends on the measurement outcome on the left.
Conditioned on the measurement outcome, the full protocol is given in~(\ref{eq:teleport_diagram}).

The state teleportation fidelity is related to the fidelity of the distillation of EPR pairs~\cite{yoshida2019disentangling},
\bea
    \figbox{0.37}{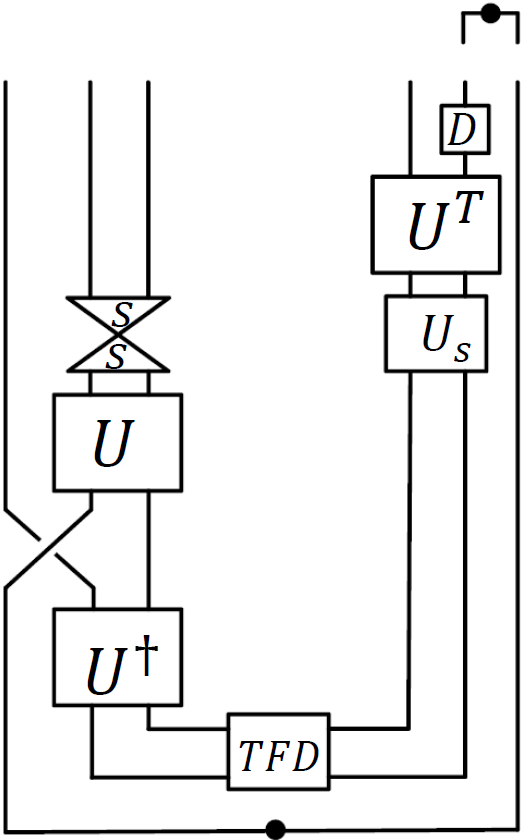}
\eea
Here, instead of inserting a state, a swap between two qubits is performed in the right side, and finally a distillation of EPR states between the swap qubit is implemented. 
The average fidelity of state teleportation is related to the fidelity of EPR distillation by, 
\bea
    \langle F_\psi \rangle = \frac{d}{d+1} F_{EPR} + \frac1{d+1},
\eea
where $d$ is the dimension of the state Hilbert space, and $\langle \cdot \rangle$ implies an average over all states.

The fidelity for distillation of an EPR state is
\bea
    F_{EPR} = \sum_{\textbf{s}_1, \textbf{s}_2}  \figbox{0.6}{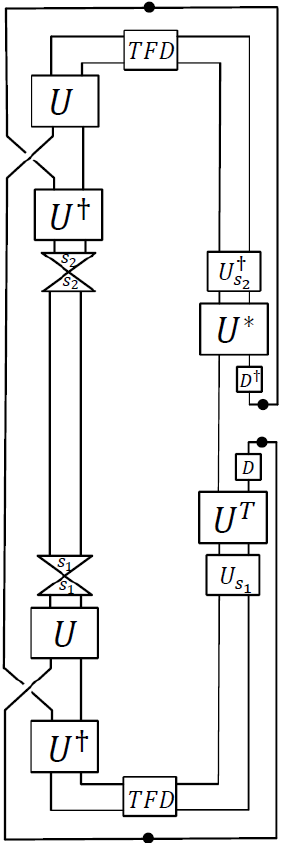} = \frac1{d^4}\sum_{\textbf{s}_1,\textbf{s}_2} \sum_{Q_1, Q_2} \figbox{0.6}{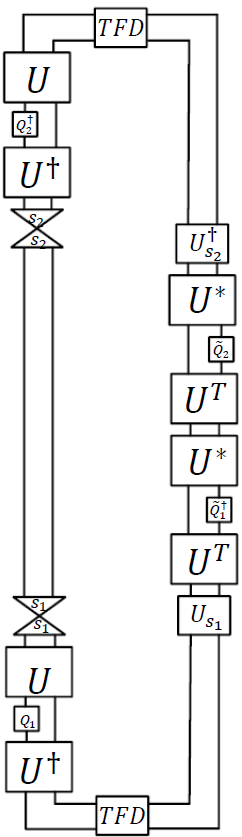} 
\eea
where in the first diagram, the projection operator enforces $\textbf s_1 = \textbf s_2$. 
The sum over all possible measurement outcomes occurs because the full fidelity is given by the sum of fidelities conditioned on each measurement outcome.
To get the second diagram, the swap operator is replaced by the following identity,
\bea
    \figbox{0.5}{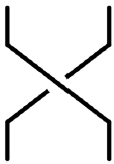} = \frac1{d} \sum_Q \figbox{0.5}{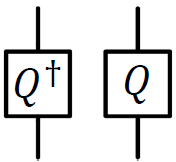}
\eea
where $Q$ is a basis operator living in the $d$ dimensional Hilbert space. 
For instance for $d=2$, $Q = \mathds 1, X, Y, Z$ the identity and three Pauli matrices.
Then we move one of the operators to the right side, and define $\tilde Q = D^\dag Q D $.
In the case of a qubit, the decoding operator is $D = Y$~\cite{brown2019quantum,nezami2021quantum,schuster2022many}. 
For the SYK model we consider in the main text, the decoding operator is trivial $D=\mathds 1$, and so we neglect it in the main text.

Thus, the distillation fidelity is lower bounded by 
\bea
    F_{EPR} = \frac1{d^4}\sum_{\textbf{s}_1, \textbf{s}_2} \sum_{Q_1, Q_2} \figbox{0.62}{Figures/fig_F_EPR_operator.png} \ge \frac1{d^4}\sum_{\textbf{s}_1, \textbf{s}_2} \sum_{Q_1, Q_2} \figbox{0.6}{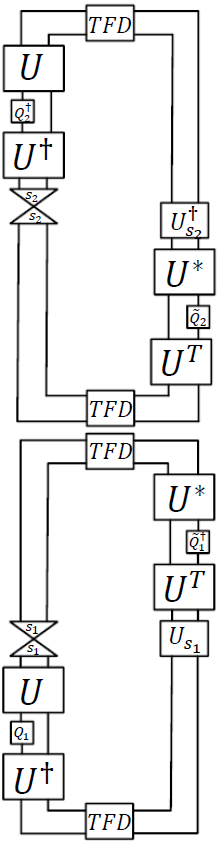} .
\eea
Namely,
\bea
    F_{EPR} \ge \left| \sum_{\textbf s, Q} \frac1{d^2} C_{\textbf s,Q}\right|^2,
\eea
where the left-right correlation function $C_{\textbf{s},Q}$ is given by
\bea
\begin{aligned}
    & C_{\textbf{s},Q} = \langle TFD | \tilde Q_R^\dag(t_0) \left[ \left( \otimes_k |s_k \rangle \langle s_k | \right) \otimes U_{\textbf s} \right] Q_L(-t_0) |TFD \rangle, \\
    & Q_L( -t_0) = (U_L Q U^\dag_L) \otimes \mathds 1, \quad \tilde Q_R(t_0) = \mathds 1 \otimes (U_R^\dag \tilde Q U_R),
\end{aligned}
\eea
where $U_{L,R}$ denotes the time evolution operator in the left and right side. 
Thus, the left-right correlation function is closely related to the fidelity bound.

\section{Symmetries of the twisted correlation function}
\label{append:correlation_function}

In this appendix, we discuss the symmetries of the twisted correlation function, which are used to simplify the calculations in the main text by reducing to a fundamental domain.
Recall the twisted correlation function is defined as
\bea
    && G_\chi(s_1, s_2) = 
    \\ && \frac{\langle L_\textbf{l} |e^{- \beta H}   \mathcal T \left[ O(\frac\beta2)   \left( \ba{cccc} \psi_{2k-1}(s_1) \psi_{2k-1}(s_2) & i l_k \psi_{2k-1}(s_1) \psi_{2k}(2\beta - s_2) \\ i l_k \psi_{2k}(2\beta - s_1) \psi_{2k-1}(s_2) &  - \psi_{2k}(2\beta - s_1) \psi_{2k}(2\beta - s_2)\ea\right) \right] | L_\textbf{l} \rangle}{\langle L_{\textbf l} |  e^{- \beta H} \mathcal T \left[ O(\frac\beta2) \right]| L_\textbf{l} \rangle }  \nn
\eea
where $\psi(\tau ) = e^{H \tau } \psi e^{- H \tau} $ is the imaginary time evolution, and $ O(\frac{\beta}2) = e^{-i \frac{\theta}q \sum_k l_k i \psi_{2k-1}(\frac\beta2) \psi_{2k}(\frac\beta2)} $ is the decoding operator. 
Because the Majorana field $\chi_k$ satisfies conventional anti-periodic boundary conditions, we have
\bea \label{eq:symm1}
    G_\chi(s_1, s_2) = - G_\chi(s_2, s_1), \quad G_\chi(s_1 , s_2) = - G_\chi(2\beta + s_1, s_2).
\eea

The twisted correlation function is real if $\theta$ is a purely imaginary number, $G_\chi(s_1 , s_2) = G_\chi(s_1, s_2)^*$. 
This is because we can find a real representation for all operators, including $O(\frac\beta2)$. 
Thus, we take $\theta$ imaginary, then analytically continue it to a real number.
If we take the complex conjugate of the twisted correlation function, we get
\bea \label{eq:symm2}
\begin{aligned}
    & G_\chi(s_1, s_2)^*  \\
    & = \frac{\langle L_\textbf{l} |  \mathcal T \left[ O(-\frac\beta2)   \left( \ba{cccc} \psi_{2k-1}(-s_2) \psi_{2k-1}(-s_1) & -i l_k \psi_{2k}(-2\beta + s_2) \psi_{2k-1}(-s_1)  \\ -i l_k \psi_{2k-1}(-s_2) \psi_{2k}(-2\beta + s_1)  &  - \psi_{2k}(-2\beta + s_2) \psi_{2k}(-2\beta + s_1)\ea\right) \right] e^{- \beta H} | L_\textbf{l} \rangle}{\langle L_{\textbf l} |   \mathcal T \left[ O(-\frac\beta2) \right] e^{- \beta H}  | L_\textbf{l} \rangle }    \\
    & = \frac{\langle L_\textbf{l} |  e^{- \beta H} \mathcal T \left[ O(\frac\beta2)   \left( \ba{cccc} \psi_{2k-1}(\beta-s_2) \psi_{2k-1}(\beta-s_1) & -i l_k \psi_{2k}(-\beta + s_2) \psi_{2k-1}(\beta-s_1)  \\ -i l_k \psi_{2k-1}(\beta-s_2) \psi_{2k}(-\beta + s_1)  &  - \psi_{2k}(-\beta + s_2) \psi_{2k}(-\beta + s_1)\ea\right) \right] | L_\textbf{l} \rangle}{\langle L_{\textbf l} |  e^{- \beta H} \mathcal T \left[ O(\frac\beta2) \right]   | L_\textbf{l} \rangle }    \\
    & = G_\chi(\beta - s_2, \beta - s_1),
    \end{aligned}
\eea
where we have used $[\psi(\tau)]^\dag = \psi(-\tau)$ to get the first line. 
In the second line, we bring the operator $e^{-\beta H}$ from the right to the left by shifting the time by $\beta$ for all operators. 
The last line can be obtained by recalling the definition of $\chi_k$ and its anti-periodic boundary conditions. 

Moreover, the twisted correlation function is invariant under $\psi_{2k-1} \rightarrow \psi_{2k}$, $\psi_{2k} \rightarrow - \psi_{2k-1}$. 
The large-$N$ action is invariant under $O(N)$ transformations, so it is invariant under this subgroup. 
Now we should check the measurement-decoding channel under this transformation. 
Both the projected state $|L_\textbf{l} \rangle$ and the decoding operator $O(\frac\beta2)$ are invariant because $-2i\psi_{2k-1} \psi_{2k}$ is invariant. 
Therefore, the twisted correlation function is invariant, which leads to the following properties
\bea \label{eq:symm3}
    G_\chi(s_1, s_2) = - G_\chi(2\beta-s_1, 2\beta-s_2) =  G_\chi(2\beta-s_2, 2\beta-s_1).
\eea

In summary, (\ref{eq:symm1}), (\ref{eq:symm2}), and (\ref{eq:symm3}) lead to the following (anti-)reflection conditions
\bea
\begin{aligned}
    & \text{anti-reflection about} \quad s_1 - s_2 =0: \quad G_\chi(s_1 ,s_2) = - G_\chi(s_2, s_1), \\
    & \text{reflection about} \quad s_1 - s_2 =\beta: \quad G_\chi(s_1,s_2) = G_\chi(s_2, s_1 - \beta), \\
    & \text{reflection about} \quad s_1 - s_2 = - \beta: \quad G_\chi(s_1, s_2) = G_\chi(s_2 - \beta, s_1 + \beta), \\
    & \text{reflection about} \quad s_1 + s_2 = 2\beta: \quad G_\chi(s_1 , s_2) = G_\chi(2\beta -s_2, 2\beta - s_1), \\
    & \text{reflection about} \quad s_1 + s_2 = 2\beta: \quad G_\chi(s_1, s_2) = G_\chi(\beta -s_2, \beta- s_1), \\
    & \text{reflection about} \quad s_1 + s_2 = \frac32 \beta: \quad G_\chi(s_1, s_2) = G_\chi(\frac32 \beta - s_2, \frac32 \beta - s_1).
\end{aligned}
\eea
They are illustrated in Fig.~\ref{fig:fundamental_domain}. 
With the help of these symmetries, we can reduce the calculation from the full domain to the fundamental domain $A$, $B$, $C$ and $D$, as discussed in the main text.

\bibliographystyle{jhep}
\bibliography{reference}

\providecommand{\href}[2]{#2}\begingroup\raggedright\begin{thebibliography}{10}

\bibitem{Maldacena:1997re}
J.~M. Maldacena, \emph{{The Large N limit of superconformal field theories and
  supergravity}}, \href{http://dx.doi.org/10.1023/A:1026654312961}{\emph{Adv.
  Theor. Math. Phys.} {\bf 2} (1998) 231--252},
  [\href{https://arxiv.org/abs/hep-th/9711200}{{\tt hep-th/9711200}}].

\bibitem{Witten:1998qj}
E.~Witten, \emph{{Anti-de Sitter space and holography}},
  \href{http://dx.doi.org/10.4310/ATMP.1998.v2.n2.a2}{\emph{Adv. Theor. Math.
  Phys.} {\bf 2} (1998) 253--291},
  [\href{https://arxiv.org/abs/hep-th/9802150}{{\tt hep-th/9802150}}].

\bibitem{Gubser:1998bc}
S.~S. Gubser, I.~R. Klebanov and A.~M. Polyakov, \emph{{Gauge theory
  correlators from noncritical string theory}},
  \href{http://dx.doi.org/10.1016/S0370-2693(98)00377-3}{\emph{Phys. Lett. B}
  {\bf 428} (1998) 105--114}, [\href{https://arxiv.org/abs/hep-th/9802109}{{\tt
  hep-th/9802109}}].

\bibitem{Aharony:1999ti}
O.~Aharony, S.~S. Gubser, J.~M. Maldacena, H.~Ooguri and Y.~Oz, \emph{{Large N
  field theories, string theory and gravity}},
  \href{http://dx.doi.org/10.1016/S0370-1573(99)00083-6}{\emph{Phys. Rept.}
  {\bf 323} (2000) 183--386}, [\href{https://arxiv.org/abs/hep-th/9905111}{{\tt
  hep-th/9905111}}].

\bibitem{Ryu2006a}
S.~Ryu and T.~Takayanagi, \emph{{Aspects of Holographic Entanglement Entropy}},
  \href{http://dx.doi.org/10.1088/1126-6708/2006/08/045}{\emph{JHEP} {\bf 08}
  (2006) 045}, [\href{https://arxiv.org/abs/hep-th/0605073}{{\tt
  hep-th/0605073}}].

\bibitem{Ryu2006b}
S.~Ryu and T.~Takayanagi, \emph{Holographic derivation of entanglement entropy
  from the anti--de sitter space/conformal field theory correspondence},
  {\emph{Physical review letters} {\bf 96} (2006) 181602}.

\bibitem{Hubeny:2007xt}
V.~E. Hubeny, M.~Rangamani and T.~Takayanagi, \emph{{A Covariant holographic
  entanglement entropy proposal}},
  \href{http://dx.doi.org/10.1088/1126-6708/2007/07/062}{\emph{JHEP} {\bf 07}
  (2007) 062}, [\href{https://arxiv.org/abs/0705.0016}{{\tt 0705.0016}}].

\bibitem{Swingle:2009bg}
B.~Swingle, \emph{{Entanglement Renormalization and Holography}},
  \href{http://dx.doi.org/10.1103/PhysRevD.86.065007}{\emph{Phys. Rev. D} {\bf
  86} (2012) 065007}, [\href{https://arxiv.org/abs/0905.1317}{{\tt
  0905.1317}}].

\bibitem{VanRaamsdonk:2010pw}
M.~Van~Raamsdonk, \emph{{Building up spacetime with quantum entanglement}},
  \href{http://dx.doi.org/10.1142/S0218271810018529}{\emph{Gen. Rel. Grav.}
  {\bf 42} (2010) 2323--2329}, [\href{https://arxiv.org/abs/1005.3035}{{\tt
  1005.3035}}].

\bibitem{Maldacena:2013xja}
J.~Maldacena and L.~Susskind, \emph{{Cool horizons for entangled black holes}},
  \href{http://dx.doi.org/10.1002/prop.201300020}{\emph{Fortsch. Phys.} {\bf
  61} (2013) 781--811}, [\href{https://arxiv.org/abs/1306.0533}{{\tt
  1306.0533}}].

\bibitem{Engelhardt:2014gca}
N.~Engelhardt and A.~C. Wall, \emph{{Quantum Extremal Surfaces: Holographic
  Entanglement Entropy beyond the Classical Regime}},
  \href{http://dx.doi.org/10.1007/JHEP01(2015)073}{\emph{JHEP} {\bf 01} (2015)
  073}, [\href{https://arxiv.org/abs/1408.3203}{{\tt 1408.3203}}].

\bibitem{Dong:2016eik}
X.~Dong, D.~Harlow and A.~C. Wall, \emph{{Reconstruction of Bulk Operators
  within the Entanglement Wedge in Gauge-Gravity Duality}},
  \href{http://dx.doi.org/10.1103/PhysRevLett.117.021601}{\emph{Phys. Rev.
  Lett.} {\bf 117} (2016) 021601},
  [\href{https://arxiv.org/abs/1601.05416}{{\tt 1601.05416}}].

\bibitem{Harlow:2016vwg}
D.~Harlow, \emph{{The Ryu\textendash{}Takayanagi Formula from Quantum Error
  Correction}},
  \href{http://dx.doi.org/10.1007/s00220-017-2904-z}{\emph{Commun. Math. Phys.}
  {\bf 354} (2017) 865--912}, [\href{https://arxiv.org/abs/1607.03901}{{\tt
  1607.03901}}].

\bibitem{Antonini:2022sfm}
S.~Antonini, G.~Bentsen, C.~Cao, J.~Harper, S.-K. Jian and B.~Swingle,
  \emph{{Holographic measurement and bulk teleportation}},
  \href{https://arxiv.org/abs/2209.12903}{{\tt 2209.12903}}.

\bibitem{numasawa2016epr}
T.~Numasawa, N.~Shiba, T.~Takayanagi and K.~Watanabe, \emph{Epr pairs, local
  projections and quantum teleportation in holography}, {\emph{Journal of High
  Energy Physics} {\bf 2016} (2016) 1--52}.

\bibitem{Sachdev:1992fk}
S.~Sachdev and J.~Ye, \emph{Gapless spin fluid ground state in a random,
  quantum heisenberg magnet},
  \href{http://dx.doi.org/10.1103/PhysRevLett.70.3339}{\emph{Phys. Rev. Lett.}
  {\bf 70} (1993) 3339}, [\href{https://arxiv.org/abs/cond-mat/9212030}{{\tt
  cond-mat/9212030}}].

\bibitem{kitaev}
A.~Kitaev, ``A simple model of quantum holography.''
  \url{http://online.kitp.ucsb.edu/online/entangled15/}, Feb. 12, April 7, and
  May 27, 2015.

\bibitem{maldacena2016remarks}
J.~Maldacena and D.~Stanford, \emph{Remarks on the sachdev-ye-kitaev model},
  {\emph{Physical Review D} {\bf 94} (2016) 106002}.

\bibitem{maldacena2016conformal}
J.~Maldacena, D.~Stanford and Z.~Yang, \emph{Conformal symmetry and its
  breaking in two-dimensional nearly anti-de sitter space}, {\emph{Progress of
  Theoretical and Experimental Physics} {\bf 2016} (2016) }.

\bibitem{maldacena2003eternal}
J.~Maldacena, \emph{Eternal black holes in anti-de sitter}, {\emph{Journal of
  High Energy Physics} {\bf 2003} (2003) 021}.

\bibitem{Maldacena:2018lmt}
J.~Maldacena and X.-L. Qi, \emph{{Eternal traversable wormhole}},
  \href{https://arxiv.org/abs/1804.00491}{{\tt 1804.00491}}.

\bibitem{Su:2020zgc}
V.~P. Su, \emph{{Variational preparation of the thermofield double state of the
  Sachdev-Ye-Kitaev model}},
  \href{http://dx.doi.org/10.1103/PhysRevA.104.012427}{\emph{Phys. Rev. A} {\bf
  104} (2021) 012427}, [\href{https://arxiv.org/abs/2009.04488}{{\tt
  2009.04488}}].

\bibitem{Teitelboim:1983ux}
C.~Teitelboim, \emph{{Gravitation and Hamiltonian Structure in Two Space-Time
  Dimensions}},
  \href{http://dx.doi.org/10.1016/0370-2693(83)90012-6}{\emph{Phys. Lett. B}
  {\bf 126} (1983) 41--45}.

\bibitem{Jackiw:1984je}
R.~Jackiw, \emph{{Lower Dimensional Gravity}},
  \href{http://dx.doi.org/10.1016/0550-3213(85)90448-1}{\emph{Nucl. Phys. B}
  {\bf 252} (1985) 343--356}.

\bibitem{kourkoulou2017pure}
I.~Kourkoulou and J.~Maldacena, \emph{Pure states in the syk model and nearly-$
  ads\_2 $ gravity}, {\emph{arXiv preprint arXiv:1707.02325} (2017) }.

\bibitem{Antonini:2021xar}
S.~Antonini and B.~Swingle, \emph{{Holographic boundary states and
  dimensionally reduced braneworld spacetimes}},
  \href{http://dx.doi.org/10.1103/PhysRevD.104.046023}{\emph{Phys. Rev. D} {\bf
  104} (2021) 046023}, [\href{https://arxiv.org/abs/2105.02912}{{\tt
  2105.02912}}].

\bibitem{cardy1989boundary}
J.~L. Cardy, \emph{Boundary conditions, fusion rules and the verlinde formula},
  {\emph{Nuclear Physics B} {\bf 324} (1989) 581--596}.

\bibitem{nezami2021quantum}
S.~Nezami, H.~W. Lin, A.~R. Brown, H.~Gharibyan, S.~Leichenauer, G.~Salton
  et~al., \emph{{Quantum Gravity in the Lab: Teleportation by Size and
  Traversable Wormholes, Part II}},
  \href{https://arxiv.org/abs/2102.01064}{{\tt 2102.01064}}.

\bibitem{chandrasekaran2022quantum}
V.~Chandrasekaran and A.~Levine, \emph{Quantum error correction in syk and bulk
  emergence}, {\emph{arXiv preprint arXiv:2203.05058} (2022) }.

\bibitem{newqes}
S.~Antonini, B.~Grado-White, S.-K. Jian and B.~Swingle, \emph{Quantum extremal
  surface formula in the syk model}, {\emph{In preparation} }.

\bibitem{gao2017traversable}
P.~Gao, D.~L. Jafferis and A.~C. Wall, \emph{Traversable wormholes via a double
  trace deformation}, {\emph{Journal of High Energy Physics} {\bf 2017} (2017)
  1--25}.

\bibitem{maldacena2017diving}
J.~Maldacena, D.~Stanford and Z.~Yang, \emph{Diving into traversable
  wormholes}, {\emph{Fortschritte der Physik} {\bf 65} (2017) 1700034}.

\bibitem{brown2019quantum}
A.~R. Brown, H.~Gharibyan, S.~Leichenauer, H.~W. Lin, S.~Nezami, G.~Salton
  et~al., \emph{Quantum gravity in the lab: teleportation by size and
  traversable wormholes}, {\emph{arXiv preprint arXiv:1911.06314} (2019) }.

\bibitem{gao2021traversable}
P.~Gao and D.~L. Jafferis, \emph{A traversable wormhole teleportation protocol
  in the syk model}, {\emph{Journal of High Energy Physics} {\bf 2021} (2021)
  1--44}.

\bibitem{zhang2020entanglement}
P.~Zhang, \emph{Entanglement entropy and its quench dynamics for pure states of
  the sachdev-ye-kitaev model}, {\emph{Journal of High Energy Physics} {\bf
  2020} (2020) 1--20}.

\bibitem{Sachdev:2015efa}
S.~Sachdev, \emph{{Bekenstein-Hawking Entropy and Strange Metals}},
  \href{http://dx.doi.org/10.1103/PhysRevX.5.041025}{\emph{Phys. Rev. X} {\bf
  5} (2015) 041025}, [\href{https://arxiv.org/abs/1506.05111}{{\tt
  1506.05111}}].

\bibitem{liu2018quantum}
C.~Liu, X.~Chen and L.~Balents, \emph{Quantum entanglement of the
  sachdev-ye-kitaev models}, {\emph{Physical Review B} {\bf 97} (2018) 245126}.

\bibitem{zhang2020subsystem}
P.~Zhang, C.~Liu and X.~Chen, \emph{Subsystem r{\'e}nyi entropy of thermal
  ensembles for syk-like models}, {\emph{SciPost Physics} {\bf 8} (2020) 094}.

\bibitem{affleck1991universal}
I.~Affleck and A.~W. Ludwig, \emph{Universal noninteger ‘‘ground-state
  degeneracy’’in critical quantum systems}, {\emph{Physical Review Letters}
  {\bf 67} (1991) 161}.

\bibitem{hamilton2006holographic}
A.~Hamilton, D.~Kabat, G.~Lifschytz and D.~A. Lowe, \emph{Holographic
  representation of local bulk operators}, {\emph{Physical Review D} {\bf 74}
  (2006) 066009}.

\bibitem{huang2019eigenstate}
Y.~Huang, Y.~Gu et~al., \emph{Eigenstate entanglement in the sachdev-ye-kitaev
  model}, {\emph{Physical Review D} {\bf 100} (2019) 041901}.

\bibitem{garcia2017analytical}
A.~M. Garc{\'\i}a-Garc{\'\i}a and J.~J. Verbaarschot, \emph{Analytical spectral
  density of the sachdev-ye-kitaev model at finite n}, {\emph{Physical Review
  D} {\bf 96} (2017) 066012}.

\bibitem{schuster2022many}
T.~Schuster, B.~Kobrin, P.~Gao, I.~Cong, E.~T. Khabiboulline, N.~M. Linke
  et~al., \emph{Many-body quantum teleportation via operator spreading in the
  traversable wormhole protocol}, {\emph{Physical Review X} {\bf 12} (2022)
  031013}.

\bibitem{Milekhin:2022bzx}
A.~Milekhin and F.~K. Popov, \emph{{Measurement-induced phase transition in
  teleportation and wormholes}},  \href{https://arxiv.org/abs/2210.03083}{{\tt
  2210.03083}}.

\bibitem{qi2019quantum}
X.-L. Qi and A.~Streicher, \emph{Quantum epidemiology: operator growth, thermal
  effects, and syk}, {\emph{Journal of High Energy Physics} {\bf 2019} (2019)
  1--28}.

\bibitem{yoshida2019disentangling}
B.~Yoshida and N.~Y. Yao, \emph{Disentangling scrambling and decoherence via
  quantum teleportation}, {\emph{Physical Review X} {\bf 9} (2019) 011006}.

\bibitem{miyaji2014boundary}
M.~{Miyaji}, S.~{Ryu}, T.~{Takayanagi} and X.~{Wen}, \emph{{Boundary states as
  holographic duals of trivial spacetimes}},
  \href{http://dx.doi.org/10.1007/JHEP05(2015)152}{\emph{Journal of High Energy
  Physics} {\bf 2015} (May, 2015) 152},
  [\href{https://arxiv.org/abs/1412.6226}{{\tt 1412.6226}}].

\bibitem{takayanagi2011holographic}
T.~Takayanagi, \emph{Holographic dual of a boundary conformal field theory},
  {\emph{Physical review letters} {\bf 107} (2011) 101602}.

\bibitem{fujita2011aspects}
M.~Fujita, T.~Takayanagi and E.~Tonni, \emph{Aspects of ads/bcft},
  {\emph{Journal of High Energy Physics} {\bf 2011} (2011) 1--40}.

\bibitem{penington2020entanglement}
G.~Penington, \emph{Entanglement wedge reconstruction and the information
  paradox}, {\emph{Journal of High Energy Physics} {\bf 2020} (2020) 1--84}.

\bibitem{penington2022replica}
G.~Penington, S.~H. Shenker, D.~Stanford and Z.~Yang, \emph{Replica wormholes
  and the black hole interior}, {\emph{Journal of High Energy Physics} {\bf
  2022} (2022) 1--87}.

\bibitem{Almheiri_2020}
A.~Almheiri, T.~Hartman, J.~Maldacena, E.~Shaghoulian and A.~Tajdini,
  \emph{Replica wormholes and the entropy of hawking radiation},
  \href{http://dx.doi.org/10.1007/jhep05(2020)013}{\emph{Journal of High Energy
  Physics} {\bf 2020} (may, 2020) }.

\bibitem{Horowitz:2003he}
G.~T. Horowitz and J.~M. Maldacena, \emph{{The Black hole final state}},
  \href{http://dx.doi.org/10.1088/1126-6708/2004/02/008}{\emph{JHEP} {\bf 02}
  (2004) 008}, [\href{https://arxiv.org/abs/hep-th/0310281}{{\tt
  hep-th/0310281}}].

\bibitem{akers2022quantum}
C.~Akers and G.~Penington, \emph{Quantum minimal surfaces from quantum error
  correction}, {\emph{SciPost Physics} {\bf 12} (2022) 157}.

\bibitem{akers2022black}
C.~Akers, N.~Engelhardt, D.~Harlow, G.~Penington and S.~Vardhan, \emph{The
  black hole interior from non-isometric codes and complexity}, {\emph{arXiv
  preprint arXiv:2207.06536} (2022) }.

\bibitem{li2019measurement}
Y.~Li, X.~Chen and M.~P. Fisher, \emph{Measurement-driven entanglement
  transition in hybrid quantum circuits}, {\emph{Physical Review B} {\bf 100}
  (2019) 134306}.

\bibitem{skinner2019measurement}
B.~Skinner, J.~Ruhman and A.~Nahum, \emph{Measurement-induced phase transitions
  in the dynamics of entanglement}, {\emph{Physical Review X} {\bf 9} (2019)
  031009}.

\bibitem{chan2019unitary}
A.~Chan, R.~M. Nandkishore, M.~Pretko and G.~Smith, \emph{Unitary-projective
  entanglement dynamics}, {\emph{Physical Review B} {\bf 99} (2019) 224307}.

\bibitem{jian2021measurement}
S.-K. Jian, C.~Liu, X.~Chen, B.~Swingle and P.~Zhang, \emph{Measurement-induced
  phase transition in the monitored sachdev-ye-kitaev model},
  \href{http://dx.doi.org/10.1103/PhysRevLett.127.140601}{\emph{Phys. Rev.
  Lett.} {\bf 127} (Sep, 2021) 140601}.

\bibitem{Bentsen:2021ukm}
G.~S. Bentsen, S.~Sahu and B.~Swingle, \emph{{Measurement-induced purification
  in large-N hybrid Brownian circuits}},
  \href{http://dx.doi.org/10.1103/PhysRevB.104.094304}{\emph{Phys. Rev. B} {\bf
  104} (2021) 094304}, [\href{https://arxiv.org/abs/2104.07688}{{\tt
  2104.07688}}].

\bibitem{kawabata2022dynamical}
K.~Kawabata, A.~Kulkarni, J.~Li, T.~Numasawa and S.~Ryu, \emph{Dynamical
  quantum phase transitions in syk lindbladians}, {\emph{arXiv preprint
  arXiv:2210.04093} (2022) }.

\bibitem{garcia2022keldysh}
A.~M. Garc{\'\i}a-Garc{\'\i}a, L.~S{\'a}, J.~J. Verbaarschot and J.~P. Zheng,
  \emph{Keldysh wormholes and anomalous relaxation in the dissipative
  sachdev-ye-kitaev model}, {\emph{arXiv preprint arXiv:2210.01695} (2022) }.

\bibitem{goto2022entanglement}
K.~Goto, M.~Nozaki, K.~Tamaoka and M.~T. Tan, \emph{Entanglement dynamics of
  the non-unitary holographic channel}, {\emph{arXiv preprint arXiv:2211.03944}
  (2022) }.

\end{thebibliography}\endgroup

\end{document}